\RequirePackage{fix-cm}
\documentclass[superscriptaddress,secnumarabic,amssymb,amsmath,nobibnotes,aps,prd,showkeys,showpacs,nofootinbib,a4paper,epjc3,twocolumn]{svjour3}
\smartqed  
\RequirePackage{graphicx}
\usepackage{graphicx}
\usepackage{placeins}
\usepackage{float}

\RequirePackage{latexsym}
\RequirePackage[numbers,sort&compress]{natbib}
\RequirePackage[colorlinks,citecolor=blue,urlcolor=blue,linkcolor=blue]{hyperref}
\journalname{Eur. Phys. J. C}
\usepackage{graphicx}
\usepackage{dcolumn}
\usepackage{bm}
\usepackage[
scale=0.7, marginratio={1:1, 2:3}, ignoreall,
text={7in,10in},centering,
margin=1.5in,
total={6.5in,8.75in}, top=1.2in, left=0.9in, includefoot,
height=10in,a5paper,hmargin={3cm,0.8in},
]{geometry}
\RequirePackage{graphicx}
\RequirePackage{mathptmx}      
\usepackage{adjustbox}
\RequirePackage{latexsym}
\RequirePackage[colorlinks,citecolor=blue,urlcolor=blue,linkcolor=blue]{hyperref}
\usepackage[english]{babel}
\usepackage{amssymb,amsmath,amsfonts,graphicx, graphics, setspace}
\usepackage{bigints}
\usepackage{anysize}
\usepackage{float}
\usepackage{fancyhdr}
\usepackage{subfigure}
\usepackage{pdflscape}
\usepackage{epsfig}
\usepackage{epsf}
\usepackage{epstopdf}
\usepackage{hyperref} 
\newtheorem{thm}{Theorem}

\newtheorem{lem}[thm]{Lemma}

\usepackage{amsmath}
\usepackage{amsfonts}
\usepackage{amssymb}
\usepackage{subfigure}
\usepackage{epstopdf}
\usepackage{epsf}
\usepackage{url} 
\usepackage{pdflscape} 
\usepackage{amsmath}
\usepackage{amssymb} 
\usepackage{epsfig} 
\usepackage{amsfonts}
\usepackage{graphicx}
\usepackage{times,fancyhdr}
\usepackage{enumitem}
\setlist[enumerate,2]{label=\roman*)}

\def\case#1/#2{\textstyle\frac{#1}{#2}}

\newcommand{\be}{\begin{equation}}
\newcommand{\ee}{\end{equation}}
\newcommand{\ben}{\begin{eqnarray}}
\newcommand{\een}{\end{eqnarray}}

\usepackage{amsmath}
\usepackage{amsfonts}
\usepackage{amssymb}
\usepackage{widetext}
\usepackage{epstopdf}
\usepackage{epsf}

\usepackage[misc,geometry]{ifsym}
\usepackage[usenames,dvipsnames,svgnames,table]{xcolor}

\begin{document}

\title{Time-averaging axion-like interacting scalar fields models}
\author{Saikat Chakraborty \thanksref{e4,addr3,addr5} \and  Esteban Gonz\'alez \thanksref{e2,addr2} \and  {Genly Leon} \thanksref{e1,addr1}  \and Bin Wang \thanksref{e3,addr3,addr4}}
\thankstext{e4}{snilch@yzu.edu.cn}
\thankstext{e1}{genly.leon@ucn.cl}  \thankstext{e2}{esteban.gonzalez@uac.cl}
\thankstext{e3}{wang\_b@sjtu.edu.cn}

\institute{Center for Gravitation and Cosmology, College of Physical Science and Technology, Yangzhou University, Yangzhou 225009, China \label{addr3} \and International Center for Cosmology, Charusat University, Anand 388421, Gujrat, India\label{addr5} \and Direcci\'on de Investigaci\'on y Postgrado, Universidad de Aconcagua, Pedro de Villagra 2265, Vitacura, 7630367 Santiago, Chile \label{addr2} \and  Departamento de  Matem\'aticas,  Universidad  Cat\'olica  del  Norte, Avda. Angamos  0610,  Casilla  1280  Antofagasta,  Chile \label{addr1} 
\and Shanghai Jiao Tong University 
800 Dongchuan RD, Minhang District, Shanghai, 200240, China\label{addr4}
}
\date{\today}

\maketitle

\begin{abstract}
In this paper, we study a cosmological model inspired in the axionic matter with two canonical scalar fields $\phi_1$ and $\phi_2$ interacting through a  term added to its potential.  Introducing novel dynamical variables, and a dimensionless time variable, the resulting dynamical system is studied. The main difficulties arising in the standard dynamical systems approach, where expansion normalized dynamical variables are usually adopted, are due to the oscillations entering the nonlinear system through the Klein-Gordon (KG) equations. This motivates the analysis of the oscillations using methods from the theory of averaging nonlinear dynamical systems. We prove that time-dependent systems, and their corresponding time-averaged versions, have the same late-time dynamics. Then, we study the time-averaged system using standard techniques of dynamical systems. We present numerical simulations as evidence of such behavior.
\end{abstract}


\maketitle

\section{Introduction}

Between 1998 and 1999 the independent projects High-z Supernova Search Team and Supernova Cosmology Project showed results that the universe went through a late-time accelerated expansion phase \cite{Riess1998, Perlmutter1999}. This behavior has been confirmed by many independent observations \cite{Komatsu2011, Planck2018, Eisentein2005}, turning it into the most fascinating puzzle in modern cosmology. The most simple model to describe this behavior of the universe is the $\Lambda$CDM model, in which the universe is currently dominated by two non-interacting fluids called dark matter (DM) and dark energy (DE). The DE component is given by the cosmological constant (CC) $\Lambda$  driving the present epoch of the accelerated expansion of the universe, and the DM component is assumed to have negligible pressure (cold DM or CDM) and it is responsible for the large scale structure formation in the universe \cite{Liddle2004, Planck2014, Planck2016}. That is, the prevailing opinion assumes DE to be a CC  whilst DM is modeled as a nonrelativistic fluid.

Even though the $\Lambda$CDM model is the most favored model by the observations \cite{Eisentein2005, Planck2018, Scolnic2018, Moresco2015}, it has the following drawbacks from the theoretical point of view: 1) the value of the CC is between 60 and 120 orders of magnitude smaller than what it was estimated in Particle Physics. This is known as the CC problem \cite{Weinberg1989},  and 2)  it suffers a ``Cosmological Coincidence Problem'' (CCP). If the energy density of DM evolves in terms of redshift as $\rho_{m} \propto (1+z)^{3}$, and the energy density of CC, $\rho_{\Lambda}=\Lambda$ is constant, why do their energy densities have the same order of magnitude today? and why in the near past ($z_{eq} \lesssim 0.55$), the matter density had dropped to the same value as the DE density? According to the Planck 2018 results \cite{Planck2018} we have the current values $\Omega_{m0}\approx 0.315, \Omega_{\Lambda0}\approx 0.685$, such that $z_{eq}\approx 0.30$.  

The remarkable fact that the energy densities of DE and DM are of the same order of magnitude around the present time seems to indicate that we are living in a very special moment of cosmic history. As we mentioned before, within the standard model where the DE density is constant and the DM density scales with the inverse third power of the cosmic scale factor, this appears to be a coincidence since it requires extremely fine-tuned initial conditions in the early universe. Both in the very early universe and in the far future universe these energy densities differ by many orders of magnitude. CCP problem was first formulated  by Steinhardt \cite{Steinhardt}, and explored in \cite{Zlatev1999,Velten2014}.

An alternative to address these problems is to consider that the DM and DE  can interact with each other (for example, see \cite{Chimento:2000kq,Zimdahl:2001ar,Chimento:2003iea,Chimento:2003sb,Cai:2004dk,Guo:2004vg,Curbelo:2005dh} and a review \cite{Wang2016}), which is a reasonable assumption considering that both fluids currently dominate the energy content of the universe \cite{Brax2006}.

Keeping in mind that the nature of the DM and DE is still an open question, it is difficult to describe these components from physical first principles. In this sense, in different descriptions of the DE that are not the CC can be treated as a fluid (for example Chaplygin gas \cite{Sen2005}), vectors field (for example Multi-Proca vector DE \cite{Gomez2021}), scalar field (for example k-essence \cite{Scherrer2004}), etc. In the scalar field description of the DM, there are several candidates defined in terms of extension of the Standard Model like the axions, which were introduced to explain the CP violation \cite{Peccei1977}.  

On the other hand, axion-like particles appear naturally when an approximate global symmetry is spontaneously broken, as in four-dimensional models \cite{Masso1995,Masso1997,Coriano2007,Coriano20072}, string theory compactifications \cite{Svrcek2006} and Kaluza-Klein theories \cite{Chang2000}. 

Both axion and axion-like are pseudoscalars fields with the same properties, like the periodicity $\phi\to\phi+2\pi C$ and the shift symmetry $\phi\to\phi+C$, but in the axion model, their mass is related to their decay constant, while for axion-like scalars their mass and their decay constant are not linked \cite{Visinelli:2011wa}, and therefore there is a family of axion-like pseudoscalar particles. Following this line, axionic models naturally appear in stringy motivated cosmology where two axionic scalar fields are employed. In particular, a generalization of this model is an assisted inflationary scenario, namely the so-called `$N$-flation'\cite{Dimopoulos:2005ac} where $N$ scalar fields are employed. The essential idea is that the `inflaton', the scalar degree of freedom that drives the inflation of the early universe, is not a single scalar field but a collection of many axionic scalar fields. More recently, such a model has been employed to construct a late time interacting DM-DE scenario \cite{DAmico:2016jbm}, where the DE and part of the DM that interacts with it were represented by two axionic scalar fields respectively.

It is important to mention that cosmology with several scalar fields has been studied for quite some time now. The theory of slow-roll inflation driven by an arbitrary number of scalar fields interacting through gravity only was developed in \cite{Starobinsky:1985ibc}. The detailed analysis of inflation driven by two massive scalar fields was presented in \cite{Polarski:1992dq}. The case of exponential potential and interaction is well studied in  \cite{Chimento:1998ju, Giacomini:2020grc}. In the pioneering paper  \cite{STAROBINSKY198099}, it was shown Einstein equations, with quantum one-loop contributions of conformally covariant matter fields, do admit a class of nonsingular isotropic homogeneous solutions, which corresponds to an early time de Sitter (inflationary) state. Several models are realizing the inflationary paradigm, most of which include scalar field(s). The scalar field models can be studied using local and global variables, providing a qualitative description of the solution space. In addition, it is possible to provide precise schemes to find analytical approximations of the solutions, as well as exact solutions or solutions in quadrature by choosing various approaches (see for example  \cite{Brans:1961sx,Guth:1980zm,Guth:1980zmb,Horndeski:1974wa,Ibanez:1995zs,Coley:1997nk,Coley:1999mj,Coley:2000zw,Coley:2000yc,Coley:2003tf,Elizalde:2004mq,Capozziello:2005tf,Curbelo:2005dh,Gonzalez:2005ie,Gonzalez:2006cj,Lazkoz:2007mx,Elizalde:2008yf,Leon:2009dt,Leon:2009rc,Leon:2009ce,Leon:2010pu,Basilakos:2011rx,Xu:2012jf,Leon:2012mt,Leon:2013qh,Fadragas:2013ina,Kofinas:2014aka,Leon:2014yua,Paliathanasis:2014yfa,DeArcia:2015ztd,Paliathanasis:2015gga,Leon:2015via,Barrow:2016qkh,Barrow:2016wiy,Cruz:2017ecg,Paliathanasis:2017ocj,Alhulaimi:2017ocb,Dimakis:2017kwx,Giacomini:2017yuk,Karpathopoulos:2017arc,DeArcia:2018pjp,Tsamparlis:2018nyo,Paliathanasis:2018vru,Basilakos:2019dof,VanDenHoogen:2018anx,Leon:2018lnd,Leon:2018skk,Leon:2019mbo,Paliathanasis:2019qch,Leon:2019jnu,Paliathanasis:2019pcl,Barrow:2018zav,Quiros:2019ktw,Shahalam:2019jgs,Nojiri:2019riz,Humieja:2019ywy,Matsumoto:2017gnx,Matsumoto:2015hua,Solomon:2015hja,Harko:2015pma,Minazzoli:2014xua,Skugoreva:2013ooa,Jamil:2012vb,Miritzis:2011zz,Hrycyna:2007gd,Copeland:1993jj,Lidsey:1995np,Copeland:1998fz,Gonzalez:2007ht,Foster:1998sk,Miritzis:2003ym,Giambo:2008ck,Leon:2014bta,Leon:2014rra,Fadragas:2014mra,Dania&Yunelsy,Leon:2008de,Giambo:2009byn,Tzanni:2014eja,vandenHoogen:1999qq,Albrecht:1999rm,Copeland:1997et,Giambo:2019ymx,Cid:2017wtf,Alho:2014fha}). In particular, relevant information about  the properties of the flow associated with an autonomous system of ordinary differential equations can be obtained by using qualitative techniques of dynamical systems. Textbooks related to qualitative theory of differential equations can be found in \cite{Coddington55,Hale69,AP,wiggins,perko,160,Hirsch,165,LaSalle,aulbach}, with some applications in cosmology in \cite{TWE,coleybook,Coley:1999uh,bassemah,LeBlanc:1994qm,Heinzle:2009zb}. 

The main difficulties that arise using standard dynamical system approaches in the study of scalar fields are due to the oscillations entering nonlinear system through the KG equations, motivating the study of theses system using perturbation and averaging methods. 

Perturbations methods and averaging methods were used, for example, in \cite{Rendall:2006cq,Alho:2015cza,Leon:2019iwj,Leon:2020ovw,Llibre:2012zz,Fajman:2020yjb,Leon:2021lct,Leon:2021rcx,Leon:2021hxc}. One idea is  to construct a time-averaged version of the original system, solving it; the oscillations of the original system are smoothed out \cite{Fajman:2020yjb}. This can be achieved  for homogeneous metrics where  the Hubble parameter  $H$   plays the role of a time dependent perturbation parameter which controls the magnitude of the error between the solutions of the  full and the time-averaged problems whenever $H$ is monotonic and sign invariant,  $H$ is positive strictly decreasing in $t$ and $\lim_{t\rightarrow \infty}H(t)=0$ \cite{Fajman:2021cli,Leon:2021lct,Leon:2021rcx}.
Therefore, it is possible to obtain information about the large-time behavior of more complicated systems via an analysis of the simpler averaged system equations using dynamical systems techniques. 

In the references \cite{Leon:2021lct,Leon:2021rcx,Leon:2021hxc} a research program ``Averaging generalized scalar field cosmologies'' was started. It consisted of using asymptotic methods and averaging theory \cite{Verhulst} to explore the solution's space of scalar field cosmologies with generalized harmonic potential in vacuum or minimally coupled to matter. As a difference with   \cite{Leon:2021lct, Leon:2021rcx}, where the Hubble parameter was used as a time-dependent perturbation parameter, in \cite{Leon:2021hxc} systems where Hubble scalar is not monotonic were studied.

In this paper we generalize the program initiated in \cite{Leon:2021lct,Leon:2021rcx,Leon:2021hxc} to two-fluid cosmological models. To this goal, we analyze an axion-like coupled scalar fields model following Ref. \cite{DAmico:2016jbm}, which have recently drawn significant interest among particle cosmologists (see Sec.\ref{model}). 
In this model there are two canonical scalar fields $\phi_1,\,\phi_2$ interacting through a  term added to its potential $V\left(\phi_1,\phi_2\right)= V_{\text{decl.}}\left(\phi_1,\phi_2\right)+     V_{\text{int}}(\phi_1, \phi_2)$. The decoupled part $V_{\text{decl.}}$,  satisfies periodicity property $\phi\to\phi+2\pi C$, shift symmetry $\phi\to\phi+C$, and time-reversal symmetry $t\to-t$. Time-reversal symmetry is not imposed to the coupling term $V_{int}$. In the uncoupled case we identify and characterize the equilibrium points but in the coupled case we need to solve transcendental equations and a little progress is made. 

Methods from the theory of averaging nonlinear dynamical systems allow us to prove that time-dependent systems and their corresponding time-averaged versions have the same late-time dynamics. The main result in this paper is that in the first approximations near $H=0$ of matter density,  normalized scalar field densities, and the phase variables   $\Phi_1$ and $\Phi_2$, defined as $\Phi_i= t \omega_i -\tan^{-1}\left(\frac{\omega_i \Psi_i}{\dot \Psi_i}\right)$, where $\Psi_1$ and $\Psi_2$ are functions of $\phi_1$ and $\phi_2$, through \eqref{lin-phi-Psi}), the full systems and their averaged values (with an averaged function to be properly defined)  have the same limit as $H\rightarrow 0$ (as $\tau\rightarrow \infty$). The averaged values of the phases, denoted by $\bar{\Phi}_i$, are zero, such that, on average, $\bar{\Psi}_i(t)= r_i \sin \left(t \omega_i\right)$ for some constants $r_i$. This is summarized in our Theorem \ref{LFZ1}.  Therefore,  simple time-averaged systems determine the future asymptotic behavior. 

The paper is organized as follow: in section \ref{model} we discuss a coupled effective axion-like model in flat FLRW cosmology. Then, in section \ref{SCT_III} we introduce the model under study. We discuss a theorem based on energy density estimates in section \ref{SECT_III_a}. Numerical solutions confirming the analytical results are discussed in section \ref{SECT_III_b}.  In section \ref{DSA} we make a dynamical system analysis using suitably defined dimensionless dynamical variables and a dimensionless time variable. In section \ref{SECTT:4} we use the averaging techniques. In particular in section \ref{Variation} we apply the variation of constants method to the model, and in section \ref{SSection5} we study the time-averaged system using standard techniques of dynamical systems. In section \ref{discussionsA} we make a discussion of our results. Finally, in section \ref{conclusions} we present our conclusions and discuss further lines of research. 
 Theorem \ref{LFZ1}  is proved in 
\ref{apppa}.
In section \ref{NUmerical} numerical simulations as evidence of such behavior are presented. 

\section{Coupled effective axion-like  model in flat FLRW cosmology}
\label{model}

In this section we introduce the coupled effective axion-like  model presented in \cite{DAmico:2016jbm}, by considering the following Lagrangian density for two  pseudoscalar fields, namely $\theta$ and $\sigma$,
\begin{equation}
    \mathcal{L}=\frac{1}{2}\partial_{\mu}\theta\partial^{\mu}\theta+\frac{1}{2}\partial_{\mu}\sigma\partial^{\mu}\sigma-V(\theta,\sigma), \label{lagrangianaxion}
\end{equation}
where the  fields interact through the non- perturbative potential \cite{Kim2005,Chatzistavrakidis2012,Pajer2013}
\begin{align}
    V(\theta,\sigma)= & \mu_{1}^{4}\left[1-\cos{\left(\frac{\theta}{g_{1}}+\frac{\sigma}{h_{1}}\right)}\right] \nonumber \\ 
    + & \mu_{2}^{4}\left[1-\cos{\left(\frac{\theta}{g_{2}}+\frac{\sigma}{h_{2}}\right)}\right], \label{potentialaxion}
\end{align}
where $g_{1,2}$ and $h_{1,2}$ are different axion-like decay constants, $\mu_{1,2}$ are some scales non-perturbatively generated, and the unity is added in order to eradicate by hand any CC.  We use units in which $M_p^{-2}= 8 \pi G=1$.

The potential \eqref{potentialaxion} is a direct generalization of the single axion-like potential $V(\theta)=\mu^{4}\left[1-\cos{\left(\theta/g\right)}\right]$, which is of interest in the context of natural inflation \cite{Freese1990}, where the inflation is identified with an axion-like particle. This is because the shift symmetry $\theta\rightarrow\theta+C$, $C$ is a constant, protects the flatness of the potential from perturbative corrections. This flatness condition for the inflaton potential is necessary to the inflation take place and this matches with the primordial perturbations indicated by the CMB \cite{Kim2005}.
It is important to note that in \cite{Balakin:2020coe} axionic DM model with a modified periodic potential for pseudoscalar field  \newline $V(\phi, \Phi_*)= \frac{m_A^2 {\Phi_*}^2}{2 \pi^2}\left[1- \cos \left(\frac{2 \pi \phi}{\Phi_*}\right)\right]$ in the framework of  axionic extension of Einstein-aether theory was studied. This periodic potential has minima at $\phi =n \Phi_*, n \in \mathbb{Z}$, whereas maxima when $n \rightarrow m +\frac{1}{2}$ are found. Near the minimum when $\phi =n \Phi_* + \psi$ and $|\psi|$ is small, $V \rightarrow \frac{m_A^2 \psi^2}{2}$ where $m_A$ the axion rests mass.

Variation of \eqref{lagrangianaxion} with respect to $\theta$ and $\sigma$ gives the Klein-Gordon equations
\begin{align}
    & \ddot{\theta}+ 3 H \dot{\theta} + \frac{\mu_1^4 \sin \left(\frac{\theta }{g_1}+\frac{\sigma
   }{h_1}\right)}{g_1}+\frac{\mu_2^4 \sin
   \left(\frac{\theta }{g_2}+\frac{\sigma
   }{h_2}\right)}{g_2}=0, \\
   & \ddot{\sigma}+ 3 H \dot{\sigma} +\frac{\mu_1^4 \sin \left(\frac{\theta }{g_1}+\frac{\sigma
   }{h_1}\right)}{h_1}+\frac{\mu_2^4 \sin
   \left(\frac{\theta }{g_2}+\frac{\sigma
   }{h_2}\right)}{h_2}=0. 
\end{align}
By neglecting higher than linear order terms we have 
\begin{align}
& \ddot{\theta}+3    H \theta= \theta  \left(-\frac{\mu_1^4}{g_1^2}-\frac{\mu_2^4}{g_2^2}\right)+\sigma  \left(-\frac{\mu_1^4}{g_1 h_1}-\frac{\mu_2^4}{g_2 h_2}\right),\\
& \ddot{\sigma} + 3 H \sigma=\theta  \left(-\frac{\mu_1^4}{g_1 h_1}-\frac{\mu_2^4}{g_2 h_2}\right) +\sigma  \left(-\frac{\mu_1^4}{h_1^2}-\frac{\mu_{2}^4}{h_2^2}\right),
\end{align}
near the minimum of the potential $(\theta, \sigma)=(0,0)$.

Potential \eqref{potentialaxion} has a mass matrix in the $(\theta,\sigma)$ basis at $(0,0)$ given by 
\begin{align*}
   \mathcal{M}=\begin{pmatrix}
    \frac{\mu_1^4}{g_1^2}+\frac{\mu_2^4}{g_2^2}  &  \frac{\mu_1^4}{g_1 h_1}+\frac{\mu_2^4}{g_2 h_2}  \\
    \frac{\mu_1^4}{g_1 h_1}+\frac{\mu_2^4}{g_2 h_2} & \frac{\mu_1^4}{h_1^2}+ \frac{\mu_{2}^4}{h_2^2}
   \end{pmatrix},
\end{align*}
 that is not diagonal, and the flat direction exist when the condition on the axion decay constants $g_{1}h_{2}=g_{2}h_{1}$ is fulfilled.
 The conditions $\left(g_1 h_1 \mu_2^4+g_2 h_2 \mu_1^4\right) \neq 0$  and  \newline \begin{small}
  $\left(g_1^2 h_1^2 \mu_2^4 (g_2-h_2) (g_2+h_2)+g_2^2 h_2^2 \mu_1^4
   (g_1-h_1) (g_1+h_1)\right)\neq 0$
 \end{small} are required to have non-zero determinant. 
Following \cite{Kim2005} we assume $g_1=g_2=g$. 
Then, the condition on the axion decay constants $g_{1}h_{2}-g_{2}h_{1}$ to be nearly zero, can be expressed as  $ h_{2}=h_{1}+ \varepsilon$, $\varepsilon \ll 1$.

Now, we impose the conditions
\begin{align*}
 & \frac{\theta}{g }+\frac{\sigma}{h_{1}} = \frac{\psi}{f_1}, \quad \frac{\theta}{g }+\frac{\sigma}{h_{2}}= \frac{\xi}{f_2}+\frac{\psi}{f_{3}},
\end{align*}
for some constants $f_1, f_2, f_3$. First, notice that when $\varepsilon=0$ and $g_2=g_1=g$, that above equations admits no solutions for $\sigma$ and $\theta$. Indeed, the linear matrix $\left(
\begin{array}{cc}
 \frac{1}{g } & \frac{1}{h_1} \\
 \frac{1}{g } & \frac{1}{h_2} \\
\end{array}
\right)$  has determinant $\frac{  h_1-  h_2}{g   h_1 h_2}=0$ if $ h_{2}- h_{1}=0$. 
Assuming $0<h_2-h_1:=\varepsilon\ll 1$, and defining 
\begin{equation*}
    f_{1}=\frac{gh_{1}}{\sqrt{g^{2}+h_{1}^{2}}},\; f_{2}=\frac{h_{2}\sqrt{g^{2}+h_{1}^{2}}}{h_{1}-h_{2}},\; f_{3}=\frac{gh_{2}\sqrt{g^{2}+h_{1}^{2}}}{g^{2}+h_{1}h_{2}},
\end{equation*} we obtain 
\begin{align*}
   & \xi = \frac{h_1 \sigma -g \theta }{\sqrt{g^2+h_1^2}}, \quad \psi = \frac{g \sigma +\theta  h_1}{\sqrt{g^2+h_1^2}},
 \end{align*}
 with inverse
 \begin{align*}
   & \theta = \frac{h_1 \psi -g \xi }{\sqrt{g^2+h_1^2}},\quad \sigma = \frac{g \psi +h_1 \xi }{\sqrt{g^2+h_1^2}}.   
 \end{align*}
Then, the potential \eqref{potentialaxion} takes the form
\begin{align}
    V(\psi,\xi)= & \mu_{1}^{4}\left[1-\cos{\left(\frac{\psi}{f_{1}}\right)}\right] \nonumber\\
    + & \mu_{2}^{4}\left[1-\cos{\left(\frac{\xi}{f_2}+\frac{\psi}{f_{3}}\right)}\right]. \label{potentialrealfull}
\end{align}
where $f_1, f_2, f_3$ were chosen to have two new fields, the heavy field $\psi$  and the light field $\xi$. That is, the field equations near the minimum becomes
\begin{align*}
   & \ddot{\xi} +3 H \xi+ \frac{\xi  \mu_2^4 (h_1-h_2)^2}{h_2^2 \left(g^2+h_1^2\right)} \nonumber \\
   & +\frac{\psi  \mu_2^4 (h_1-h_2) \left(g^2+h_1 h_2\right)}{g h_2^2
   \left(g^2+h_1^2\right)} =0, \\
   & \ddot{\psi}+3 H \psi+\frac{\psi  \left(\frac{\mu_1^4
   \left(g^2+h_1^2\right)^2}{h_1^2}+\frac{\mu_2^4 \left(g^2+h_1 h_2\right)^2}{h_2^2}\right)}{g^2 \left(g^2+h_1^2\right)} \nonumber \\
   & + \frac{\xi  \mu_2^4 (h_1-h_2) \left(g^2+h_1 h_2\right)}{g h_2^2 \left(g^2+h_1^2\right)}=0,
 \end{align*}
 where higher order terms in $\xi$ and $\psi$ were dropped out. 
Note that we can adjust $h_1-h_2$ to make the effective axion decay constant $f_{2}$ arbitrary large, and to make the real scalar field $\psi$  a heavy field, whose evolution is dominated only by the first term in the potential \eqref{potentialrealfull} if $\mu_1\gg \mu_2$ and $f_2\gg 1$, while the real scalar field $\xi$ to be a light field, whose evolution is dominated only by the second term in the potential \eqref{potentialrealfull} with $\psi\approx 0$. 
For $\mu_1\gg \mu_2$
we have 
\begin{align}
   & \ddot{\psi} +  3 H \psi+   m^{2}_{\psi} \psi +  \mathcal{O} \left((h_1-h_2)\right)=0, \\  & \ddot{\xi} +3 H \xi+m^{2}_{\xi}  \xi + \mathcal{O}\left((h_1-h_2)^3\right)=0, \label{Meq17}
\end{align}
as $h_1-h_2\rightarrow 0$, where 
\begin{equation} 
  m^{2}_{\psi}=\left(\frac{1}{g^{2}}+\frac{1}{h_{1}^{2}}\right)(\mu_{1}^{4}+\mu_{2}^{4}) \sim \left(\frac{1}{g^{2}}+\frac{1}{h_{1}^{2}}\right) \mu_{1}^{4}, \label{defofphi1}
\end{equation}
\begin{equation}
 m^{2}_{\xi}=\frac{(h_{1}-h_{2})^{2}}{h_{2}^{2}(g^{2}+h_{1}^{2})}\mu_{2}^{4}.  \label{defofphi2}
\end{equation}
To deduce eq. \eqref{Meq17}, we assume $\psi = \mathcal{O}\left( (h_1-h_2)^2\right) \approx 0.$
Neglecting $\psi$ in the second term of \eqref{potentialrealfull} and renaming $\psi=\phi_1$ and $\xi=\phi_2$, we obtain the decoupled effective potential 
\begin{align}
& V_{\text{decl.}}\left(\phi_1,\phi_2\right)= V_1(\phi_1)+ V_2(\phi_2),\label{pot_phi}
\end{align}
where 
\begin{align}
  & V_1(\phi_1):=   \mu_1^4\left[1-\cos\left(\frac{\phi_1}{f_1}\right)\right], \\
  & V_2(\phi_2):= \mu_2^4\left[1-\cos\left(\frac{\phi_2}{f_2}\right)\right]. 
\end{align}
In this paper, we study a modification of the potential \eqref{pot_phi} generated by the scalar fields $\phi_1$ and $\phi_2$, following the references \cite{DAmico:2016jbm,Kaloper2016}, in which the potential is written as
\begin{align}
& V\left(\phi_1,\phi_2\right)= V_{\text{decl.}}\left(\phi_1,\phi_2\right)+     V_{\text{int}}(\phi_1, \phi_2), \label{pot}
\end{align}
where the interaction term between $\phi_1$ and $\phi_2$ is
\begin{equation}
    V_{\text{int}}\left(\phi_1,\phi_2\right)=\mu_3^4\left[1-\cos\left(\frac{\phi_1}{f_1}-n\frac{\phi_2}{f_2}\right)\right]. \label{V_int}
\end{equation}
The heavy field is $\phi_1$  and the light field is $\phi_2$, which interact through the third term presented in the potential \eqref{pot}. The interaction is turned on when $n\neq 0$. When $n=0$ the first and third terms are merged by replacing $\mu_1^4+\mu_3^4 \rightarrow \mu_1^4$ and potential \eqref{pot_phi} is recovered.

To obtain the minima/maxima of $V(\phi_1, \phi_2)$ we have to solve the transcendental equations
\begin{subequations}
\label{maxima-minima}
\begin{align}
  \frac{\mu_1^4}{f_1}\sin\left(\frac{\phi_1^*}{f_1}\right) +\frac{\mu_3^4}{f_1}\sin\left(\frac{\phi_1^*}{f_1}-n\frac{\phi_2^*}{f_2}\right)=0, \\
-\frac{\mu_2^4}{f_2}\sin\left(\frac{\phi_2^*}{f_2}\right)  +n\frac{\mu_3^4}{f_2}\sin\left(\frac{\phi_1^*}{f_1}-n\frac{\phi_2^*}{f_2}\right)=0. 
\end{align}
\end{subequations}
The second partial derivatives test classifies the point as a local maximum or local minimum. Define the second derivative test discriminant as
\begin{align}
D	& =	\frac{\partial^2 V}{\partial^2 \phi_1} 	\frac{\partial^2 V}{\partial^2 \phi_2}-\left(\frac{\partial^2 V}{\partial\phi_1 \partial \phi_2} \right)^2,	    
\end{align}
then
\begin{enumerate}
    \item If $D>0$ and $\frac{\partial^2 V}{\partial^2 \phi_1} (\phi_1^*,\phi_2^*)>0$ the point is a local minimum. 
    \item If $D>0$ and $\frac{\partial^2 V}{\partial^2 \phi_1} (\phi_1^*,\phi_2^*)<0$, the point is a local maximum.
    
    \item If $D<0$, the point is a saddle point.
    \item If $D=0$, higher-order tests must be used.
\end{enumerate}
The conditions
\begin{align}
   &  f_1^2 f_2^2 \left(\mu_2^4 \mu_3^4+\mu_1^4 \left(\mu_2^4+\mu_3^4 n^2\right)\right)>0, \\
   & f_1^2 \left(\mu_1^4+\mu_3^4\right)>0, \quad f_2^2 \left(\mu_2^4+\mu_3^4
   n^2\right)>0,
\end{align}
implies that the origin is a global minimum of $V(\phi_1, \phi_2)$. 
 
Potential \eqref{pot} has a masses matrix in the $(\phi_1,\phi_2)$ basis at $(0,0)$ given by 
\begin{align*}
\mathcal{M}=\begin{pmatrix}
 \frac{\mu_3^4+\mu_1^4}{f_1^2} & -\frac{n \mu_3^4}{f_1 f_2} \\
 -\frac{n \mu_3^4}{f_1 f_2} & \frac{n^2 \mu_3^4+\mu_2^4}{f_2^2} 
\end{pmatrix},
\end{align*}
 that is not diagonal if $n\neq 0$ in which case the interaction is switched on. For $n=0$ the mass matrix is diagonal and the scalar fields are decoupled (do not interact). The case $\mu_3$ reduces the the decoupled potential \eqref{pot_phi}. 
To diagonalize $\mathcal{M}$ when $n\neq0, \mu_3\neq 0$, we introduce the transformation 
\begin{align}
\label{lin-phi-Psi}
    \begin{pmatrix}
    \Psi_1 \\
    \Psi _2
    \end{pmatrix}= \mathbf{R}     \begin{pmatrix}
    \phi_1 \\
    \phi _2
    \end{pmatrix}, \quad 
\mathbf{R} =  \left(
\begin{array}{cc}
 -\frac{\sqrt{\frac{c^2}{c^2+1}}}{c} & \sqrt{\frac{c^2}{c^2+1}} \\
 c \sqrt{\frac{1}{c^2+1}} & \sqrt{\frac{1}{c^2+1}} \\
\end{array}
\right), 
\end{align}
with determinant $-1$,
such that the linearized KG equations near  $(\phi_1, \phi_2)=(0,0)$
\begin{equation*}
  \begin{pmatrix}
    \ddot{\phi_1} \\
    \ddot{\phi _2}
    \end{pmatrix}+ 3 H     \begin{pmatrix}
    \dot{\phi_1} \\
    \dot{\phi _2}
    \end{pmatrix}+ \mathcal{M}  \begin{pmatrix}
    \phi_1 \\
    \phi _2
    \end{pmatrix}= \begin{pmatrix}
   0\\
   0
    \end{pmatrix},
\end{equation*}
are transformed to 
\begin{equation*}
    \begin{pmatrix}
    \ddot{\Psi_1} \\
    \ddot{\Psi _2}
    \end{pmatrix}+ 3 H     \begin{pmatrix}
    \dot{\Psi_1} \\
    \dot{\Psi _2}
    \end{pmatrix}+ \mathbf{R}  \mathcal{M} \mathbf{R} ^{-1} \begin{pmatrix}
    \Psi_1 \\
    \Psi _2
    \end{pmatrix}= \begin{pmatrix}
   0\\
   0
    \end{pmatrix},
\end{equation*}
Then, by choosing
\begin{equation}
    \label{c}
   c = \frac{f_1^2 \left(\mu_3^4
   n^2+\mu_2^4\right)-f_2^2 \left(\mu_3^4+ \mu_1^4\right)+\sqrt{\Delta}}{2 f_1 f_2 \mu_3^4 n},
\end{equation}
where \newline $\Delta=4 f_1^2 f_2^2 \mu_3^8 n^2+\left(f_2^2
   \left(\mu_3^4+\mu_1^4\right)-f_1^2 \left(\mu_3^4
   n^2+\mu_2^4\right)\right)^2$, 
the mass matrix in the  $(\Psi_1,\Psi_2)$ basis becomes diagonal  
\begin{equation*}
   \mathbf{R}    \mathcal{M} \mathbf{R} ^{-1}= \begin{pmatrix}
   \omega_1^2 & 0 \\
   0 & \omega_2^2
   \end{pmatrix},
\end{equation*}
with 
 \begin{equation}
     \omega_{1,2}^2= \frac{f_1^2 \left(\mu_3^4
   n^2+\mu_2^4\right)+f_2^2 \left(\mu_3^4+\mu_1^4\right)\pm\sqrt{\Delta}}{2 f_1^2 f_2^2}, \label{omega}
 \end{equation}
The masses satisfy $\omega_{1,2}^2 \geq 0$ for $(n|\mu_1|\mu_2|\mu_3)\in \mathbb{R}, f_1\neq 0,  f_2\neq 0$. \newline 
With 
$0\leq \varepsilon: = 4 f_1^2 f_2^2 \left(\mu_2^4 \left( \mu_{3}^4+\mu_1^4\right)+\mu_3^4 n^2 \mu_1^4\right)\ll 1$ we  have $ \omega_1^2 \gg  \omega_2^2$ such that $\Psi_1$ is a massive scalar field and  $\Psi_2$ a lighter one. 
Then we have 
   \begin{equation}
        \omega_1^2 \approx \frac{f_1^2 \left(\mu_3^4
   n^2+\mu_2^4\right)+f_2^2 \left(\mu_3^4+\mu_1^4\right)}{ f_1^2 f_2^2}, \quad
   \omega_2^2 \approx 0,  
   \end{equation}
   as $\varepsilon\rightarrow 0$. 
   That is, we can keep the masses hierarchy in the formulation of \cite{Kim2005}. 
 
\section{The model}
\label{SCT_III}
The complete action of the model is given by 
\begin{equation}
S=S_g(g_{\mu\nu}) + S_{\phi}(\phi_1,\phi_2) + S_m, \nonumber
\end{equation}
where $S_g$ is the Einstein Hilbert action, $S_m$ is the action of the non-interacting barotropic CDM and the interacting axion-like  part of the action is given by,
\begin{align*}
 & S_{\phi}(\phi_1,\phi_2)\nonumber \\
 & =-\int d^4x\sqrt{-g}  \Big(\frac{1}{2}(\partial\phi_1)^2 + \frac{1}{2}(\partial\phi_2)^2  + V(\phi_1,\phi_2)\Big),
\end{align*}
where $V\left(\phi_1,\phi_2\right)$ is defined by \eqref{pot}. 
That is, the ingredients of the model are a heavy field $\phi_1$, and a lighter field $\phi_2$, which interacts through the third term in \eqref{pot}. Finally, $\rho$ corresponds to dust. 

We get the total energy momentum tensor consisting of the non-interacting part of the CDM and the two fields $\phi_1,\,\phi_2$ from the variation of $S_m + S_{\phi}(\phi_1,\phi_2)$ with respect to the metric, that is, 
\begin{equation*}
T_{\mu\nu}=-\frac{2}{\sqrt{-g}}\frac{\delta(S_m+S_{\phi})}{\delta g^{\mu\nu}}\,.
\end{equation*}

The equations governing the evolution of an FLRW universe 
for the potential \eqref{pot},  are as follows.

Friedmann constraint equation is given by:  
\begin{align}
& {3H^2} = \rho + \frac{1}{2}\dot{\phi_1}^2 + \frac{1}{2}\dot{\phi_2}^2 \nonumber \\
&  + 2\mu_1^4\sin^2\left(\frac{\phi_1}{2f_1}\right)   +2\mu_2^4\sin^2\left(\frac{\phi_2}{2f_2}\right) \nonumber \\
& +2\mu_3^4\sin^2\left(\frac{\phi_1}{2f_1}-n\frac{\phi_2}{2f_2}\right)\,. \label{Friedmann}
\end{align}
Raychaudhuri equation is given by: 
\begin{equation}
\label{Raych}
2\dot{H}=-\rho-\dot{\phi_1}^2 - \dot{\phi_2}^2\,. 
\end{equation}
KG equations are given by: 
\begin{subequations}
\label{KG}
\begin{align}
&\ddot{\phi_1}+3H\dot{\phi_1}+\frac{\mu_1^4}{f_1}\sin\left(\frac{\phi_1}{f_1}\right) \nonumber \\ & +\frac{\mu_3^4}{f_1}\sin\left(\frac{\phi_1}{f_1}-n\frac{\phi_2}{f_2}\right)=0, \label{KG1}\\
&\ddot{\phi_2}+3H\dot{\phi_2}+\frac{\mu_2^4}{f_2}\sin\left(\frac{\phi_2}{f_2}\right) \nonumber \\ &  -n\frac{\mu_3^4}{f_2}\sin\left(\frac{\phi_1}{f_1}-n\frac{\phi_2}{f_2}\right)=0. \label{KG2}
\end{align}
\end{subequations}
Continuity equation is given by: 
\begin{equation}\label{cont-eq}
\dot{\rho}+3H\rho=0\,.
\end{equation}
Introducing the auxiliary functions $\chi_1= \dot{\phi_1}$ and $\chi_2= \dot{\phi_2}$ we obtain the dynamical system 
\begin{subequations}
\label{Non_min2}
\begin{align}
\dot{H}=& -\frac{1}{2}\left(\rho+\chi_1^2+ \chi_2^2\right),\label{Rachd2} \\
\dot{\rho}=& -3 H \rho, \\
\dot{\chi_1}=& - 3H \chi_{1}-\frac{\mu_1^4}{f_1}\sin\left(\frac{\phi_1}{f_1}\right) \nonumber \\
& -\frac{\mu_3^4}{f_1}\sin\left(\frac{\phi_1}{f_1}-n\frac{\phi_2}{f_2}\right), \\
\dot{\chi_2}=& -3H\chi_{2}-\frac{\mu_2^4}{f_2}\sin\left(\frac{\phi_2}{f_2}\right) \nonumber \\
& +n\frac{\mu_3^4}{f_2}\sin\left(\frac{\phi_1}{f_1}-n\frac{\phi_2}{f_2}\right), \\
\dot{\phi_1}= & \chi_1,\\
\dot{\phi_2}= & \chi_2,
\end{align}
\end{subequations}
defined on the phase space
\begin{align}
 \mathbf{X}=  & \Bigg\{ (H, \rho, \chi_1, \chi_2, \phi_1, \phi_2)\in \mathbf{R}^6: \nonumber\\
 & {3H^2} = \rho + \frac{1}{2}\chi_1^2 + \frac{1}{2}\chi_1^2 \nonumber \\
&  + 2\mu_1^4\sin^2\left(\frac{\phi_1}{2f_1}\right)   +2\mu_2^4\sin^2\left(\frac{\phi_2}{2f_2}\right) \nonumber \\
& +2\mu_3^4\sin^2\left(\frac{\phi_1}{2f_1}-n\frac{\phi_2}{2f_2}\right) \Bigg\}\,. \label{conds}
\end{align}
The set  $\left\{(H, \rho, \chi_1, \chi_2, \phi_1, \phi_2)\in \mathbf{X}: H=0\right\}$  is invariant for the flow of  \eqref{Non_min2} and $H$ does not change sign.  On the contrary, if there is an orbit with  $H(0)>0$ and $H(t_1)<0$ for some $t_1>0$, this solution passes through the origin violating the existence and uniqueness of the solutions of a $C^1$ flow. 

\subsection{Local energy estimates}
\label{SECT_III_a}
Using local energy estimates, we can prove the following theorem
\begin{thm}
\label{local-estimate}
Let $O^+(x_0)$ be the positive orbit that passes  through the regular point 
 \begin{align}
    x_0\in
\left\{(H, \rho, \chi_1, \chi_2, \phi_1, \phi_2)\in \mathbf{X}: H>0\right\}, \label{interior}
 \end{align} at the time $t=t_0$. Then, we have 
 \begin{align}
&  \lim_{t\rightarrow \infty} \left( \rho(t), \chi_1(t), \chi_2(t)\right)=  (0,0, 0),   \label{theo1_a}
\\
 &    \lim_{t\rightarrow \infty}  \Bigg \{3H(t)^2 -\Bigg[ \mu_1^4\sin^2\left(\frac{\phi_1(t)}{2f_1}\right)   +2\mu_2^4\sin^2\left(\frac{\phi_2(t)}{2f_2}\right) \nonumber \\
& +2\mu_3^4\sin^2\left(\frac{\phi_1(t)}{2f_1}-n\frac{\phi_2(t)}{2f_2}\right)\Bigg]\Bigg\}=0, \label{theo1_b}
 \end{align}
 along the positive orbit $O^+(x_0)$.
\end{thm}
\textbf{Proof.} Let $O^+(x_0)$ be the positive orbit that passes  through the regular point  $x_0$ defined as in \eqref{interior}. 
Since $H$ is positive and decreases along $O^+(x_0)$, the limit $\lim_{t\rightarrow \infty} H(t)$ exists, and it is a non-negative number $\eta$. Furthermore, $H(t)\leq H(t_0)$ for all $t>t_0$. Then, 
 $\rho(t)+\frac{1}{2}\chi_1(t)^2+\frac{1}{2}\chi_2(t)^2+V(\phi_1(t), \phi_2(t))= 3H(t)^2\leq 3 H(t_0)^2$ for all $t>t_0$. All above terms are non-negative,  so it follows that $\rho(t)$, $\frac{1}{2}\chi_1(t)^2$,  $\frac{1}{2}\chi_2(t)^2,$ and \newline $V(\phi_1(t), \phi_2(t))$ are bounded by  $3 H(t_0)^2$  for all $t>t_0$.
Defining the set 
\begin{align}
A=\{(\phi_1, \phi_2)\in \mathbb{R}^2: V(\phi_1, \phi_2)\leq 3 H(t_0)^2\} .   
\end{align}
Then, the orbit $O^+(x_0)$ is such that $(\phi_1, \phi_2)$  remains at the interior of $A$ for all $t>t_0$. By integration of \eqref{Rachd2}, it follows that: 
\begin{equation*}
H(t_0)-H(t)=\int_{t_0}^t \left( \rho(s) +\frac{1}{2} \chi_1(s)^2  +\frac{1}{2} \chi_2(s)^2\right) \, ds.
\end{equation*}
Taking the limit as $t\rightarrow +\infty$, it is obtained 
\begin{align*}
H(t_0)-\eta & = H(t_0) -\lim_{t\rightarrow \infty} H(t) \nonumber \\
& =\int_{t_0}^\infty \left( \rho(s) +\frac{1}{2} \chi_1(s)^2  +\frac{1}{2} \chi_2(s)^2\right) \, ds.
\end{align*}
From this equation, the improper integral is convergent:
\begin{equation*}
\int_{t_0}^\infty \left( \rho(s) +\frac{1}{2} \chi_1(s)^2  +\frac{1}{2} \chi_2(s)^2\right) \, ds<\infty.
   \end{equation*}
Defining $f(t)= \rho(t) +\frac{1}{2} \chi_1(t)^2  +\frac{1}{2} \chi_2(t)^2$, and taking the $t$ derivative, it follows\\
\begin{align}
& \frac{d}{dt} f(t)  =  -\frac{2 \chi_1 \left( \mu_{3}^4 \sin \left(\frac{ \phi_{1}}{f_1}-\frac{n \phi_2}{f_2}\right)+\mu_1^4 \sin \left(\frac{\phi_1}{f_1}\right)\right)}{f_1} \nonumber \\
   & +\frac{2 \chi_2
   \left(\mu_3^4 n \sin \left(\frac{\phi_1}{f_1}-\frac{n \phi_2}{f_2}\right)-\mu_2^4 \sin \left(\frac{\phi_2}{f_2}\right)\right)}{f_2} \nonumber \\
   & -3 H \rho -6 H \chi_1^2-6 H
   \chi_2^2 .  
\end{align} 
Using the results that $\rho(t), \frac{1}{2}\chi_1(t)^2$ and $\frac{1}{2}\chi_2(t)^2$ are bounded by $3 H(t_0)^2$ and $|\sin(\cdot)|\leq 1$  for all $t>t_0$, it follows 
\begin{align}
\left|\frac{d}{dt} f(t)\right|\leq    & \frac{2 \sqrt{6}H_0 (\mu_1^4+ \mu_3^4)}{f_1}  + \frac{2 \sqrt{6}H_0 (\mu_2^4+ n \mu_3^4)}{f_2} \nonumber\\
   & + 3 \sqrt{3} H_0^2 + 24 H_0^3, 
\end{align}
for all $t>t_0$ along the positive orbit $O^+(x_0)$.  Summarizing, the function $f(t)$ is non-negative, it has a bounded derivative through the orbit $O^+(x_0)$, and  \newline $\int_{t_0}^\infty f(s) \, ds$ is convergent.  Hence, $\lim_{t\rightarrow \infty} f(t)=0$, from which, along with the non-negativeness of each term of $f(t)$,  we have $\lim_{t\rightarrow \infty} \left( \rho, \chi_1, \chi_2\right)=(0,0, 0)$. Finally, from \eqref{conds},
\begin{align}
3\eta^2  
= \lim_{t\rightarrow \infty} \Bigg[ & 2\mu_1^4\sin^2\left(\frac{\phi_1(t)}{2f_1}\right)   +2\mu_2^4\sin^2\left(\frac{\phi_2(t)}{2f_2}\right) \nonumber \\
& +2\mu_3^4\sin^2\left(\frac{\phi_1(t)}{2f_1}-n\frac{\phi_2(t)}{2f_2}\right)\Bigg]. 
\end{align}
$\qed $

\subsection{Numerical solutions}
\label{SECT_III_b}

We integrate the equations \eqref{Friedmann}, \eqref{Raych},  \eqref{KG} using the redshift $z$ instead of the cosmic time $t$ as the independent variable. Quantities $t$ and $z$ are related through the expression 
\begin{align}
\frac{df}{dt}=-H_0 E (1+z)\frac{df}{dz}.    
\end{align}
Moreover, 
\begin{small}
\begin{align}
& \frac{d^2 f}{dt^2} =   H_0^2 E^2 \left[(1+z)^2 \frac{d^2 f}{dz^2} +  (1+z) (q+2) \frac{df}{dz}\right],\\
&  \frac{d^2 f}{dt^2}+3 H \frac{df}{dt}  =  H_0^2 E^2 \left[(1+z)^2 \frac{d^2 f}{dt^2} +  (1+z) (q-1) \frac{df}{dz}\right], 
 \end{align}
 \end{small}
 where the deceleration parameter
 \begin{align}
 q=-1+(1+z) \frac{d \ln H}{d z}.
\end{align}
Hence, eqs. \eqref{KG1}- \eqref{KG2}
becomes
\begin{subequations}
\label{NNKG}
\begin{small}
\begin{align}
&  E^2 \left[(1+z)^2 \frac{d^2 \phi_1}{dz^2} +  (1+z) (q-1) \frac{d\phi_1}{dz}\right] \nonumber \\ & +\frac{\mu_1^4}{H_0^2 f_1}\sin\left(\frac{\phi_1}{f_1}\right) +\frac{\mu_3^4}{H_0^2f_1}\sin\left(\frac{\phi_1}{f_1}-n\frac{\phi_2}{f_2}\right)=0, \label{NKG1}\\
& E^2 \left[(1+z)^2 \frac{d^2 \phi_1}{dz^2} +  (1+z) (q-1) \frac{d\phi_1}{dz}\right] \nonumber \\ &+\frac{\mu_2^4}{H_0^2 f_2}\sin\left(\frac{\phi_2}{f_2}\right)   -n\frac{\mu_3^4}{f_2}\sin\left(\frac{\phi_1}{f_1}-n\frac{\phi_2}{H_0^2 f_2}\right)=0. \label{NKG2}
\end{align}
\end{small}
Raychaudhuri equation becomes
\begin{equation}
    (1+z)\frac{d E}{d z}=(q+1)E. \label{NeqE}
\end{equation}
where  $q$ can be expressed as 
\begin{small}
\begin{align}
  q= \frac{1}{2} \Bigg[ 1 & + \frac{(1+z)^2}{2} \left( \left(\frac{d\phi_1}{dz}\right) ^2 + \left( \frac{d\phi_2}{dz} \right)^2\right)  \nonumber \\
   & -\frac{2 \mu_1^4
   \sin ^2\left(\frac{\phi_1}{2
   f_1}\right)}{E^2 H_0^2}  -\frac{2 \mu_2^4
   \sin ^2\left(\frac{\phi_2}{2
   f_2}\right)}{E^2 H_0^2} \nonumber \\
  & -\frac{2 \mu_3^4 \sin ^2\left(\frac{\phi_1}{2
   f_1}-\frac{n \phi_2}{2
   f_2}\right)}{E^2 H_0^2}\Bigg].  
\end{align}
\end{small}
 \end{subequations}
Then, we obtain a system of differential equations for $\phi_{1}(z)$, $\phi_{2}(z)$, and $E(z)$ given by \eqref{NKG1}, \eqref{NKG2} and \eqref{NeqE} and integrate in terms of redshift $z$, from $z=100$ to $z=-1$.
The parameter values $f_{1}=0.1$,  $f_{2}=0.1$, $\mu_{1}^{4}=1.1 H_0^2$, $\mu_{2}^{4}=10.75 H_0^2$,          $\mu_{3}^{4}=1.07$ and  $n=9$ are chosen. As initial conditions for the fields we use   $\phi_{1}|_{z=100}=0.155$,  $\phi_{2}|_{z=100}=0.7835$,
$\frac{d\phi_{1}}{d z}_{z=100}=0$ and $\frac{d\phi_{2}}{d z}_{z=100}=0$, such that $V(\phi_1(z), \phi_2(z))|_{z=100}= 11.6351 H_0^2$. As an initial condition for the Hubble parameter when $z=100$, we take as an ``educated guess'' the value obtained for the $\Lambda$CDM model at $z=100$. That is, we use the expression 
\begin{equation}
    E(z)=\sqrt{\Omega_{r0}(1+z)^{4}+\Omega_{m0}(1+z)^{3}+\Omega_{\Lambda0}}, \label{HLambdaCDM}
\end{equation}
where $\Omega_{r0}=2.469\times 10^{-5}h^{-2}(1+0.2271N_{\text{eff}})$ and $\Omega_{\Lambda0}=1-\Omega_{r0}-\Omega_{m0}$. For these parameters we consider $H_{0}=67.4\;km\;s^{-1}\;Mpc^{-1}$, $\Omega_{m0}=0.315$ and $N_{\text{eff}}=2.99$ according to the Planck 2018 results \cite{Planck2018}.
\begin{figure*}
    \centering
    \subfigure[\label{sima} Evolution of the axion-like fields density parameter $\Omega_{\phi}$ and the matter density parameter $\Omega$ of our model and the density parameters $\Omega_m$, $\Omega_\Lambda$ and $\Omega_r$ of $\Lambda$CDM as a function of the redshift $z$.]{\includegraphics[scale = 0.35]{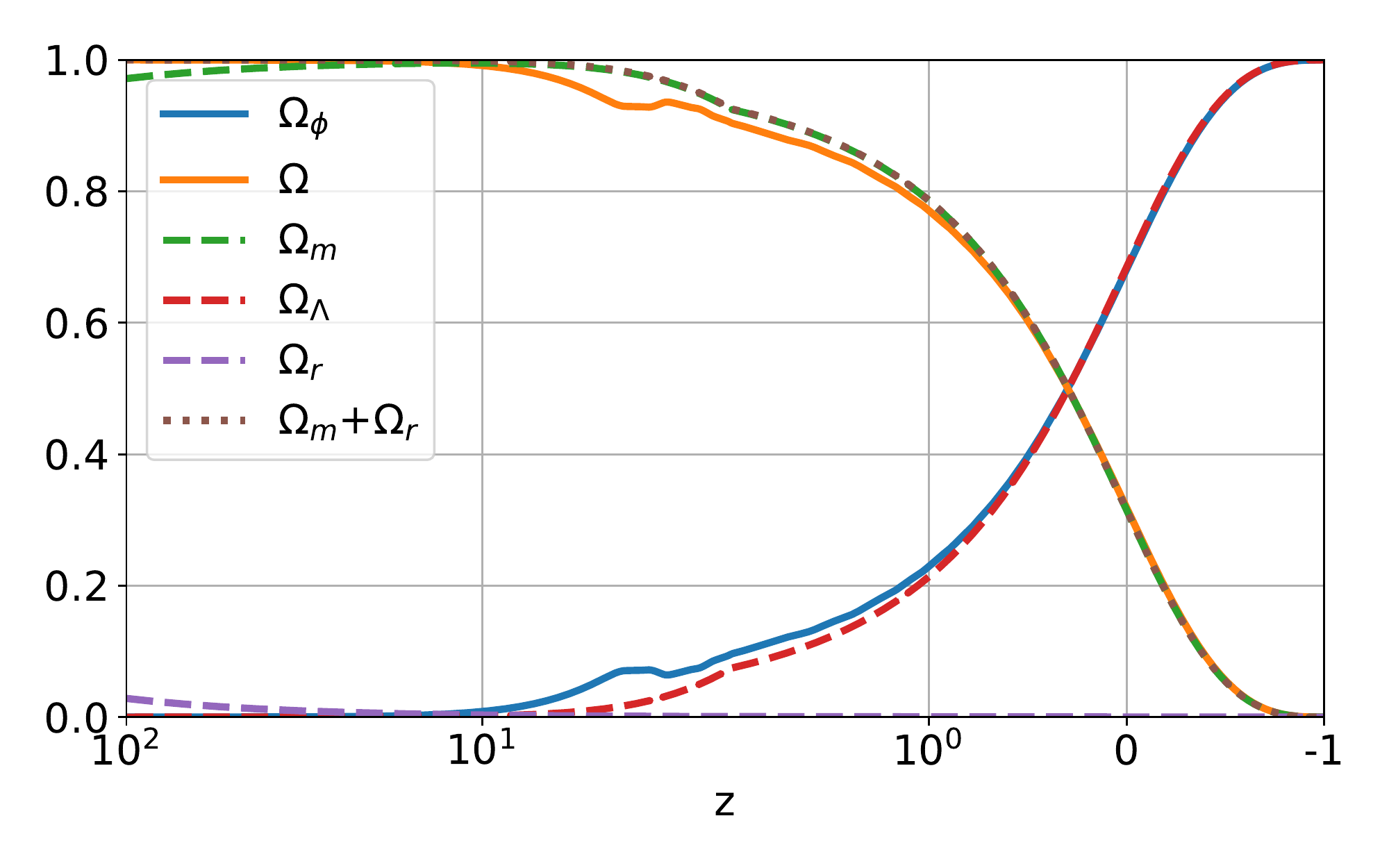}}
    \hspace{0.35cm}
     \subfigure[\label{simb} Evolution of the effective barotropic index $\omega_{\phi}$ associated to the axion-like fields as a function of the redshift $z$ and $\omega_\Lambda =-1$ corresponding to $\Lambda$CDM.]{\includegraphics[scale = 0.35]{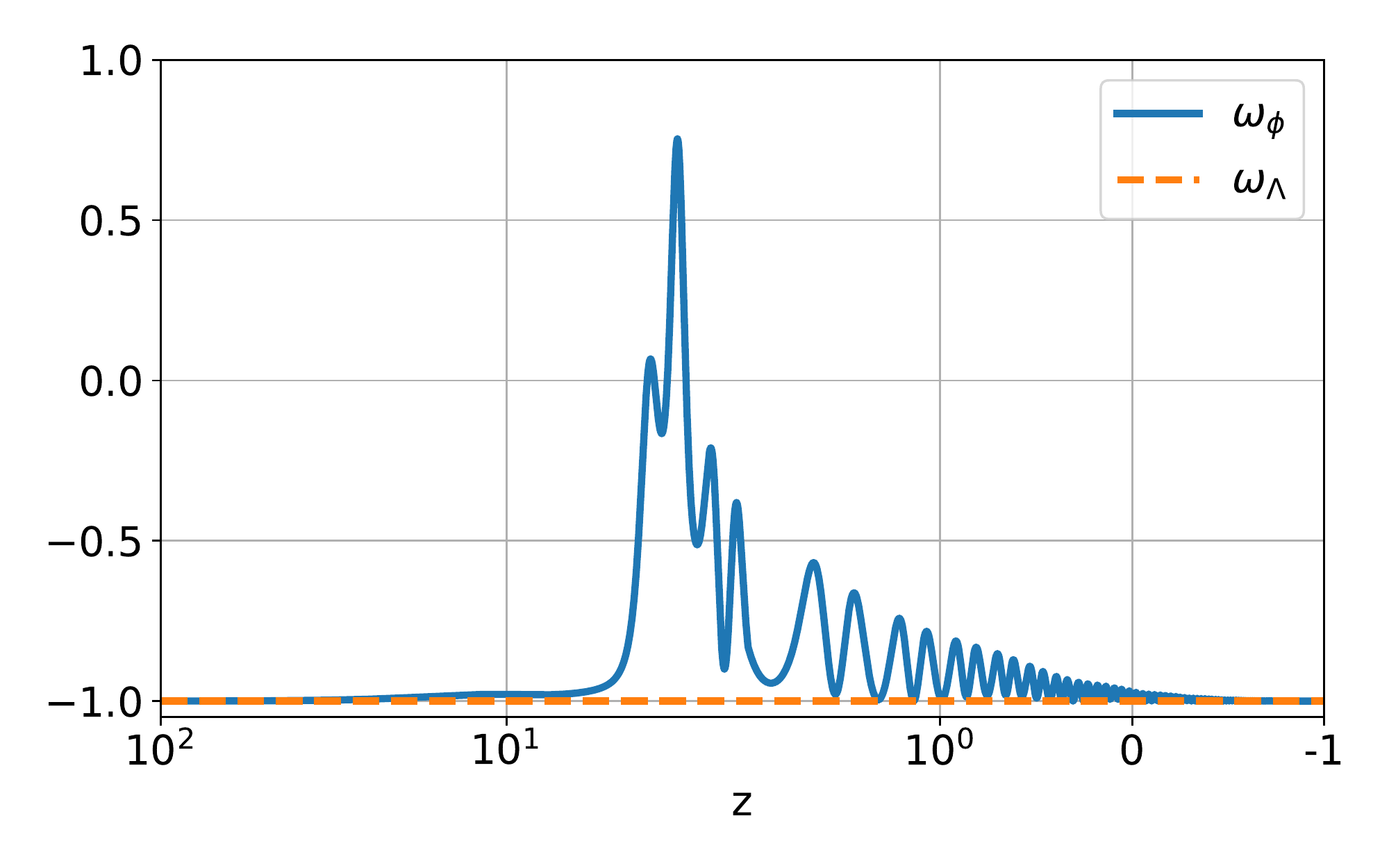}}
      \hspace{0.35cm}
         \subfigure[\label{Fig2a} Evolution of the kinetic, potential and interaction terms of $\Omega_\phi$ given by \eqref{kin-pot} as  functions of redshift $z$.]{\includegraphics[scale = 0.35]{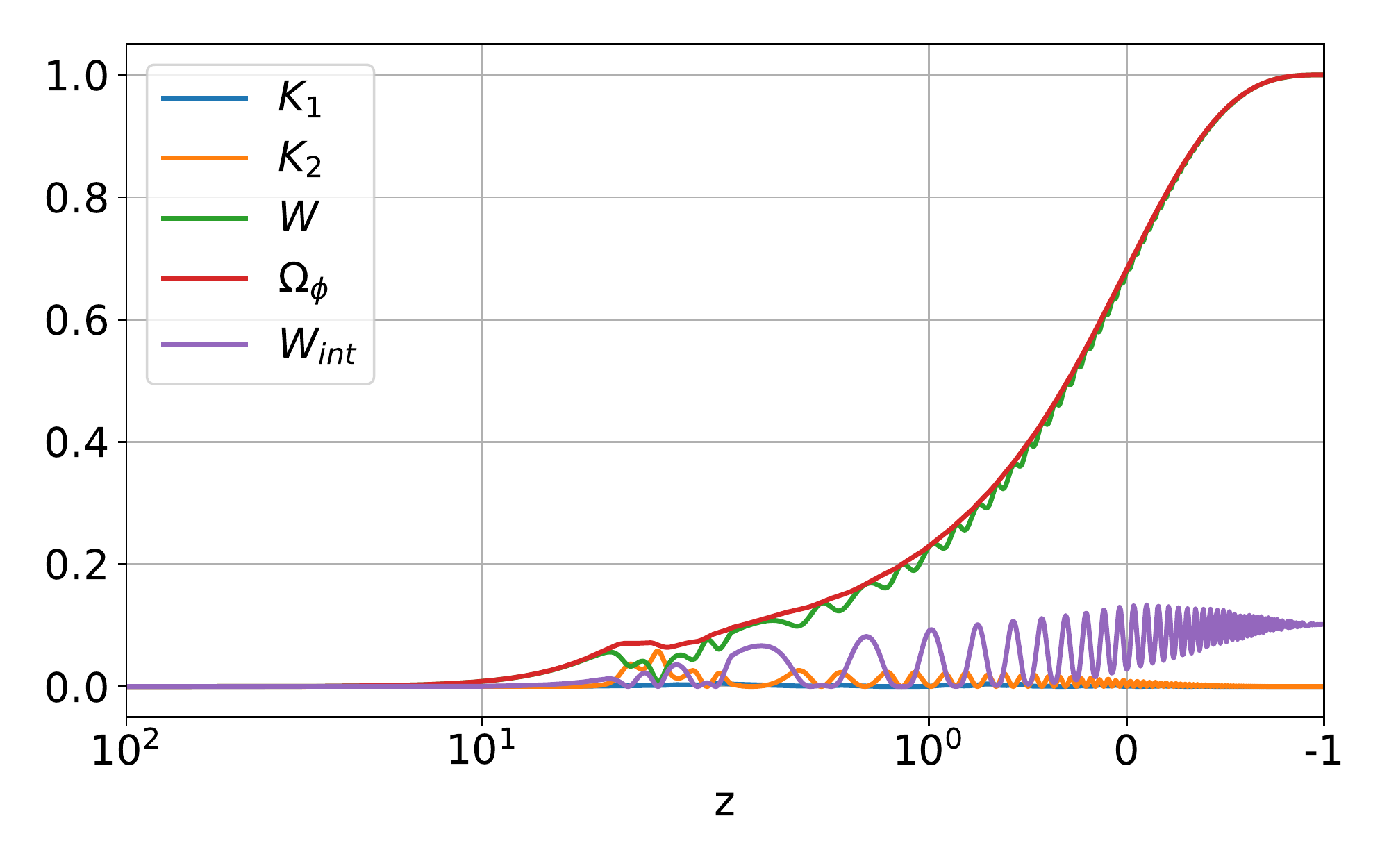}}
    \hspace{0.35cm}
    \subfigure[\label{Fig2b} Zoom of kinetic terms \eqref{kin1} and \eqref{kin2} as  functions of redshift $z$.]{\includegraphics[scale = 0.35]{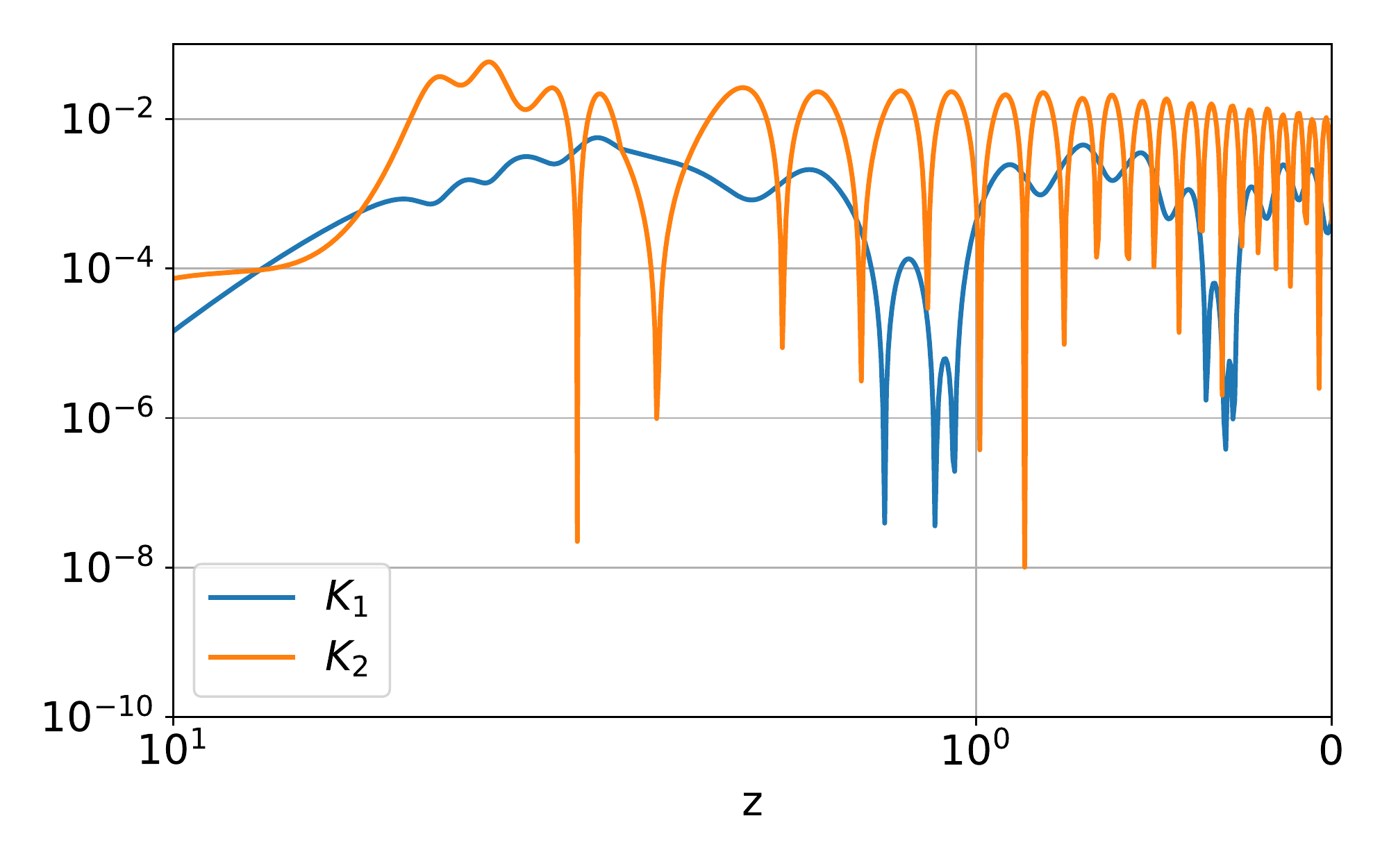}}
     \subfigure[\label{simd} Evolution of the ratios $ \Omega/\Omega_{\phi}$ and $ \Omega/\Omega_{\Lambda}$ as a function of the redshift $z$. According  Planck 2018 results \cite{Planck2018}, the current value of $r$ is $ \Omega_{m0}/\Omega_{\Lambda_0}= 63/137$. ]{\includegraphics[scale = 0.35]{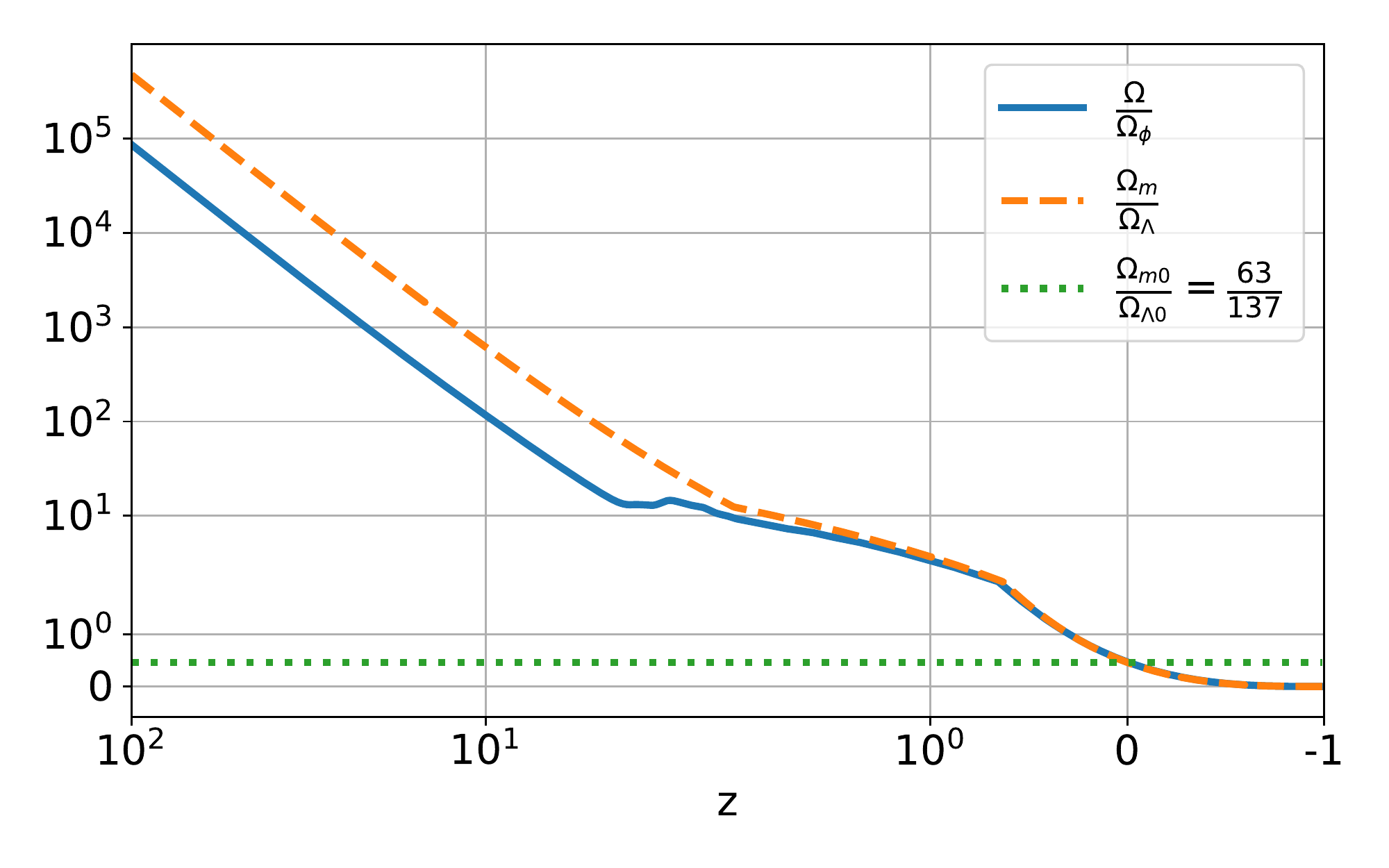}}
     \hspace{0.35cm}
    \subfigure[\label{sime} Evolution of the dimensionless Hubble parameter $E=H/H_0$ for our model and for $\Lambda$CDM as a function of the redshift $z$.]{\includegraphics[scale = 0.35]{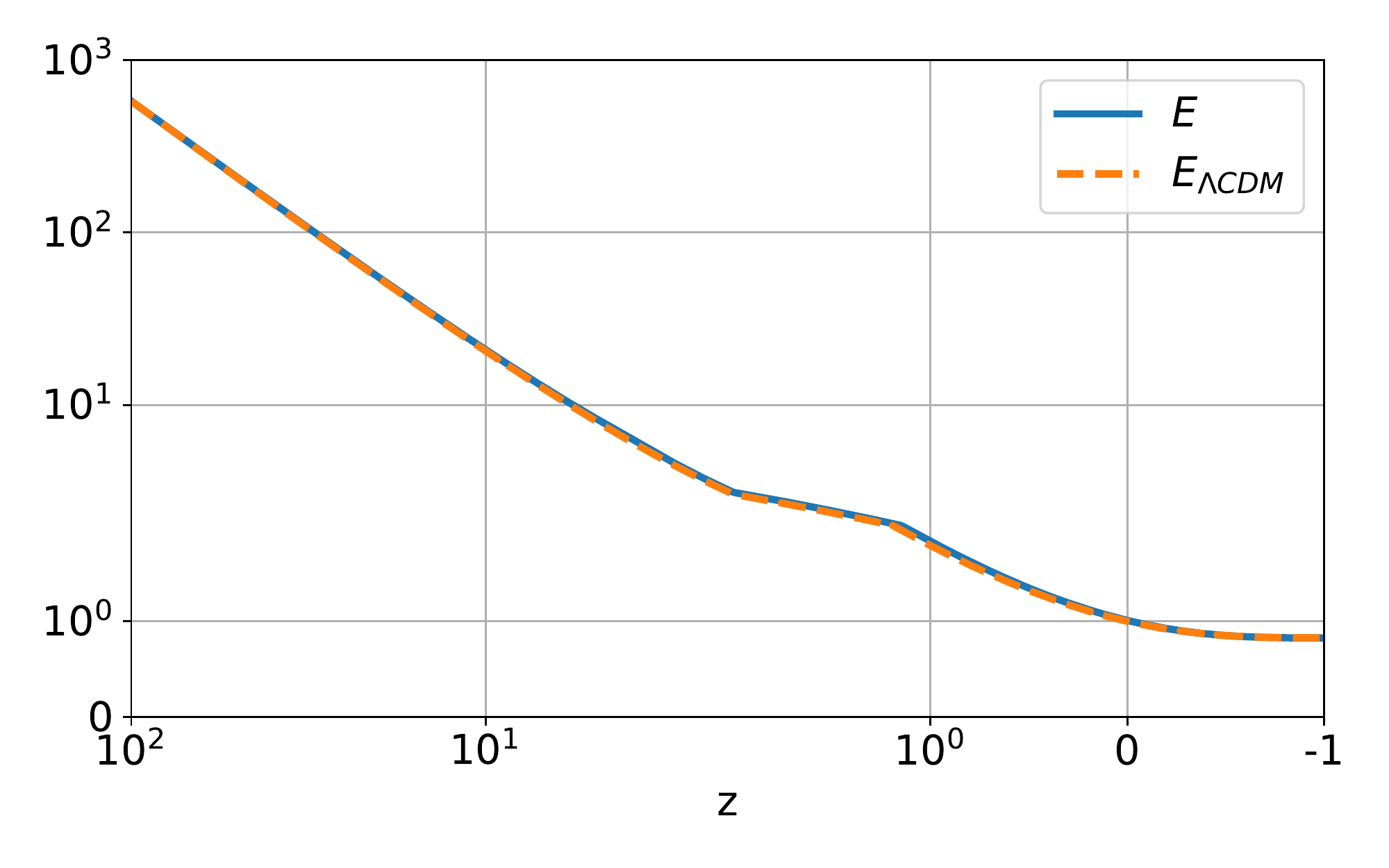}}
    \hspace{0.35cm}
    \subfigure[\label{simf} Evolution of the deceleration parameter defined by  $q:=-1-\dot{H}/H^2$ for our model and for $\Lambda$CDM as a function of the redshift $z$.]{\includegraphics[scale = 0.35]{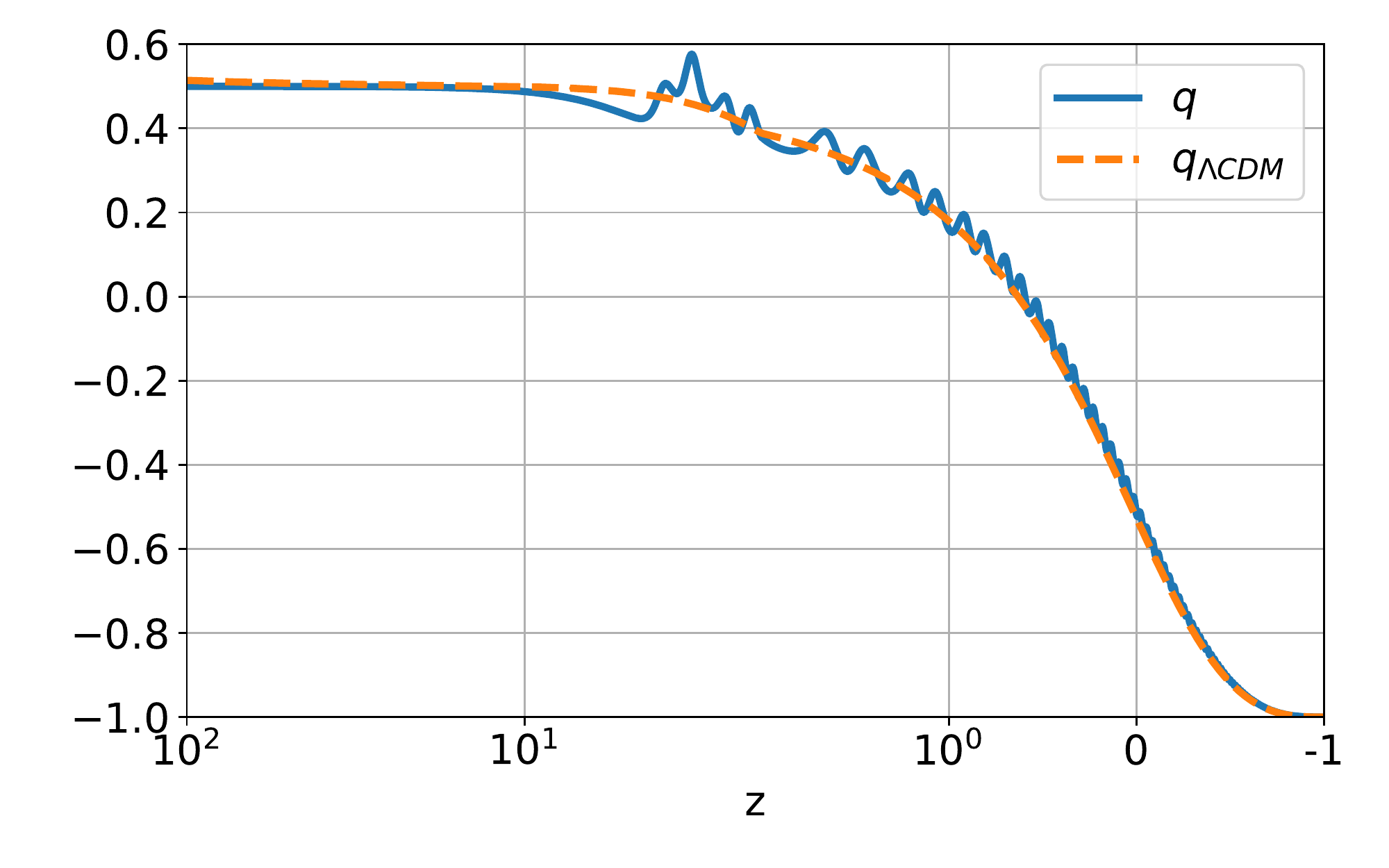}}
    \hspace{0.35cm}
     \subfigure[\label{simc} Evolution of the axion-like fields $\phi_{1}$ and $\phi_{2}$ as a function of the redshift $z$.]{\includegraphics[scale = 0.35]{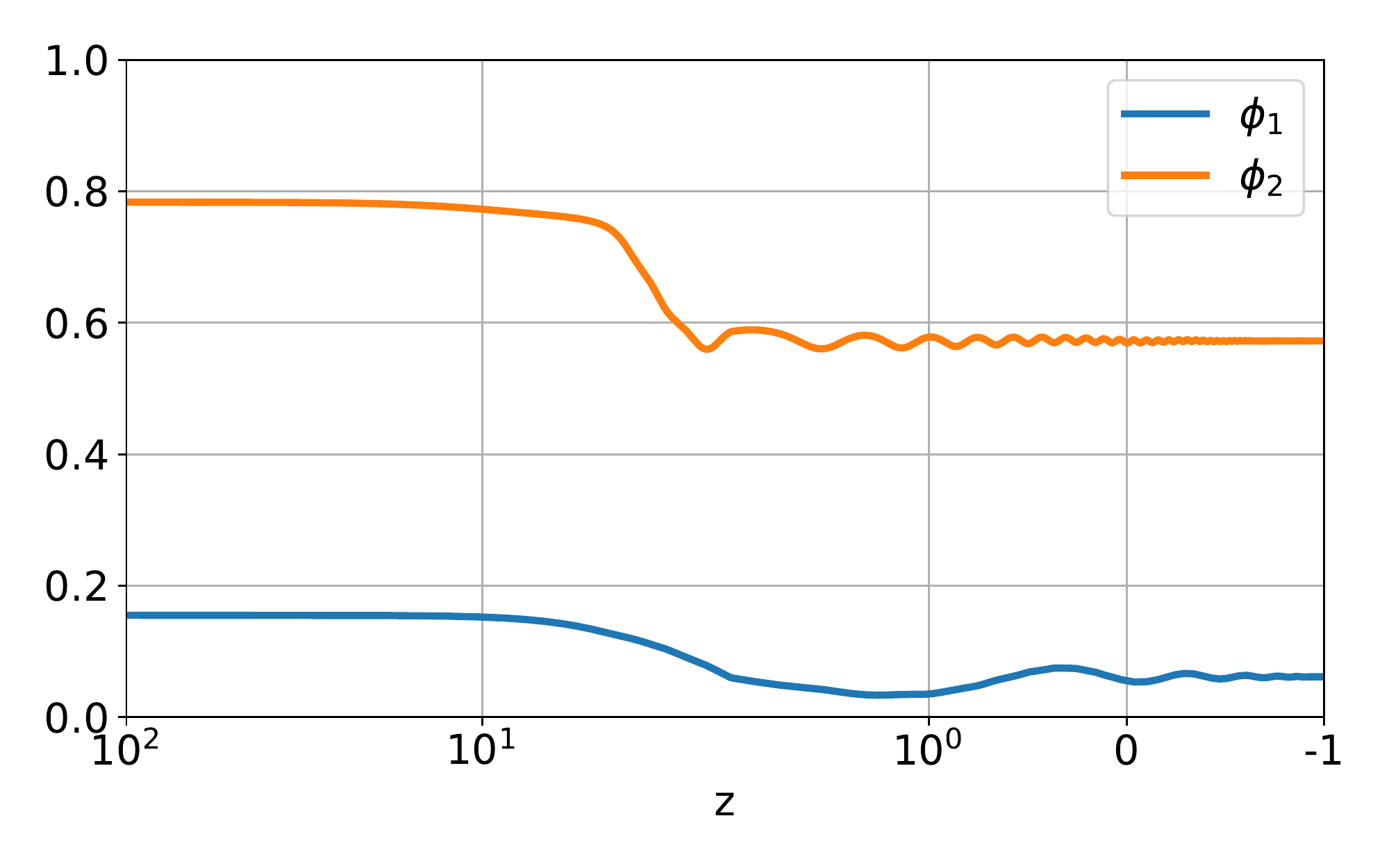}}
         \caption{Numerical simulation of the system \eqref{Raych}-\eqref{KG} with initial conditions $\phi_{1}|_{z=100}=0.155$,  $\phi_{2}|_{z=100}=0.7835$,
        $\frac{d\phi_{1}}{d z}_{z=100}=0$ and $\frac{d\phi_{2}}{d z}_{z=100}=0$. The initial value $H|_{z=100}$  is estimated from expression \eqref{HLambdaCDM}. The exact solutions for the $\Lambda$CDM model  are superimposed for a comparison.}
    \label{fig:fieldssimulations}
\end{figure*}
\begin{figure}
\centering
\subfigure[Surface $V(\phi_1, \phi_2)/(3H_0^2)$.]{\includegraphics[scale = 0.3]{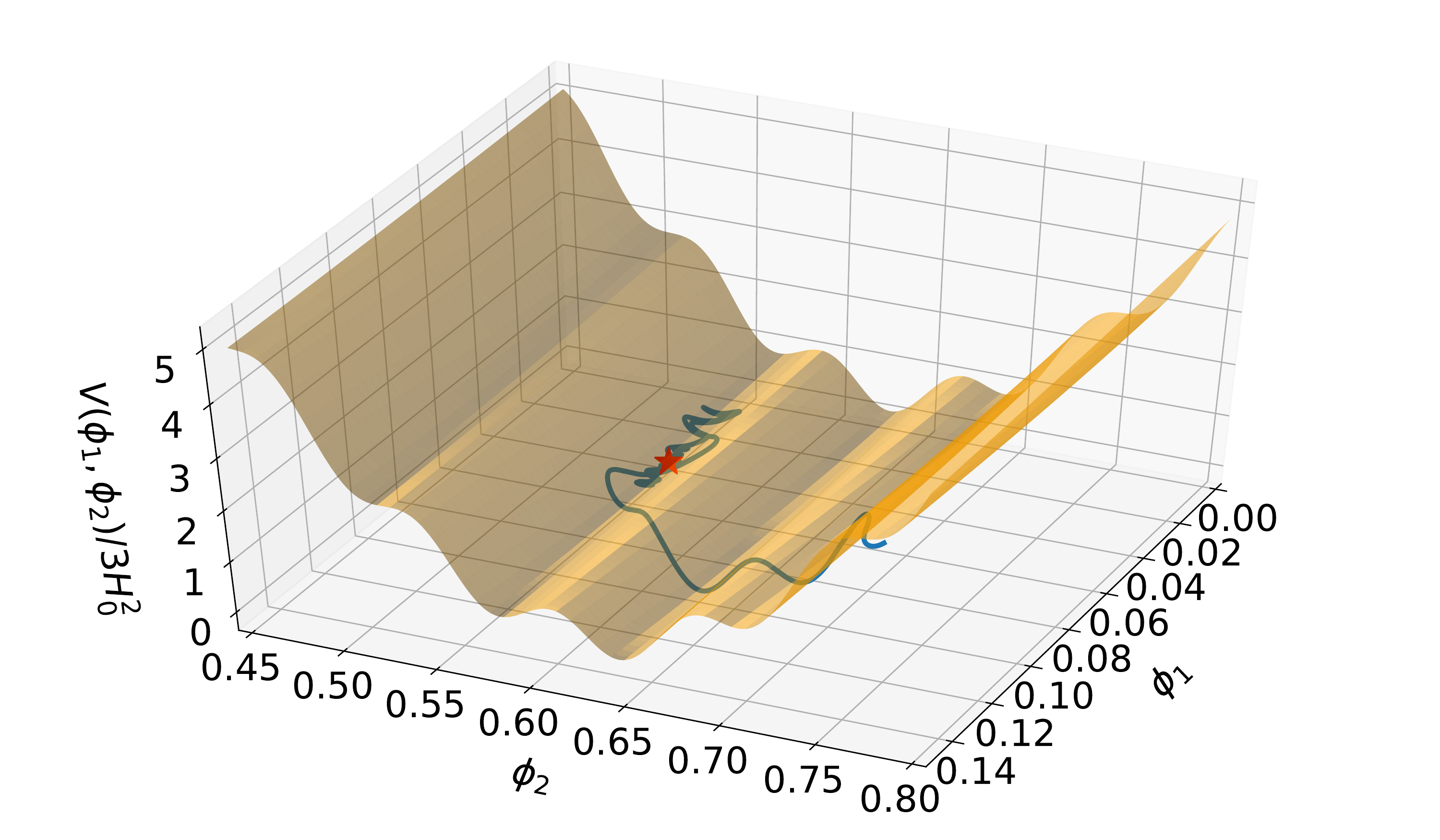}}
 \subfigure[Surface $V(\phi_1, \phi_2)/(3H_0^2)$ (basin of attraction of $\phi^*$).]{\includegraphics[scale = 0.3]{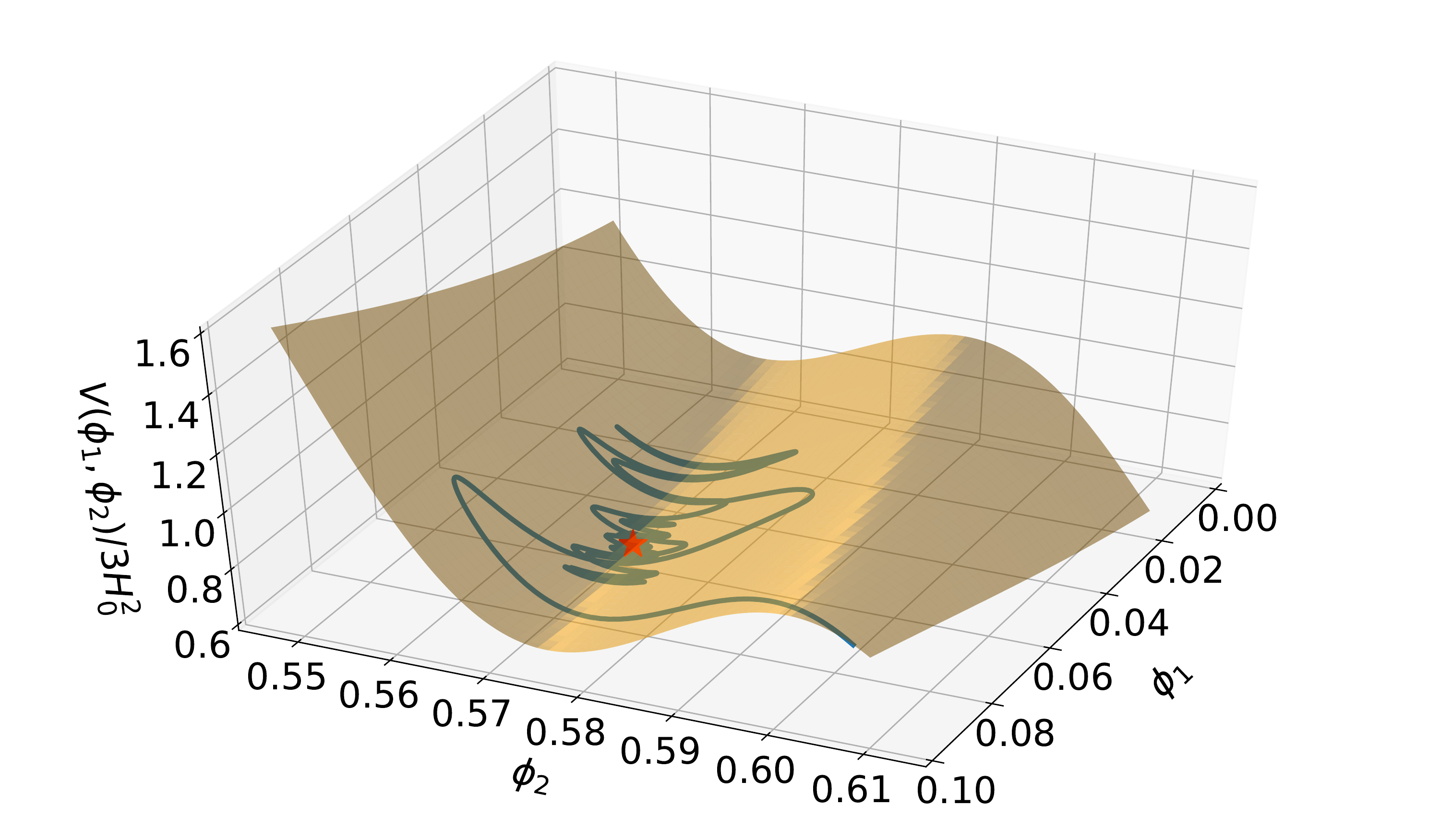}}
    \caption{\label{Fig2c} Surface $V(\phi_1, \phi_2)/(3H_0^2)$ with the local minimum of $V(\phi_1, \phi_2)/(3H_0^2)$  at $\phi^*:=(\phi_{1}, \phi_{2})=(0.0614165, 0.572375)$ with minimum value of $\frac{\Lambda_{\text{eff}}}{3 H_0^2}=0.682603$ (denoted by a red star). The parametric curve (denoted by a blue line) $\frac{V(\phi_1(z), \phi_2(z))}{3H_0^2}, z\in[-1,100]$, obtained by evaluating the solution which converges to $\phi^*$, is attached to the surface.}
    \end{figure}
    
For the analysis we have defined $\Omega_{\phi}=\rho_{\phi}/3H^{2}$ and $\omega_{\phi}=p_{\phi}/\rho_{\phi}$, where
\begin{align}
    \rho_{\phi}=\frac{1}{2}\dot{\phi_{1}}^{2}+\frac{1}{2}\dot{\phi_{2}}^{2}+V(\phi_{1},\phi_{2}), \\
    p_{\phi}=\frac{1}{2}\dot{\phi_{1}}^{2}+\frac{1}{2}\dot{\phi_{2}}^{2}-V(\phi_{1},\phi_{2}),
\end{align}
with $V(\phi_{1},\phi_{2})$ given by equation \eqref{pot}, and from equation \eqref{Friedmann} we obtain the constraint $\Omega=1-\Omega_{\phi}$, where $\Omega=\rho/3H^{2}$. 
To compare with $\Lambda$CDM, we represent with dashed lines the functions \begin{subequations}
\begin{align}
& \Omega_\Lambda=\frac{1-\Omega_{m0}-\Omega_{r0}}{\Omega_{r0}(1+z)^{4}+\Omega_{m0}(1+z)^{3}+\Omega_{\Lambda0}}, \\
& \Omega_m= \frac{\Omega_{m0}(1+z)^3}{\Omega_{r0}(1+z)^{4}+\Omega_{m0}(1+z)^{3}+\Omega_{\Lambda0}},\\
& \Omega_r=\frac{\Omega_{r0}(1+z)^4}{\Omega_{r0}(1+z)^{4}+\Omega_{m0}(1+z)^{3}+\Omega_{\Lambda0}}, 
\end{align}
\begin{align}
& \omega_{\Lambda}=-1, \\
& r=\Omega_m/\Omega_{\Lambda}.
 \end{align}
\end{subequations}
Using \eqref{HLambdaCDM}, the deceleration parameter for $\Lambda$CDM model is 
\begin{align}
    q = -1 + \frac{(z+1)^3 (3 \Omega_{m 0}+4 \Omega_{r 0} (z+1))}{2
   \left(\Omega_{\Lambda 0}+(z+1)^3 (\Omega_{m 0}+\Omega_{r 0}(1+z)\right)}. 
\end{align}

We define the kinetic and potential terms of $\Omega_\phi$ as \begin{subequations}
\label{kin-pot}
\begin{align}
   & K_1:= \frac{\dot{\phi_1}^2}{6 H^2} = \frac{1}{6}(1+z)^2 \left(\frac{d\phi_1}{d z}\right)^2, \label{kin1}\\
   & K_2:= \frac{\dot{\phi_2}^2}{6 H^2} = \frac{1}{6}(1+z)^2 \left(\frac{d\phi_2}{d z}\right)^2, \label{kin2}\\
   & W:= \frac{V(\phi_1, \phi_2)}{3 H^2}= \frac{V(\phi_1, \phi_2)}{3 H_0^2 E^2}, \label{FullPotential}
\\
& W_{\text{int}}:= \frac{\mu_3^4}{3 H_0^2 E^2}\left[1-\cos\left(\frac{\phi_1}{f_1}-n\frac{\phi_2}{f_2}\right)\right], \label{Int}
\end{align}
\end{subequations}
and 
 \begin{equation}
    \frac{ \Lambda_{\text{eff}}}{3 H_0^2}= \lim_{z\rightarrow -1} \frac{V(\phi_1(z), \phi_2(z))}{3H_0^2}= \lim_{z\rightarrow -1} E^2 W.
 \end{equation}
Using theorem \ref{local-estimate}, it can be obtained 
\begin{align}
    \lim_{z \rightarrow -1} W(z)=1. 
\end{align} 
Based on this, we obtain the effective CC constant $\Lambda_{\text{eff}}$:
\begin{small}
\begin{align}
 \frac{\Lambda_{\text{eff}}}{3H_0^2}=\frac{1}{150}\lim_{z\rightarrow -1}  \Big[ & 107 \sin ^2(5 (\phi_1(z)-9
   \phi_2(z))) \nonumber \\ & +110 \sin ^2(5 \phi_1(z)) + 1075 \sin ^2(5
   \phi_2(z))\Big]\nonumber \\
   = &\lim_{z\rightarrow -1}  E^2(z), \end{align}
\end{small}
\noindent where $\lim_{z\rightarrow -1} \phi_1(z) = \phi_1^*$ and $\lim_{z\rightarrow -1} \phi_2(z) = \phi_2^*$ should satisfy the conditions of extreme point
\begin{small}
\begin{subequations}
\label{extrem_cond}
\begin{align}
& 107 \sin (10 (\phi_1^*-9 \phi_2^*))+110 \sin
   (10 \phi_1^*) =0,\\
   &  963 \sin (10 (\phi_1^*-9
   \phi_2^*))-1075 \sin (10 \phi_2^*) =0,
\end{align}
\end{subequations}
\end{small}
and conditions for local minima, which are 
\begin{small}
\begin{subequations}
\label{second_derivative_test_minima}
\begin{align}
& H_0^2 \left[107 \cos (10 (\phi_1^*-9 \phi_2^*))+110
   \cos (10 \phi_1^*)\right]>0\\
   & \frac{5}{2}H_0^4\Bigg[190674 \cos (20 \phi_1^*-90 \phi_2^*)+23005 \cos (10 (\phi_1^*-10 \phi_2^*)) \nonumber \\
   & +23005 \cos (10 (\phi_1^*-8 \phi_2^*))+23650 \cos
   (10 (\phi_1^*-\phi_2^*)) \nonumber \\
   &+23650 \cos (10 (\phi_1^*+\phi_2^*))+190674 \cos (90 \phi_2^*)\Bigg]>0. 
    \end{align}
    \end{subequations}
    \end{small}
In Fig. \ref{fig:fieldssimulations} we present some numerical results obtained from the integration of the equations \eqref{NKG1}, \eqref{NKG2} and \eqref{NeqE}. 
In figure \ref{sima}, we present the axion-like fields density parameter $\Omega_{\phi}$ and the matter density parameter $\Omega$ of our model and the density parameters $\Omega_m$, $\Omega_\Lambda$ and $\Omega_r$ of $\Lambda$CDM as a function of the redshift $z$. In figure \ref{simb}, we present the of the effective barotropic index $\omega_{\phi}$ associated to the axion-like fields and $\omega_\Lambda=-1$ of $\Lambda$CDM as a function of the redshift $z$. In figure \ref{Fig2a} the evolution of the kinetic and potential terms of $\Omega_\phi$ given by \eqref{kin-pot} are shown as  functions of redshift $z$. In figure \ref{Fig2b} a zoom of the evolution of kinetic terms \eqref{kin1} and \eqref{kin2} are presented  as  functions of redshift $z$. In figure \ref{simd}, we present the evolution of the ratios $ \Omega/\Omega_{\phi}$ and $ \Omega/\Omega_{\Lambda}$ as a function of the redshift $z$. In figure \ref{sime}, we present the evolution of the dimensionless Hubble parameter $E=H/H_0$ for our model and for $\Lambda$CDM as a function of the redshift $z$. Finally, in figure \ref{simf}, we present the evolution of the deceleration parameter $q:=-1-\dot{H}/H^2$ for our model and for $\Lambda$CDM as a function of the redshift $z$. In figure \ref{simc} we present the evolution  of the axion-like fields $\phi_{1}$ and $\phi_{2}$ as a function of the redshift $z$. According to figure \ref{simc} the values $\phi_1 = 0.061$ and $\phi_2 = 0.572$ can be used as initial points for finding numerically the roots of \eqref{extrem_cond} which are found to be $\phi_1^*= 0.0614165$, and $\phi_2^*= 0.572375$. 
  Evaluating \eqref{second_derivative_test_minima}  we have $176.099 H_0^2>0$, and $788143 H_0^4>0$. Then, as shown in figure \ref{Fig2c}, a local minimum with minimum value $\lim_{z\rightarrow -1} E(z)^2=\frac{\Lambda_{\text{eff}}}{3 H_0^2}=0.682603$ is obtained. In figure \ref{Fig2c} the surface $\frac{V(\phi_1, \phi_2)}{3H_0^2}$ is plotted.  By evaluating the solution which converges to $\phi^*$, the parametric curve  $\frac{V(\phi_1(z), \phi_2(z))}{3H_0^2}, z\in[-1,100]$, was obtained  and attached to the surface, and $\lim_{z \rightarrow -\infty} W_{\text{int}}=  0.101566$. The parameter values and initial conditions were chosen for obtaining a ratio of the DM and DE densities at $z=0$ equal to 
  $r|_{z=0}=\frac{0.315}{0.685}= \frac{63}{137} \approx 0.459854$. A late-times $r\rightarrow 0$, since $\Omega_\phi$ dominates, in particular $\sqrt{\frac{V(\phi_1^*,\phi_2^*)}{3 \lim_{z\rightarrow -1}H(z)^2}}= \sqrt{\frac{\Lambda_{\text{eff}}}{3 H_0^2 \lim_{z\rightarrow -1} E(z)}}=1$.

\section{Dynamical systems analysis using $H_0$-normalized variables}
\label{DSA}

Although expansion normalized dynamical variables are commonly used for dynamical analysis of late time cosmologies, it turns out that for this model they do not lead to an autonomous dynamical system. To make an autonomous dynamical system for this model, we introduce the following $H_0$-normalized variables
\begin{align}
 & E=\frac{H}{H_0}, \;   \Omega_0=\frac{\rho}{3H_0^2}, \;   x_1=\frac{\dot{\phi_1}}{2H_0 f_1}, \;   x_2=\frac{\dot{\phi_2}}{2H_0 f_2},  \end{align}
together with the scalar field redefinition 
\begin{align}
 y_1=\frac{\phi_1}{2f_1}, \;   y_2=\frac{\phi_2}{2f_2},
\end{align}
the dimensionless time variable 
\begin{equation}
    \tau=H_0 t,
\end{equation}
and the following parameters
\begin{align}
& a_1=\frac{1}{3} f_1^2, \,\,a_2=\frac{1}{3} f_2^2, \nonumber\\ & b_1=\frac{\mu_1^4}{3H_0^2}, \,\,b_2=\frac{\mu_2^4}{3H_0^2}, \,\,b_3=\frac{\mu_3^4}{3H_0^2}\,.
\end{align}
In the above $H_0$ is the present value of the Hubble parameter. We will denote the derivative with respect to $\tau$ with a prime: $(\cdot)^{\prime}\equiv d/d\tau$. In terms of the above variables and parameters we can write down the dynamical system as
\begin{subequations}
\label{dyn_sys_axn}
\begin{align}
& E^2 -2a_1x_1^2 -2a_2x_2^2  -2b_1\sin^2 y_1 \nonumber\\
& -2b_2\sin^2 y_2 -2b_3\sin^2(y_1-n y_2)=\Omega_0, \\
& E^{\prime}=-\frac{3}{2}E^2 -3a_1x_1^2 -3a_2x_2^2 +3b_1\sin^2 y_1 \nonumber\\
& +3b_2\sin^2 y_2 +3b_3\sin^2(y_1-n y_2), 
\\
& x_1^{\prime}=-3Ex_1-\frac{b_1}{2a_1}\sin(2y_1) - \frac{b_3}{2a_1}\sin(2y_1 - 2n y_2), \\
& x_2^{\prime}=-3Ex_2-\frac{b_2}{2a_2}\sin(2y_2)   + n\frac{b_3}{2a_2}\sin(2y_1 - 2n y_2), \\
& y_1^{\prime}=x_1, \\
& y_2^{\prime}=x_2.\,
\end{align}
\end{subequations}
We note that the dynamical system \eqref{dyn_sys_axn} is periodic in both $y_1,y_2$ with a period of $\pi$ as long as $n$ is an integer. This means that for the decoupled limit ($n=0$) of the model, as well as the interacting cases when $n$ is an integer, we can confine our attention within the range $-\pi/2\leq y_1,y_2\leq\pi/2$. 

\begin{table*}[t]
    \centering
      \caption{\label{tab:fixed_pts} Some isolated fixed points of the system \eqref{dyn_sys_axn} for $n\in\mathbb{Z}$ over $(y_1, y_2)\in \left[-\frac{\pi}{2}, \frac{\pi}{2}\right]\times \left[-\frac{\pi}{2}, \frac{\pi}{2}\right]$.}
   \begin{tabular}{lcccc}\hline
        Fixed point & Coordinates ($E,x_1,x_2,y_1,y_2$) & Existence & Stability & Solution \\
        \hline
        $P_{(\pm1,\pm1)}$ & $\sqrt{2 (b_1+b_2+b_3)},0,0,\pm  \frac{\pi}{2}, \pm \frac{\pi}{2}$ & always & Saddle for $n=0$, & de Sitter \\
         &  &  & depends on $n$ otherwise &  \\
        \hline
        $P_{(0,\pm1)}$ & $\sqrt{2b_2},0,0,0, \pm \frac{\pi}{2}$ & always & Saddle for $n=0$, & de Sitter \\
         &  &  & depends on $n$ otherwise &  \\
        \hline
        $P_{(\pm1,0)}$ &  $\sqrt{2(b_1+b_3)},0,0,\pm \frac{\pi }{2},0$ & always & Saddle for $n=0$, & de Sitter \\
        &  &  & depends on $n$ otherwise &  \\
        \hline
        $P_{(0,0)}$ & $0,0,0,0,0$ & always & Stable for all $n$ & Minkowski \\
        \hline
    \end{tabular}
  \end{table*}
   \begin{figure*}
    \centering
    \includegraphics[scale=0.5]{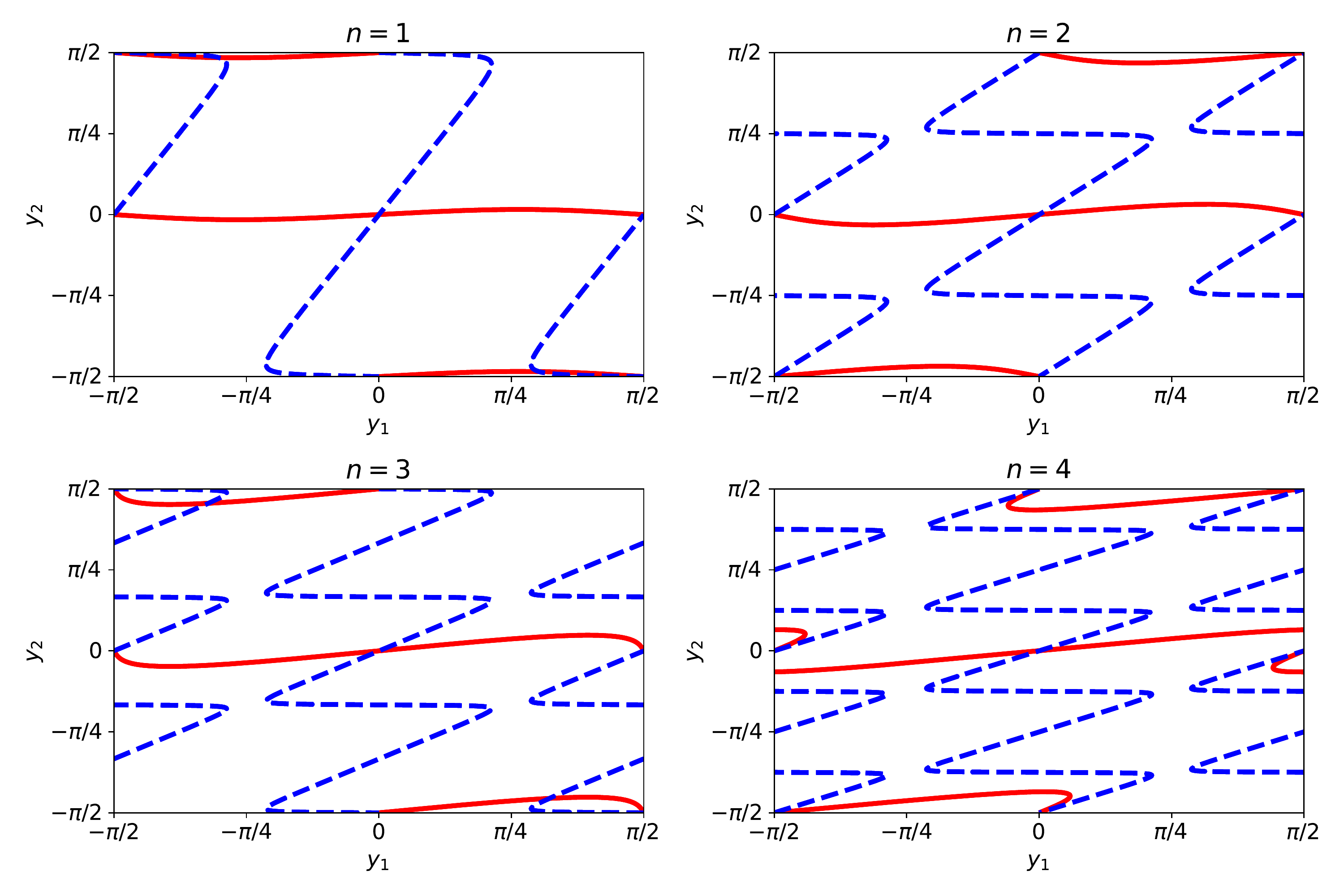}
    \caption{Contour domains of restrictions \eqref{EQ3.7} (solid red line) and
\eqref{EQ3.8} (dashed blue line). At the points where the two lines coincides are found the equilibrium points of the system}
    \label{fig:plot1}
\end{figure*}
 The fixed points/lines of the above system are given by the conditions
\begin{subequations}
\label{critconds}
\begin{align}
& x_1=x_2=0, \\
& -n b_1\sin(2y_1)= b_2\sin(2y_2)=n b_3\sin(2y_1-2ny_2), \label{critconds2}\\
& E^2=2b_1\sin^2y_1 + 2b_2\sin^2y_2 + 2b_3\sin^2(y_1-ny_2). \label{critconds3}
\end{align}
\end{subequations}
The isolated fixed points of the decoupled version of the model and their stability are listed in table \ref{tab:fixed_pts}. In the coupled case $n \neq 0$, the transcendental equations \eqref{critconds2} have to be solved numerically. 

For a given  $(y_1, y_2, E)=(y_1^*, y_2^*, E^*)$ which  satisfies \eqref{critconds}, it can be found univocally the values $b_1,b_2$ and $b_3$: 
\begin{align}
& b_1= \frac{{h^*}^2 \csc ({y_1^*})}{2 \sin (n {y_2^*}) \sec ({y_1^*}-n {y_2^*})-2 n \cos ({y_1^*}) \tan ({y_2^*})},\\
& b_2= \frac{{h^*}^2 n \csc ^2({y_2^*})}{2 n-2 \sec ({y_1^*}) \cot ({y_2^*})
   \sin (n {y_2^*}) \sec ({y_1^*}-n {y_2^*})},\\
  & b_3=\frac{{h^*}^2 \csc ({y_1^*}-n {y_2^*})}{2 (n \tan ({y_2^*}) \cos ({y_1^*}-n {y_2^*})-\sec ({y_1^*}) \sin (n {y_2^*}))}.
\end{align}
However, to proceed further is a hard task and little progress can be made.  

\subsection{Uncoupled case $n=0$}

\begin{enumerate}
    \item  For the four points $P_{(\pm 1,\pm 1)}$: $(E, x_1, x_2, \Omega_0, y_1, y_2)= \left(\sqrt{2} \sqrt{b_1+b_2+b_3},  0,  0,  0,  \pm \frac{\pi }{2}+ 2 \pi  c_1,  \pm \frac{\pi }{2} + 2 \pi  c_2\right)$, \newline $c_1, c_2 \in \mathbb{Z}$, module $2\pi$,  the eigenvalues are \\$ \Big\{-3 \sqrt{2}
   \sqrt{b_1+b_2+b_3}$,\newline
   $-\frac{3 \sqrt{b_1+b_2+b_3}}{\sqrt{2}}\pm \frac{\sqrt{a_2 (2 b_2+9 a_2 (b_1+b_2+b_3))}}{\sqrt{2} a_2}$,\newline
   $-\frac{3 \sqrt{b_1+b_2+b_3}}{\sqrt{2}}\pm \frac{\sqrt{a_1 ((9
   a_1+2) b_1+2 b_3+9 a_1 (b_2+b_3))}}{\sqrt{2} a_1}\Big\}.$
   
   \item In the first quadrant ($y_1\geq 0, y_2 \geq 0$) there are two representations of $P_{(0,\pm 1)}$ module $2 \pi$: \newline
 $(E, x_1, x_2, \Omega_0, y_1, y_2)$\newline
   $= \left(\sqrt{2} \sqrt{b_2},  0,  0,  0,  2 \pi  c_1,  \pm \frac{\pi }{2} + 2 \pi  c_2\right) $ and  \newline $\left(\sqrt{2} \sqrt{b_2},  0,  0,  0,  \pi + 2 \pi  c_1,  \pm \frac{\pi }{2} + 2 \pi  c_2\right)$, \newline $c_1, c_2 \in \mathbb{Z}$. The eigenvalues are \newline $\Big\{-3 \sqrt{2} \sqrt{b_2},-\frac{3
   \sqrt{b_2}}{\sqrt{2}} \pm \frac{\sqrt{a_2 (9 a_2+2) b_2}}{\sqrt{2} a_2}$,\newline
   $ -\frac{3
   \sqrt{b_2}}{\sqrt{2}}\pm \frac{\sqrt{a_1 (-2 b_1+9 a_1 b_2-2 b_3)}}{\sqrt{2} a_1}\Big\}$.

\item  In the first quadrant ($y_1\geq 0, y_2 \geq 0$) there are two representations of $P_{(\pm 1,0)}$ module $2 \pi$: \newline 
 $(E, x_1, x_2, \Omega_0, y_1, y_2)$,\newline
   $= \left(\sqrt{2} \sqrt{b_1+b_3}, 0, 0, 0,\pm \frac{\pi }{2}+ 2 \pi  c_1, 0 + 2 \pi  c_2\right)$ and  \newline $\left(\sqrt{2} \sqrt{b_1+b_3}, 0, 0, 0,\pm \frac{\pi }{2}+ 2 \pi  c_1, \pi + 2 \pi  c_2\right)$, \newline $c_1, c_2 \in \mathbb{Z}$,  with eigenvalues \newline
   $\Big\{-3 \sqrt{2} \sqrt{b_1+b_3},-\frac{3
   \sqrt{b_1+b_3}}{\sqrt{2}}\pm \frac{\sqrt{a_2 (9 a_2 (b_1+b_3)-2 b_2)}}{\sqrt{2} a_2}$,\newline
   $-\frac{3 \sqrt{b_1+b_3}}{\sqrt{2}}\pm \frac{\sqrt{a_1 (9 a_1+2) (b_1+b_3)}}{\sqrt{2} a_1}\Big\}$.

\item In the first quadrant ($y_1\geq 0, y_2 \geq 0$) there are four representations of $P_{(0,0)}$ module $2 \pi$: \newline
$(E, x_1, x_2, \Omega_0, y_1, y_2)=\left(0, 0,  0,  0, 0 + 2  \pi c_1 , 0 + 2  \pi c_2 \right) $,  \newline
$\left(0, 0,  0,  0, 0+ 2 \pi  c_1, \pi+  2 \pi  c_2\right)$,  \newline
$\left(0, 0, 0, 0, \pi + 2 \pi  c_1, 2 \pi  c_2\right)$ and \newline  
$\left(0, 0, 0, 0, \pi + 2 \pi  c_1, \pi+ 2 \pi  c_2\right)$, $c_1, c_2 \in \mathbb{Z}$,
with eigenvalues \newline $\left\{0,-i \sqrt{\frac{b_2}{a_2}},i \sqrt{\frac{b_2}{a_2}},-i
\sqrt{\frac{b_1+b_3}{a_1}},i \sqrt{\frac{b_1+b_3}{a_1}}\right\}$.

Corresponding to the point $P_{(0,0)}$ we get a Minkowski solution because we have chosen the potential \eqref{pot} to have the minimum value zero.
Adding a cosmological constant term and provided it is positive, the stable fixed point $P_{(0,0)}$ will corresponds to a de Sitter solution.
\end{enumerate}
 
\subsection{Coupled case $n\neq 0$}

For $n\neq 0$, the equilibrium points discussed in table \ref{tab:fixed_pts} are equilibrium points of \eqref{dyn_sys_axn}, satisfying conditions \eqref{critconds}. 
The other possible equilibrium points are found by reducing the conditions \eqref{critconds}, by solving the transcendental equations \eqref{critconds2}  numerically using the parameter values:  $a_1=\frac{1}{300}, \,\,a_2=\frac{1}{300}, \,\,b_1=\frac{11}{30}, \,\,b_2=\frac{43}{12}, \,\,b_3=\frac{107}{300}$.

 \begin{figure*}
   \centering
\includegraphics[scale=0.6]{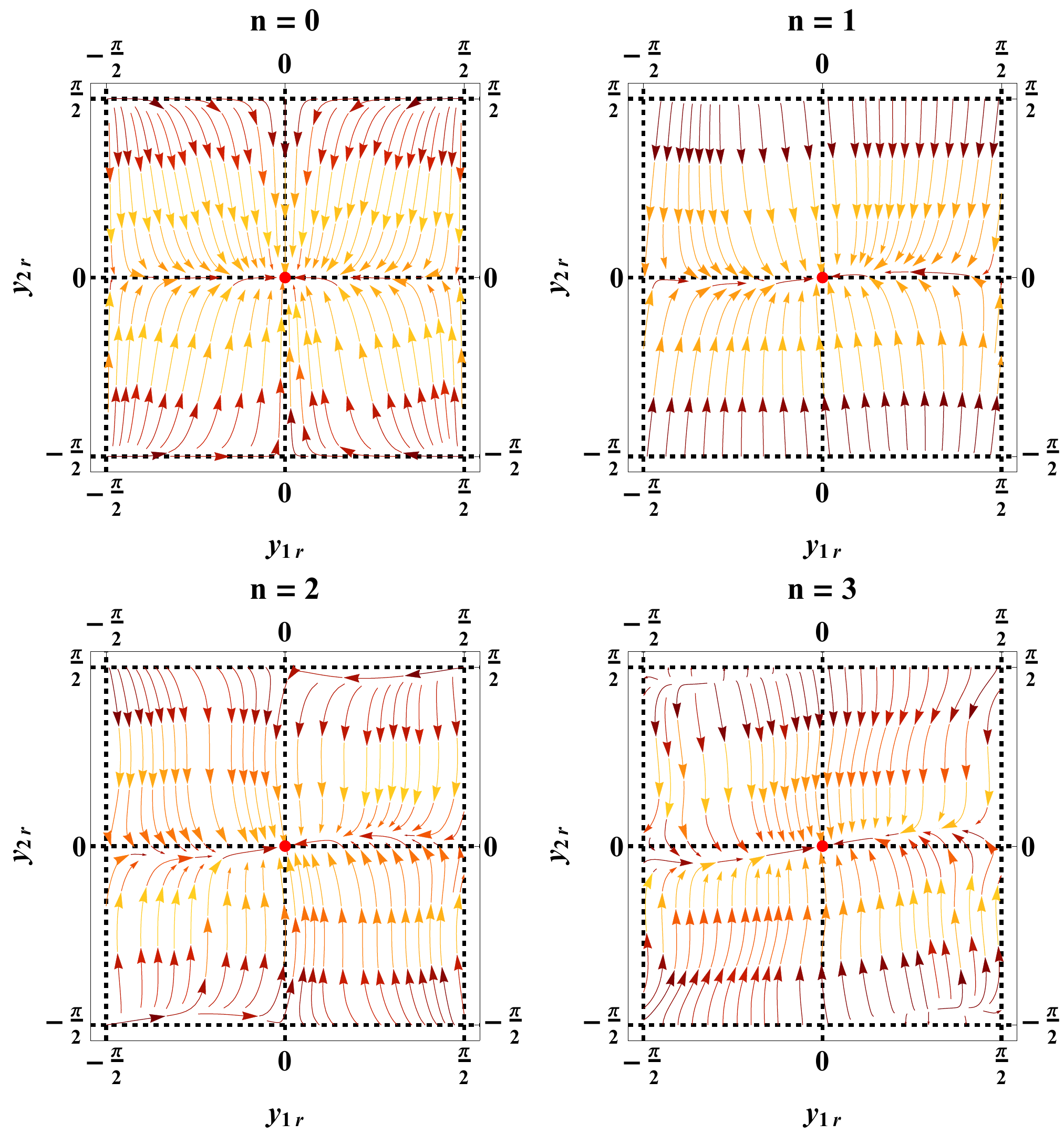}
   \caption{Projection of the phase portrait for $n=0$ (decoupled limit) and for $n=1, 2, 3$, on the 2-dimensional field space shown within the domain $-\pi\leq \frac{\phi_1}{f_1},\frac{\phi_2}{f_2}\leq \pi$. This domain is taken to show the periodic nature of the phase flow with a period $\frac{\pi}{2}$ for $a_1=\frac{1}{300}, \,\,a_2=\frac{1}{300}, \,\,b_1=\frac{11}{30}, \,\,b_2=\frac{43}{12}, \,\,b_3=\frac{107}{300}$}
    \label{fig:my_label}
\end{figure*}

The conditions \eqref{critconds2} and \eqref{critconds3} 
becomes 
\begin{small}
\begin{align}
    & 107 n \sin (2(y_1- n y_2))-1075 \sin (2 y_2)=0, \label{EQ3.7}\\
    & 107 \sin (2 (y_1-n y_2))+110 \sin (2 y_1)=0, \label{EQ3.8}\\
    & E^2=\frac{107}{150} \sin ^2(y_1-n y_2)  +\frac{11}{15} \sin ^2(y_1)+\frac{43}{6} \sin
   ^2(y_2).
\end{align}
\end{small}
Therefore, in the coupled case we need to solve transcendental equations and a little progress is made.

In Fig. \ref{fig:plot1} are presented the contour domains of the restrictions given by the equations  \eqref{EQ3.7} and \eqref{EQ3.8} for the coupled case with $n=1,2,3$ and $4$. At the points where the solid red line and the dashed blue lines coincide are found the equilibrium points of the system.

\subsubsection{Reduced dynamical system}
All the fixed points of the system \eqref{dyn_sys_axn} must necessarily reside on the 3-dimensional hypersurface specified by $(x_1^{\prime},x_2^{\prime})=(0,0)$, where we can express $x_1,x_2$ in terms of the other variables as 
\begin{subequations}
\begin{small}
\begin{align}
& x_{1r}=-\frac{b_1}{6 E_r a_1}\sin(2y_{1r}) - \frac{b_3}{6 E_r a_1}\sin(2y_{1r} - 2n y_{2r}), \\
& x_{2r}=-\frac{b_2}{6 E_r a_2}\sin(2y_{2r}) + n\frac{b_3}{6 E_r a_2}\sin(2y_{1r} - 2n y_{2r}), 
\end{align}
\end{small}
\end{subequations} 
for $x_{1r}\neq 0, x_{2r}\neq  0, E_r \neq 0$, 
where we use the subscript  $r$  to denote the value of a variable on the 3-surface. Defining $\eta = \int \frac{d\tau}{6E_r}$, one can reduce the original 5-dimensional system \eqref{dyn_sys_axn} to a 3-dimensional system on this 3-surface:
\begin{subequations}
\begin{small}
\begin{align}
&  \frac{d E_{r}}{d\eta}=6 E_r\Big[-\frac{3}{2}E_r^2 -3a_1x_{1r}^2 -3a_2x_{2r}^2 +3b_1\sin^2 y_{1r} \nonumber \\
& +3b_2\sin^2 y_{2r} +3b_3\sin^2(y_{1r}-n y_{2r})\Big], \\
& \frac{dy_{1r}}{d\eta}=-\frac{b_1}{a_1}\sin(2y_{1r}) - \frac{b_3}{ a_1}\sin(2y_{1r} - 2n y_{2r}), \\
& \frac{dy_{2r}}{d\eta}=-\frac{b_2}{ a_2}\sin(2y_{2r}) + n\frac{b_3}{ a_2}\sin(2y_{1r} - 2n y_{2r}). 
\end{align}
\end{small}
\end{subequations}
In the figure   \ref{fig:my_label} is is represented a projection of the phase portrait for $n=0$ (decoupled limit) and for $n=1$, $n=2$ and $n=3$, on the 2-dimensional field space shown within the domain $-\pi\leq \frac{\phi_1}{f_1},\frac{\phi_2}{f_2}\leq \pi$. The red dot at the center being the origin corresponds to Minkowski solution. This extended domain is taken to show the periodic nature of the phase flow with a period $\pi$. 
 The parameter values used for the plots are  $a_1=\frac{1}{300}, \,\,a_2=\frac{1}{300}, \,\,b_1=\frac{11}{30}, \,\,b_2=\frac{43}{12}, \,\,b_3=\frac{107}{300}\,
$.
Asymptotically we have  $E^{\prime}=-\frac{3}{2}E^2. $
Using the fact that $t^{\prime}(\tau)=H_0^{-1}$ and  defining $\tau_0$ as the current value associated to $t=t_a$ (``age of the universe''), we have
\begin{small}
\begin{align*}
& E(\tau )= \frac{2}{3 \tau -3  \tau_{0}+2},  \quad  t(\tau
   )= \frac{t_a H_0+\tau -\tau_{0}}{H_0}.  
\end{align*}
\end{small}
Finally, as $t\rightarrow \infty$, we have 
\begin{small}
\begin{align*}
& E= \frac{2}{2 +3 H_0(t- t_a)}, \quad  H= \frac{2 H_0}{2 +3 H_0(t- t_a)} \rightarrow 0.
\end{align*}
\end{small}
\section{Oscillatory behavior: averaging}
\label{SECTT:4}
Let us assume $n\neq 0$. Defining the new fields \eqref{lin-phi-Psi}
the field equations become
\begin{subequations}
\begin{small}
\begin{align*}
   & \ddot{\Psi_1} + 3 H \dot{\Psi_1}\nonumber \\
   & =-\frac{\sqrt{3} b_3 H_0^2 \left(\sqrt{a_1} c n+\sqrt{a_2}\right) \sin \left(\frac{\Psi_1 \left(\sqrt{a_1} c n+\sqrt{a_2}\right)+\Psi_2 \left(\sqrt{a_1}
   n-\sqrt{a_2} c\right)}{\sqrt{3} \sqrt{a_1} \sqrt{a_2} \sqrt{c^2+1}}\right)}{\sqrt{a_1} \sqrt{a_2} \sqrt{c^2+1}} \nonumber \\
   & -\frac{\sqrt{3} b_1 H_0^2 \sin \left(\frac{\Psi_1-c
   \Psi_2}{\sqrt{3} \sqrt{a_1} \sqrt{c^2+1}}\right)}{\sqrt{a_1} \sqrt{c^2+1}}-\frac{\sqrt{3} b_2 c H_0^2 \sin \left(\frac{c \Psi_1+\Psi_2}{\sqrt{3} \sqrt{a_2}
   \sqrt{c^2+1}}\right)}{\sqrt{a_2} \sqrt{c^2+1}}, 
\end{align*}
\begin{align*}
   &     \ddot{\Psi_2} + 3 H \dot{\Psi_{2}} \nonumber \\
   & = \frac{\sqrt{3} b_3 H_0^2 \left(\sqrt{a_2} c-\sqrt{a_1} n\right) \sin \left(\frac{\Psi_1 \left(\sqrt{a_1} c n+\sqrt{a_2}\right)+\Psi_2 \left(\sqrt{a_1}
   n-\sqrt{a_2} c\right)}{\sqrt{3} \sqrt{a_1} \sqrt{a_2} \sqrt{c^2+1}}\right)}{\sqrt{a_1} \sqrt{a_2} \sqrt{c^2+1}}\nonumber \\
   & +\frac{\sqrt{3} b_1 c H_0^2 \sin \left(\frac{\Psi_1-c
   \Psi_2}{\sqrt{3} \sqrt{a_1} \sqrt{c^2+1}}\right)}{\sqrt{a_1} \sqrt{c^2+1}}-\frac{\sqrt{3} b_2 H_0^2 \sin \left(\frac{c \Psi_1+\Psi_2}{\sqrt{3} \sqrt{a_2}
   \sqrt{c^2+1}}\right)}{\sqrt{a_2} \sqrt{c^2+1}},
\end{align*}
\end{small}
The Friedmann equation becomes
\begin{small}
\begin{align*}
    & -6 b_3 H_0^2 \sin ^2\left(\frac{\Psi_1 \left(\sqrt{a_1} c n+\sqrt{a_2}\right)+\Psi_2 \left(\sqrt{a_1} n-\sqrt{a_2} c\right)}{2 \sqrt{3} \sqrt{a_1} \sqrt{a_2}
   \sqrt{c^2+1}}\right)\nonumber \\
   & -6 b_1 H_0^2 \sin ^2\scriptscriptstyle\left(\frac{\Psi_1-c \Psi_2}{2 \sqrt{3} \sqrt{a_1} \sqrt{c^2+1}}\right) \nonumber \\
   & -6 b_2 H_0^2 \sin ^2\left(\frac{c \Psi_1+\Psi_2}{2 \sqrt{3} \sqrt{a_2} \sqrt{c^2+1}}\right)\nonumber \\
   & +3 H^2-\rho -\frac{1}{2} {\dot{\Psi_1}}^2-\frac{1}{2} {\dot{\Psi_2}}^2=0.
\end{align*}
\end{small}
and the deceleration parameter is 
\begin{small}
\begin{align*}
  & q:= \nonumber \\
  & =\frac{1}{2} -\frac{3 {b_1} {H_0}^2 \sin ^2\left(\frac{{\Psi_1}-c {\Psi_2}}{\sqrt{3} \sqrt{{a_1}} \sqrt{c^2+1}}\right)}{{H}^2} \nonumber \\
  & -\frac{3 {b_2} {H_0}^2 \sin ^2\left(\frac{c {\Psi_1}+{\Psi_2}}{2 \sqrt{c^2+1} {f_2}}\right)}{{H}^2} \nonumber \\
   & -\frac{3 {b_3} {H_0}^2 \sin ^2\left(\frac{\frac{\sqrt{3} ({\Psi_1}-c {\Psi_2})}{\sqrt{{a_1}}}+\frac{3 n (c {\Psi_1}+{\Psi_2})}{{f_2}}}{3 \sqrt{c^2+1}}\right)}{{H}^2} \nonumber \\
   & +\frac{{\omega_1}^2 \tan ^2({\theta_1}) {\Psi_1}^2}{4 {H}^2}+\frac{{\omega_2}^2 \tan ^2({\theta_2}) {\Psi_2}^2}{4
   {H}^2}.
\end{align*}
\end{small}
\end{subequations}
Imposing the conditions 
\begin{small}
\begin{subequations}
\label{eq:33}
\begin{align}
  & c= \frac{a_1 \left(b_2+b_3 n^2\right)-a_2
   (b_1+b_3)+\sqrt{\Delta}}{2 \sqrt{a_1} \sqrt{a_2} b_3 n},\\
  & \frac{\omega_1^2}{H_0^2}=  \frac{a_1 \left(b_2+b_3 n^2\right)+a_2 (b_1+b_3)+\sqrt{\Delta}}{2
   a_1 a_2},\\
   & \frac{\omega_2^2}{H_0^2}=  \frac{a_1 \left(b_2+b_3 n^2\right)+a_2 (b_1+b_3)-\sqrt{\Delta}}{2
   a_1 a_2},
\end{align}
\end{subequations}
\end{small}
where \begin{small}
$ \Delta=\left(a_2 (b_1+b_3)-a_1 \left(b_2+b_3 n^2\right)\right)^2+4 a_1 a_2 b_3^2 n^2$ 
\end{small}, 
we obtain the decoupled oscillators
\begin{align*}
& \ddot \Psi_1+ 3 H \dot\Psi_1+ \omega_1^2 \Psi_{1}=0, \;  \ddot \Psi_2+ 3 H \dot\Psi_2+\omega_2^2 \Psi_{2}=0, 
\end{align*}
in the limit $\Psi_i\rightarrow 0$.  We can impose  additional conditions  $0\leq 8 a_1 a_2 \left(b_2 (b_1+b_3)+2
   b_1 b_3 n^2\right)\ll 1$ to have $\omega_2^2\approx 0$
   and $ \frac{\omega_1^2}{H_0^2}\approx  \frac{a_1 \left(b_2+b_3 n^2\right)+a_2 (b_1+b_3)}{a_1 a_2}$ such that $\Psi_1$ is a massive scalar field and  $\Psi_2$ a lighter one. 
   
\noindent   
 Defining the new variables $\theta_1, \theta_2$ and $T$,  through
 \begin{small}
\begin{align*}
    &\Psi_1=r_1 \cos(\theta_1) , \dot\Psi_1= r_1 \omega_1 \sin(\theta_1), \\
    & \Psi_2=r_2 \cos(\theta_2) , \dot\Psi_2= r_2 \omega_2 \sin(\theta_2), 
   \\
    & \theta_1= \tan ^{-1}\left(\frac{\dot{\Psi_{1}}}{ \omega_{1} \Psi_{1}}\right),  \;  \theta_2= \tan ^{-1}\left(\frac{\dot{\Psi_{2}}}{ \omega_{2} \Psi_{2}}\right), \;  T=\frac{H_0}{H+H_0}, 
    \end{align*} 
    \end{small}
    and the time derivative
    \begin{align*}
    & \frac{d f}{d \bar{\tau}}=\frac{1}{H_0+ H} \frac{d f}{d t},
\end{align*}
we obtain in the limit  $\Psi_i\rightarrow 0$ the reduced 
dynamical system 
\begin{subequations}
\label{Theta-T}
\begin{align}
& \frac{d \theta_1}{d \bar{\tau}}=-T  \bar{\omega}_1 - \frac{3}{2} (1-T) \sin ( 2\theta_{1}) , \\
& \frac{d \theta_2}{d \bar{\tau}}=-T  \bar{\omega}_2 - \frac{3}{2} (1-T) \sin ( 2\theta_{2}), \\
& \frac{d T}{d \bar{\tau}}=\frac{3}{2} T(1-T)^2, 
\end{align}
\end{subequations}
where we denote $\bar{\omega}_1= \frac{\omega_{1}}{H_0}$, $\bar{\omega}_2= \frac{\omega_{2}}{H_0}$
and we have used the fact that $q\rightarrow \frac{1}{2}$  as $ \Psi_i\rightarrow 0$.

That means that the decoupled oscillators mimic dust matter. The dynamical system \eqref{Theta-T} is regular. 

In the asymptotic regime $T=0$, which corresponds to $H\rightarrow \infty$, the equilibrium points are given by the equations $\sin ( 2\theta_{1})=0,  \;   \sin ( 2\theta_{2})=0. $
That is,  $\theta_{1}= l \frac{\pi}{2}, \;  \theta_{1}= m \frac{\pi}{2}, \;  l, m \in \mathbb{Z}.$ In the asymptotic regime $T=1$, which corresponds to $H\rightarrow 0$, the equilibrium points are given by the equations  $\frac{d \theta_1}{d \bar{\tau}}=-   \bar{\omega}_1, \; 
 \frac{d \theta_2}{d \bar{\tau}}=-\bar{\omega}_2. $ Eliminating $H_0$ we have $\theta_1 = -\omega_1 t,   \;   \theta_2 = -\omega_2 t$
as $H\rightarrow 0.$ 
Therefore, as $H\rightarrow 0$, we obtain  the decoupled oscillators
\begin{align}
\label{decoupled:oscillators}
\ddot \Psi_1+ \omega_1^2 \Psi_{1}=0, \; 
\ddot \Psi_2+\omega_2^2 \Psi_{2}=0.
\end{align}
The solutions of \eqref{decoupled:oscillators} can be written as
\begin{subequations}
\label{sol-oscillators}
\begin{align}
   & \Psi_i(t)= r_i \sin \left(t \omega_i-\Phi_{i}\right), \\
   & \dot \Psi_i(t)=r_i \omega_i \cos \left(t
   \omega_i- \Phi_{i}\right), i=1,2,
\end{align}
\end{subequations}
where $r_i$ and $\Phi_{i}$ are integration constants. 

\subsection{Variation of constants}
\label{Variation}

According to the Raychaudhuri equation \eqref{Raych}, $H$ is a monotonic decreasing function. Additionally, as the minimum of $V(\phi_1, \phi_2)$ in $(\phi_1, \phi_2)=(0,0)$ is approached, $H\rightarrow 0$. 
Therefore, as $H\rightarrow 0$, we obtain  the decoupled oscillators \eqref{decoupled:oscillators}. Motivated by the solution \eqref{sol-oscillators}, we use the variation of constants to propose the solution of the full KG system as 
\begin{subequations}
\begin{align*}
   & \Psi_i(t)= r_i(t) \sin \left(t \omega_i-\Phi_{i}(t)\right), \\
   & \dot \Psi_i=r_i(t)  \omega_i \cos \left(t
   \omega_i- \Phi_{i}(t) \right),  i=1,2,
\end{align*}
\end{subequations}
with inverse functions
\begin{align*}
& r_i^2={\Psi_i^2+\left(\frac{\dot \Psi_i}{\omega_i}\right)^2}, \; \Phi_i= t \omega_i -\tan^{-1}\left(\frac{\omega_i \Psi_i}{\dot \Psi_i}\right), i=1, 2,
\end{align*}
where $c$ and $\omega_1, \omega_2$ are undetermined constants.

The full system is given by 
\begin{widetext}
   \begingroup\makeatletter\def\f@size{8}\check@mathfonts
\begin{subequations}
\begin{align*}
   & \dot{r_1}= -3 r_1 H \cos ^2(\Phi_1-t \omega_1)-\frac{1}{2} r_1 \omega_1 \sin (2 (\Phi_1-t  \omega_{1}))  -\frac{\sqrt{3} \sqrt{a_2} b_1 H_0^2 \cos (\Phi_1-t \omega_1) \sin \left(\frac{c r_2 \sin (\Phi_2-t \omega_2)-r_1 \sin (\Phi_1-t  \omega_{1})}{\sqrt{3} \sqrt{a_1 \left(c^2+1\right)}}\right)}{\sqrt{a_1} \omega_1 \sqrt{a_2 \left(c^2+1\right)}} \nonumber \\
    & +\frac{\sqrt{3} b_2 c H_0^2 \cos
   (\Phi_1-t \omega_1) \sin \left(\frac{c r_1 \sin (\Phi_1-t \omega_1)+r_2 \sin (\Phi_2-t \omega_2)}{\sqrt{3} \sqrt{a_2
   \left(c^2+1\right)}}\right)}{\omega_1 \sqrt{a_2 \left(c^2+1\right)}} \nonumber \\
   & -\frac{\sqrt{3} b_3 H_0^2 \left(\sqrt{a_1} c n+\sqrt{a_2}\right) \cos
   (\Phi_1-t \omega_1) \sin \left(\frac{r_1 \left(-\sqrt{a_1} c n-\sqrt{a_2}\right) \sin (\Phi_1-t \omega_1)+r_2 \left(\sqrt{a_2} c-\sqrt{a_1} n\right) \sin
   (\Phi_2-t \omega_2)}{\sqrt{3} \sqrt{a_1} \sqrt{a_2 \left(c^2+1\right)}}\right)}{\sqrt{a_1} \omega_1 \sqrt{a_2 \left(c^2+1\right)}},
\end{align*}
\begin{align*}
 & \dot{r_2}=-3 r_2 H \cos ^2(\Phi_2-t \omega_2)-r_2 \omega_2 \sin (\Phi_2-t \omega_2) \cos
   (\Phi_2-t \omega_2)   + \frac{\sqrt{3} \sqrt{a_2} b_1 c H_0^2 \cos (\Phi_2-t \omega_2) \sin \left(\frac{c r_2 \sin (\Phi_2-t \omega_2)-r_1 \sin (\Phi_1-t \omega_1)}{\sqrt{3} \sqrt{a_1 \left(c^2+1\right)}}\right)}{\sqrt{a_1} \omega_2 \sqrt{a_2 \left(c^2+1\right)}}  \nonumber \\
   & +\frac{\sqrt{3} b_2 H_0^2 \cos
   (\Phi_2-t \omega_2) \sin \left(\frac{c r_1 \sin (\Phi_1-t \omega_1)+r_2 \sin (\Phi_2-t \omega_2)}{\sqrt{3} \sqrt{a_2
   \left(c^2+1\right)}}\right)}{\omega_2 \sqrt{a_2 \left(c^2+1\right)}}
   \nonumber \\
   & +\frac{\sqrt{3} b_3 H_0^2 \left(\sqrt{a_2} c-\sqrt{a_1} n\right) \cos
   (\Phi_2-t \omega_2) \sin \left(\frac{r_1 \left(-\sqrt{a_1} c n-\sqrt{a_2}\right) \sin (\Phi_1-t \omega_1)+r_2 \left(\sqrt{a_2} c-\sqrt{a_1} n\right) \sin
   (\Phi_2-t \omega_2)}{\sqrt{3} \sqrt{a_1} \sqrt{a_2 \left(c^2+1\right)}}\right)}{\sqrt{a_1} \omega_2 \sqrt{a_2 \left(c^2+1\right)}},  
\end{align*}
\begin{align*}
&\dot{\Phi_1}=\frac{3}{2} H \sin (2 (\Phi_1-t \omega_1))+\omega_1 \sin ^2(\Phi_1-t \omega_1)   -\frac{\sqrt{3} b_2 c H_0^2 \sin (\Phi_1-t \omega_1) \sin \left(\frac{c r_1 \sin (\Phi_1-t \omega_1)+r_2 \sin (\Phi_2-t \omega_2)}{\sqrt{3} \sqrt{a_2 \left(c^2+1\right)}}\right)}{r_1 \omega_1 \sqrt{a_2 \left(c^2+1\right)}} \nonumber \\
& \frac{\sqrt{3} \sqrt{a_2} b_1 H_0^2 \sin (\Phi_1-t \omega_1) \sin \left(\frac{c r_2 \sin (\Phi_2-t \omega_2)-r_1 \sin (\Phi_1-t \omega_1)}{\sqrt{3} \sqrt{a_1 \left(c^2+1\right)}}\right)}{\sqrt{a_1} r_1 \omega_1 \sqrt{a_2 \left(c^2+1\right)}} \nonumber\\
& +\frac{\sqrt{3} b_3 H_0^2 \left(\sqrt{a_1} c
   n+\sqrt{a_2}\right) \sin (\Phi_1-t \omega_1) \sin \left(\frac{r_1 \left(-\sqrt{a_1} c n-\sqrt{a_2}\right) \sin (\Phi_1-t \omega_1)+r_2 \left(\sqrt{a_2}
   c-\sqrt{a_1} n\right) \sin (\Phi_2-t \omega_2)}{\sqrt{3} \sqrt{a_1} \sqrt{a_2 \left(c^2+1\right)}}\right)}{\sqrt{a_1} r_1 \omega_1 \sqrt{a_2
   \left(c^2+1\right)}}, 
\end{align*}
\begin{align*}
   & \dot{\Phi_2}= \frac{3}{2} H \sin (2 (\Phi_2-t \omega_2))+\omega_2 \sin ^2(\Phi_2-t \omega_2)   -\frac{\sqrt{3} \sqrt{a_2} b_1 c H_0^2 \sin (\Phi_2-t \omega_2) \sin \left(\frac{c r_2 \sin (\Phi_2-t \omega_2)-r_1 \sin (\Phi_1-t \omega_1)}{\sqrt{3} \sqrt{a_1 \left(c^2+1\right)}}\right)}{\sqrt{a_1} r_2 \omega_2 \sqrt{a_2 \left(c^2+1\right)}} \nonumber \\
   & -\frac{\sqrt{3}
   b_2 H_0^2 \sin (\Phi_2-t \omega_2) \sin \left(\frac{c r_1 \sin (\Phi_1-t \omega_1)+r_2 \sin (\Phi_2-t \omega_2)}{\sqrt{3} \sqrt{a_2
   \left(c^2+1\right)}}\right)}{r_2 \omega_2 \sqrt{a_2 \left(c^2+1\right)}}\nonumber \\
   & -\frac{\sqrt{3} b_3 H_0^2 \left(\sqrt{a_2} c-\sqrt{a_1}
   n\right) \sin (\Phi_2-t \omega_2) \sin \left(\frac{r_1 \left(-\sqrt{a_1} c n-\sqrt{a_2}\right) \sin (\Phi_1-t \omega_1)+r_2 \left(\sqrt{a_2} c-\sqrt{a_1}
   n\right) \sin (\Phi_2-t \omega_2)}{\sqrt{3} \sqrt{a_1} \sqrt{a_2 \left(c^2+1\right)}}\right)}{\sqrt{a_1} r_2 \omega_2 \sqrt{a_2 \left(c^2+1\right)}},
\end{align*}
\begin{align*}
    & 3 H^2-\frac{1}{2} r_1^2  \omega_{1}^2 \cos ^2(\Phi_1-t \omega_1)-\frac{1}{2} r_2^2 \omega_2^2 \cos ^2(\Phi_2-t \omega_2)-\rho \nonumber \\
   & -6 b_1 H_0^2 \sin ^2\left(\frac{c r_2 \sin (\Phi_2-t \omega_2)-r_1 \sin (\Phi_1-t \omega_1)}{2 \sqrt{3} \sqrt{a_1 \left(c^2+1\right)}}\right)   -6 b_2 H_0^2 \sin ^2\left(\frac{c r_1 \sin (\Phi_1-t \omega_1)+r_2 \sin (\Phi_2-t \omega_{2})}{2 \sqrt{3} \sqrt{a_2 \left(c^2+1\right)}}\right) \\
   & -6 b_3 H_0^2 \sin ^2\left(\frac{r_2 \left(\sqrt{a_2} c-\sqrt{a_1} n\right) \sin (\Phi_2-t \omega_2)-r_1 \left(\sqrt{a_1} c n+\sqrt{a_2}\right) \sin ( \Phi_{1}-t \omega_1)}{2 \sqrt{3} \sqrt{a_1} \sqrt{a_2 \left(c^2+1\right)}}\right)=0.
\end{align*}
\end{subequations}
\endgroup
\end{widetext}
It is worth noticing that by conveniently choosing $c, \omega_1, \omega_2$ as in \eqref{eq:33}, the equations up to linear order in $r_1, r_2$ can be reduced to the usual equations for two decoupled harmonic oscillators. Let us define 
\begin{equation}
\label{eq:59}
    \Omega_i=\frac{ r_i^2 \omega_i}{6 H^2}, \;  \Omega=\frac{ \rho}{3 H^2}. 
\end{equation}
We obtain the full system: 
\begin{subequations}
\begin{widetext}
   \begingroup\makeatletter\def\f@size{8}\check@mathfonts
\begin{align}
\dot{\Omega_1}& = H \left(6 \Omega_1^2 \cos ^2(\Phi_1-t \omega_1)+\Omega_1 \left(6 \Omega_2 \cos ^2(\Phi_2-t \omega_2)+3 \left(\Omega -2 \cos
   ^2(\Phi_1-t \omega_1)\right)\right)\right) -\omega_1 \Omega_1 \sin (2 (\Phi_1-t \omega_1)) \nonumber \\
   & -\frac{\sqrt{2} \sqrt{a_2} b_1 \sqrt{\Omega_1} \cos (\Phi_1-t \omega_1) \sin \left(\frac{\sqrt{2} H \left(c \omega_1 \sqrt{\Omega_2} \sin
   (\Phi_2-t \omega_2)-\sqrt{\Omega_1}  \omega_{2} \sin (\Phi_1-t \omega_1)\right)}{\sqrt{a_1 \left(c^2+1\right)} \omega_1 \omega_{2}}\right)
   H_0^2}{\sqrt{a_1} \sqrt{a_2 \left(c^2+1\right)} H} \nonumber \\
   & + \frac{\sqrt{2} b_2 c \sqrt{\Omega_1} \cos (\Phi_1-t \omega_1) \sin \left(\frac{\sqrt{2} H \left(c \sqrt{\Omega_1}  \omega_{2} \sin (\Phi_1-t \omega_{1})+\omega_1 \sqrt{\Omega_2} \sin (\Phi_2-t \omega_2)\right)}{\sqrt{a_2 \left(c^2+1\right)} \omega_1 \omega_{2}}\right) H_0^2}{\sqrt{a_2
   \left(c^2+1\right)} H}\nonumber \\
   & +\frac{\sqrt{2} b_3 \left(\sqrt{a_1} c n+\sqrt{a_2}\right) \sqrt{\Omega_1} \cos (\Phi_1-t \omega_1) \sin
   \left(\frac{\sqrt{2} H \left(\left(\sqrt{a_1} c n+\sqrt{a_2}\right) \sqrt{\Omega_1}  \omega_{2} \sin (\Phi_1-t \omega_1)+\left(\sqrt{a_1} n-\sqrt{a_2} c\right)
   \omega_1 \sqrt{\Omega_2} \sin (\Phi_2-t \omega_2)\right)}{\sqrt{a_1} \sqrt{a_2 \left(c^2+1\right)} \omega_1  \omega_{2}}\right) H_0^2}{\sqrt{a_1}
   \sqrt{a_2 \left(c^2+1\right)} H},
\end{align}
\begin{align}
  \dot{\Omega_2} & =H \left(6 \Omega_1 \Omega_2 \cos ^2(\Phi_1-t \omega_1)+6
   \Omega_2^2 \cos ^2(\Phi_2-t \omega_2)+3 \Omega_2 \left(\Omega -2 \cos ^2(\Phi_2-t \omega_2)\right)\right)   - \omega_{2} \Omega_2 \sin (2 (\Phi_2-t  \omega_2))\nonumber \\
   & + \frac{\sqrt{2} \sqrt{a_2} b_1 c \sqrt{\Omega_2} \cos
   (\Phi_2-t \omega_2) \sin \left(\frac{\sqrt{2} H \left(c  \omega_{1} \sqrt{\Omega_2} \sin (\Phi_2-t \omega_2)-\sqrt{\Omega_1}  \omega_{2} \sin (\Phi_1-t
    \omega_{1})\right)}{\sqrt{a_1 \left(c^2+1\right)}  \omega_{1}  \omega_{2}}\right) H_0^2}{\sqrt{a_1} \sqrt{a_2 \left(c^2+1\right)} H} \nonumber \\
   & +\frac{\sqrt{2} b_2 \sqrt{\Omega_2} \cos (\Phi_2-t
    \omega_{2}) \sin \left(\frac{\sqrt{2} H \left(c \sqrt{\Omega_1}  \omega_{2} \sin (\Phi_1-t \omega_1)+ \omega_{1} \sqrt{\Omega_2} \sin (\Phi_2-t \omega_{2})\right)}{\sqrt{a_2 \left(c^2+1\right)}  \omega_{1}  \omega_{2}}\right) H_0^2}{\sqrt{a_2 \left(c^2+1\right)} H} \nonumber \\
   & +\frac{\sqrt{2} b_3 \left(\sqrt{a_1}
   n-\sqrt{a_2} c\right) \sqrt{\Omega_2} \cos (\Phi_2-t \omega_2) \sin \left(\frac{\sqrt{2} H \left(\left(\sqrt{a_1} c n+\sqrt{a_2}\right) \sqrt{\Omega_1}  \omega_{2}
   \sin (\Phi_1-t \omega_1)+\left(\sqrt{a_1} n-\sqrt{a_2} c\right)  \omega_{1} \sqrt{\Omega_2} \sin (\Phi_2-t \omega_2)\right)}{\sqrt{a_1} \sqrt{a_2
   \left(c^2+1\right)}  \omega_{1}  \omega_{2}}\right) H_0^2}{\sqrt{a_1} \sqrt{a_2 \left(c^2+1\right)} H},
\end{align}
\begin{align}
\dot{\Omega}  &    = 3 H \Omega  (\Omega +\Omega_{1}+ \Omega_{2}+ \Omega_{1} \cos (2 (\Phi_1-t  \omega_{1}))+ \Omega_{2} \cos (2 (\Phi_2-t \omega_2))-1),
\end{align}
\begin{align}
   \dot{\Phi_1} &=\frac{3}{2} H \sin (2 (\Phi_1-t \omega_1))+ \omega_{1} \sin ^2(\Phi_1-t \omega_1) \nonumber \\
   & + \frac{\sqrt{a_2} b_1 \sin (\Phi_1-t \omega_1) \sin \left(\frac{\sqrt{2} H \left(c  \omega_{1} \sqrt{\Omega_2} \sin (\Phi_2-t \omega_{2})-\sqrt{\Omega_1}  \omega_{2} \sin (\Phi_1-t \omega_1)\right)}{\sqrt{a_1 \left(c^2+1\right)} \omega_{1} \omega_{2}}\right) H_0^2}{\sqrt{2} \sqrt{a_1}
   \sqrt{a_2 \left(c^2+1\right)} H \sqrt{\Omega_1}} \nonumber \\
   & -\frac{b_2 c \sin (\Phi_1-t \omega_1) \sin \left(\frac{\sqrt{2} H \left(c \sqrt{\Omega_1}  \omega_{2} \sin (\Phi_1-t \omega_1)+ \omega_{1}
   \sqrt{\Omega_2} \sin (\Phi_2-t \omega_2)\right)}{\sqrt{a_2 \left(c^2+1\right)}  \omega_{1}  \omega_{2}}\right) H_0^2}{\sqrt{2} \sqrt{a_2 \left(c^2+1\right)} H
   \sqrt{\Omega_1}}\nonumber \\
   & -\frac{b_3 \left(\sqrt{a_1} c n+\sqrt{a_2}\right) \sin (\Phi_1-t \omega_1) \sin \left(\frac{\sqrt{2} H
   \left(\left(\sqrt{a_1} c n+\sqrt{a_2}\right) \sqrt{\Omega_1}  \omega_{2} \sin (\Phi_1-t \omega_1)+\left(\sqrt{a_1} n-\sqrt{a_2} c\right)  \omega_{1}
   \sqrt{\Omega_2} \sin (\Phi_2-t \omega_2)\right)}{\sqrt{a_1} \sqrt{a_2 \left(c^2+1\right)}  \omega_{1} \omega_{2}}\right) H_0^2}{\sqrt{2} \sqrt{a_1}
   \sqrt{a_2 \left(c^2+1\right)} H \sqrt{\Omega_1}},
\end{align}
\begin{align}
 \dot{\Phi_2}  &=\frac{3}{2} H \sin (2 (\Phi_2-t \omega_2)) + \omega_{2} \sin ^2(\Phi_2-t
 \omega_{2}) \nonumber \\
   & -\frac{\sqrt{a_2} b_1
   c \sin (\Phi_2-t \omega_2) \sin \left(\frac{\sqrt{2} H \left(c  \omega_{1} \sqrt{\Omega_2} \sin (\Phi_2-t \omega_2)-\sqrt{\Omega_1}  \omega_{2} \sin
   (\Phi_1-t \omega_1)\right)}{\sqrt{a_1 \left(c^2+1\right)}  \omega_{1}  \omega_{2}}\right) H_0^2}{\sqrt{2} \sqrt{a_1} \sqrt{a_2 \left(c^2+1\right)} H \sqrt{ \Omega_{2}}} \nonumber \\
 & -\frac{b_2 \sin ( \Phi_{2}-t \omega_2) \sin \left(\frac{\sqrt{2} H \left(c \sqrt{\Omega_1}  \omega_{2} \sin (\Phi_1-t \omega_1)+ \omega_{1} \sqrt{\Omega_2} \sin (\Phi_2-t
   \omega_{2})\right)}{\sqrt{a_2 \left(c^2+1\right)}  \omega_{1} \omega_{2}}\right) H_0^2}{\sqrt{2} \sqrt{a_2 \left(c^2+1\right)} H \sqrt{\Omega_2}}\nonumber \\
   & +\frac{b_3 \left(\sqrt{a_2} c-\sqrt{a_1} n\right) \sin (\Phi_2-t \omega_2) \sin \left(\frac{\sqrt{2} H \left(\left(\sqrt{a_1} c n+\sqrt{a_2}\right) \sqrt{\Omega_1}
    \omega_{2} \sin (\Phi_1-t \omega_1)+\left(\sqrt{a_1} n-\sqrt{a_2} c\right) \omega_{1} \sqrt{\Omega_2} \sin (\Phi_2-t \omega_2)\right)}{\sqrt{a_1}
   \sqrt{a_2 \left(c^2+1\right)} \omega_1 \omega_{2}}\right) H_0^2}{\sqrt{2} \sqrt{a_1} \sqrt{a_2 \left(c^2+1\right)} H \sqrt{\Omega_2}},
   \end{align}
\begin{align}
   & \dot{H}=-(1+q)H^2, 
\end{align}  
\endgroup 
\end{widetext}
\end{subequations} 

where the deceleration parameter is expressed as 
\begin{small}
\begin{align}
q:= -1 + \frac{3 \Omega }{2} + 3 \Omega_1 \cos ^2( t \omega_1-\Phi_1)+3 \Omega_2 \cos ^2(t \omega_2- \Phi_2).
\end{align}
 \end{small}

The Friedman equation is transformed to 
\begin{small}
\begin{align}
& 1= \Omega+{\Omega_1} \cos ^2(t {\omega_1}-{\Phi_1})+{\Omega_2} \cos ^2(t {\omega_2}-{\Phi_2}) \nonumber \\
& +\frac{2b_1 H_0^2}{H^2} \sin^2 \scriptscriptstyle\left(\frac{\frac{\sqrt{6} H \sqrt{{\Omega_1}} \sin (t {\omega_1}-{\Phi_1})}{{\Omega_1}}-\frac{\sqrt{6} c H \sqrt{{\Omega_2}}
   \sin (t {\omega_2}-{\Phi_2})}{{\Omega_2}}}{2\sqrt{c^2+1} {f_1}}\right) \nonumber\\
   & +\frac{2b_2 H_0^2}{H^2} \sin^2 \scriptscriptstyle\left(\frac{\frac{\sqrt{6} c H \sqrt{{\Omega_1}} \sin (t {\omega_1}-{\Phi_1})}{{\Omega_1}}+\frac{\sqrt{6} H \sqrt{{\Omega_2}} \sin (t {\omega_2}-{\Phi_2})}{{\Omega_2}}}{2\sqrt{c^2+1} {f_2}}\right)  \nonumber \\
   & + \frac{2b_3 H_0^2}{H^2}\sin^2 \scriptscriptstyle \left(\frac{\frac{\sqrt{6} H \sqrt{{\Omega_1}} (c {f_1} n+{f_2}) \sin (t {\omega_1}-{\Phi_1})}{{\Omega_1}}+\frac{\sqrt{6} H \sqrt{{\Omega_2}}
   ({f_1} n-c {f_2}) \sin (t {\omega_2}-{\Phi_2})}{{\Omega_2}}}{2\sqrt{c^2+1} {f_1} {f_2}}\right). 
\end{align} 
\end{small}

All the terms in the above equations are non-negative.  Then, it follows that $0\leq \Omega, \Omega_1, \Omega_2\leq 1$.

Imposing the conditions \eqref{eq:33}, or alternatively, imposing the conditions
\begin{small}
\begin{subequations}
\label{parameters}
\begin{align}
 & -\sqrt{a_1} \sqrt{a_2} b_3 \left(c^2-1\right) n+a_1 c \left(b_2+b_3 n^2\right)-a_2 c (b_1+b_3)=0,\\
 &\frac{\omega_{1}^2}{H_0^2 }= \frac{\left(2 \sqrt{a_1} \sqrt{a_2} b_3 c n+a_1 c^2 \left(b_2+b_3 n^2\right)+a_2 (b_1+b_3)\right)}{a_1 a_2 \left(c^2+1\right)},\\
 &\frac{\omega_2^2}{H_0^2 }= \frac{\left(-2 \sqrt{a_1} \sqrt{a_2} b_3
   c n+a_1 \left(b_2+b_3 n^2\right)+a_2 c^2 (b_1+b_3)\right)}{a_1 a_2 \left(c^2+1\right)},
\end{align}
\end{subequations}
\end{small}
which admit as special solution \eqref{eq:33}, the undesired terms of zero order in $H$ are eliminated. 
That is, by defining $x= \left(\Omega_1, \Omega_2 , \Omega, \Phi_1, \Phi_2\right)^T$, by expanding in Taylor's series around $H=0$ we have the $6$- dimensional system 
\begin{subequations}
\label{Nonstandardform2}
\begin{align}
& \dot x = H F^{[1]}(t, x)  +\mathcal{O}\left(H^2\right), \;  x(0)=x_0, \;  t \geq 0, \\
& \dot H= -  G^{[2]}(t,x) H^2, \label{expdotepsilon}
\end{align}
\end{subequations}
as $H\rightarrow 0$,  
where 
   \begingroup\makeatletter\def\f@size{7.5}\check@mathfonts
\begin{align}
\label{def1}
& F^{[1]}(t,x)= \nonumber \\
& \scriptscriptstyle\left( \begin{array}{c}  
   -3 \Omega_1 \left(2 (1-\Omega_1) \cos ^2(t \omega_1-  \Phi_{1})-2
  \Omega_{2} \cos ^2(t  \omega_{2}- \Phi_{2})- \Omega \right)\\
  -3  \Omega_{2} \left(-2
   \Omega_1 \cos ^2(t \omega_1-  \Phi_{1})+2 (1-\Omega_{2}) \cos ^2(t  \omega_{2}- \Phi_{2})-\Omega \right) \\
-  3 \Omega  \left(1-2 \Omega_{1} \cos ^2(t \omega_1-  \Phi_{1})-2 \Omega_{2} \cos
   ^2(t  \omega_{2}- \Phi_{2}))-\Omega \right)\\
-\frac{3}{2}\sin (2 (t \omega_{1}-\Phi_1))
\\
-\frac{3}{2}\sin (2 (t \omega_{2}-\Phi_2))
\\
\end{array}\right), \\
& G^{[2]}(t,x)= 3 {\Omega_1} \cos ^2({\Phi_1}-t {\omega_1})+3 {\Omega_2} \cos ^2({\Phi_2}-t {\omega_2})+\frac{3 \Omega }{2}.
\end{align}
\endgroup
Following arguments stressed in \cite{Fajman:2021cli} but for the two-scalar field setting, since we obtain the equations of motion of two decoupled oscillators \eqref{decoupled:oscillators}  as $H\rightarrow 0$, we can study the full system as perturbed harmonic oscillators and apply averaging techniques for vector functions $f_n(t,x)$ which have two independent periods $T_n, \;  n =1, 2$, where $H$ is considered as a time-dependent and itself is governed by the evolution equation \eqref{expdotepsilon}. A surprising feature of such an approach is the possibility of exploiting the fact that $H$ is strictly decreasing and goes to zero, therefore, it can be promoted to a time-dependent perturbation parameter; controlling the magnitude of the error between solutions of full and time-averaged problems. 
Hence, with strictly decreasing $H$ the error should decrease as well. Therefore, it is possible to obtain information about the large-time behavior of a more complicated full system via an analysis of simpler averaged system equations. 
Indeed,  from $G^{[2]}(t,x)\geq 0$, it follows that $H$ is a decreasing function of time, that allows defining a decreasing sequence of parameters to construct asymptotic expansions. Therefore, although $H$ is a function of time and  it is not properly a ``constant parameter'', as $t \rightarrow 0$, we can choose a sequence $t_n$ and  for $n$ large enough such that for $H(t_n)= H_n$, which is actually a parameter. This procedure is supported by numerical simulations (see related work \cite{Fajman:2020yjb}). Similar arguments can be provided using the tools of \cite{Alho:2015cza} or, as is the case in this research, by implementing an analogous program as in the references \cite{Leon:2021lct, Leon:2021rcx, Leon:2021hxc} but for averaging multi-scalar fields cosmologies. 

\subsubsection{Regular asymptotic expansions}

In this section we study the system \eqref{Nonstandardform2} with the definitions \eqref{def1}. We propose the following expansions: 
\begin{subequations}
\begin{align}
 &\Omega_{i} \equiv \Omega_i (t)= \Omega_{i0}(t)+{H}(t)\Omega_{i1}(t) + \mathcal{O}(H^2), \\
 &\Omega\equiv \Omega (t)= \Omega_{0}(t)+H(t)\Omega_{01}(t) + \mathcal{O}(H^2),\\
 & \Phi_i \equiv  \Phi_i(t)=  \Phi_{i0}(t)+H(t)\Phi_{i1}(t) + \mathcal{O}(H^2),
\end{align}
\end{subequations}
for $i=1,2$. 
Applying the chain rule and  using the fact that  $f \frac{d H}{dt}=\mathcal{O}(H^2)$, $f=\{\Omega_{01}, \Omega_{11}, \Omega_{21}, \Phi_{11}, \Phi_{21}\}$, according \eqref{expdotepsilon}, we obtain  
\begin{subequations}
\begin{align}
 &\frac{d \Omega_i}{d t} =\frac{d \Omega_{i0}}{d t}+ H \frac{d \Omega_{i1}}{d t}  + \mathcal{O}(H^2), \\
&\frac{d \Omega}{d t} =\frac{d \Omega_{0}}{d t}+ H \frac{d \Omega_{01}}{d t}  + \mathcal{O}(H^2), \\
  &\frac{d \Phi_i}{d t} =\frac{d \Phi_{i0}}{d t}+ H \frac{d \Phi_{i1}}{d t}  + \mathcal{O}(H^2).
\end{align}
\end{subequations}
By integrating order by order: 
\begin{align}
        & O(H^0):   \left\{
\begin{array}{c}
\frac{d \Omega_{10}}{d t}=\frac{d \Omega_{20}}{d t}=\frac{d\Omega_0}{d t}=\frac{d \Phi_{10}}{d t}=\frac{d \Phi_{20}}{d t}=0,
\end{array}\right.
\end{align}
\begin{widetext}
\begin{small}
\begin{align}
& O(H):   \left\{
\begin{array}{c}
\frac{d \Omega_{11}}{dt}=3 \Omega_{10} ((\Omega_{10}-1) \cos (2 \Phi_{10}-2
   t \omega_{1})+\Omega_{20} \cos (2 \Phi_{20}-2 t  \omega_{2})+ \Omega _{0}+\Omega_{10}+\Omega_{20}-1),\\\\
\frac{d \Omega_{21}}{dt}=3 \Omega_{20} (\Omega_{10} \cos (2
   \Phi_{10}-2 t \omega_{1})+(\Omega_{20}-1) \cos (2 \Phi_{20}-2 t  \omega _{2})+ \Omega _{0}+\Omega_{10}+\Omega_{20}-1),\\\\
\frac{d \Omega_{01}}{dt}=  3  \Omega_{0}
   (\Omega_{10} \cos (2 \Phi_{10}-2 t  \omega_{1})+\Omega_{20} \cos (2 \Phi_{20}-2 t  \omega_{2})+ \Omega_{0}+\Omega_{10}+\Omega_{20}-1),\\\\
 \frac{d \Phi_{11}}{dt}= \frac{3}{2} \sin (2 (\Phi_{10}-t
   \omega_{1})),\\\\
 \frac{d \Phi_{21}}{dt}= \frac{3}{2} \sin (2 (\Phi_{20}-t \omega_{2})).
\end{array}\right.
\end{align}
\end{small}
we obtain 
\begin{subequations}
\begin{small}
\begin{align*}
& \Phi_{10}(t)= \Phi_{10},\Phi_{11}(t)= c_1+\frac{3 \cos (2 (\Phi_{10}-t  \omega_{1}))}{4  \omega _{1}},\Phi_{20}(t)= \Phi_{20}, \Phi_{21}(t)=
   c_2+\frac{3 \cos (2 (\Phi_{20}-t  \omega_{2}))}{4  \omega_{2}},\\
   &\Omega_{0}(t)=  \Omega_{0},   \Omega_{10}(t)= \Omega_{10},\Omega_{20}(t)= \Omega_{20}, 
\\
   & \Omega_{01}(t)= c_3-\frac{3  \Omega_{0} \Omega_{10} \sin (2 (\Phi_{10}-t  \omega_{1}))}{2  \omega_{1}}-\frac{3  \Omega_{0} \Omega_{20} \sin (2 (\Phi_{20}-t \omega_{2}))}{2
    \omega_{2}}+3 t  \Omega_{0} (\Omega_{0}+\Omega_{10}+\Omega_{20}-1), \\
   &\Omega_{11}(t)= c_4-\frac{3 (\Omega_{10}-1) \Omega_{10} \sin (2 (\Phi_{10}-t   \omega_{1}))}{2  \omega_{1}}-\frac{3 \Omega_{10} \Omega_{20} \sin (2 (\Phi_{20}-t  \omega_{2}))}{2  \omega_{2}}+3 t \Omega_{10} ( \Omega_{0}+\Omega_{10}+\Omega_{20}-1),\\
   & \Omega_{21}(t)= c_5-\frac{3 \Omega_{10} \Omega_{20} \sin (2 (\Phi_{10}-t \omega_{1}))}{2  \omega_{1}}-\frac{3 (\Omega_{20}-1) \Omega_{20} \sin (2 (\Phi_{20}-t  \omega_{2}))}{2  \omega_{2}}+3 t \Omega_{20} (\Omega_{0}+\Omega_{10}+\Omega_{20}-1).
\end{align*}
\end{small}
\end{subequations}
\end{widetext}
Additionally, two compatibility equations must be satisfied which come from the Hubble equation expanded in series of $H$, say, 
\begin{align*}
& \Omega_0 + \Omega_{10} +\Omega_{20} =1,\quad   \Omega_{01}+\Omega_{11}+\Omega_{21}=0.
\end{align*}
Using the first condition, the secular terms proportional to $t$ are eliminated, and using the second condition we have $c_3+c_4+c_5=0$. These conditions, however, limit the applicability of the model.

In general, the regular asymptotic expansion fails in presence of resonant terms. One alternative can be using the method of multiples scales, e.g., a Poincarè-Lindstedt's - like method, where we set $t_1= (1+ \omega_1 h + \omega_2 h^2 + \ldots) t$, $t_2= h t$, where $h=H/H_0 \ll 1$. This method would determine solutions of perturbed oscillators by suppressing resonant forcing terms that would yield spurious secular terms in the asymptotic expansions. The $t_1$ and $t_2$ time variables are introduced to keep a well ordered expansion, 
where $t_1$ is the regular (or ``fast'') time variable and $t_2$ is a new variable describing the ``slow-time'' dependence of the solution. The idea is to use any freedom that is in the  $t$-dependence of $t_1$ and $t_2$ to minimize the approximation's error, and whenever is possible to remove unbounded or secular terms. 
To our knowledge, this method has not been implemented yet in the cosmological setup. However, basic examples of oscillators show that by implementing a time-averaged version of the model instead of multiple scales, the issue of secular terms is overcome; getting the same accuracy as in the two-timing method. We elaborate more on averaging techniques in subsection \ref{SECT:II}.

\subsubsection{Time-averaging}
\label{SECT:II}
 If we have vector functions $f_n(t,x)$ which have $N$ independent periods $T_n, \;  n =1, \ldots, N$, we take the averaging
\begin{equation}
\dot y= \varepsilon f^{(0)}(y), \;  y(0)=x_0, 
\end{equation}
where 
\begin{equation}
f^{(0)}(y)=\sum_{n=1}^{N} \frac{1}{T_n} \int_{0}^{T_n} f_n(t,y) d t, 
\end{equation}
where $y$ is considered as a parameter that is kept constant during integration.

Assuming $\omega_1\neq \omega_2$, with  $H$ playing the role of $\varepsilon$, $N=2$, $T_1=\frac{2 \pi}{\omega_1}$ and  $T_2=\frac{2 \pi}{\omega_2}$ we can use the following averaging procedure: 
\begin{align}
\label{timeavrg}
f^{(0)}(y)= \frac{\omega_1}{2\pi} \int_{0}^{\frac{2 \pi}{\omega_1}} f_1(t,y) d t+ \frac{\omega_2}{2\pi} \int_{0}^{\frac{2 \pi}{\omega_2}} f_2(t,y) d t,
\end{align}
where  the vector field $f(t,y)$ in the right-hand-side of the equation must be the sum of two vector functions $f_1(t,y)$ and  $f_2(t,y)$ where each of them is periodic with one period. 

The averaged system obtained using such approach  is
\begin{subequations}
\label{avergsystem}
\begin{align}
& \left( \begin{array}{c}
     \partial_t{\bar{\Omega}_1}  \\
     \partial_t{\bar{\Omega}_2}  \\
     \partial_t{\bar{\Omega}}\\
     \partial_t{\bar{\Phi}_1}\\
     \partial_t{\bar{\Phi}_2}
\end{array}\right)= - 3 H \left( \begin{array}{c}  
 \bar{\Omega}_1 \left(1- \bar{\Omega}_1- \bar{\Omega}_2 -\bar{\Omega}\right) \\
 \bar{\Omega}_2 \left(1- \bar{\Omega}_1- \bar{\Omega}_2 -\bar{\Omega} \right) \\
 \bar{\Omega} \left(1- \bar{\Omega}_1- \bar{\Omega}_2 -\bar{\Omega} \right)\\
0
\\
0
\\
\end{array}\right), \label{equx}\\
&\dot H= -\frac{3}{2}H^2 \left(\bar{\Omega}_1+ \bar{\Omega}_2 +\bar{\Omega} \right). \label{EQ:81b}
\end{align}
\end{subequations}
Proceeding in an analogous way as in references \cite{Alho:2015cza,Alho:2019pku,Leon:2021lct,Leon:2021rcx} we implement a local nonlinear transformation:   
\begin{align}
&\mathbf{x}_0:=\left({\Omega}_{10}, {\Omega}_{20}, {\Omega}_0, {\Phi}_{10},  {\Phi}_{20}\right)^T  \nonumber \\
& \mapsto \mathbf{x}:=\left({\Omega}_{1}, {\Omega}_{2}, {\Omega}, {\Phi}_{1},  {\Phi}_{2}\right)^T \nonumber \\
& \mathbf{x}=\psi(\mathbf{x}_0):=\mathbf{x}_0 + H \mathbf{g}(H, \mathbf{x}_0,t), \label{Appquasilinear211}
\\
& \mathbf{g}(H, \mathbf{x}_0,t)= \left(\begin{array}{c}
    g_1(H ,{\Omega}_{10}, {\Omega}_{20}, {\Omega}_0, {\Phi}_{10},  {\Phi}_{20}, t)\\
    g_2(H , {\Omega}_{10}, {\Omega}_{20}, {\Omega}_0, {\Phi}_{10},  {\Phi}_{20}, t)\\
    g_3(H , {\Omega}_{10}, {\Omega}_{20}, {\Omega}_0, {\Phi}_{10},  {\Phi}_{20}, t)\\
    g_4(H , {\Omega}_{10}, {\Omega}_{20}, {\Omega}_0, {\Phi}_{10},  {\Phi}_{20}, t)\\
		g_5(H , {\Omega}_{10}, {\Omega}_{20}, {\Omega}_0, {\Phi}_{10},  {\Phi}_{20}, t)\\
 \end{array}\right).   \label{eqT55}
\end{align}
Taking time derivative in both sides of \eqref{Appquasilinear211} with respect to $t$ we obtain 
\begin{align}
    & \dot{\mathbf{x}_0}+ \dot{H} \mathbf{g}(H, \mathbf{x}_0,t)  \nonumber \\
    & + H \Bigg(\frac{\partial }{\partial t} \mathbf{g}(H, \mathbf{x}_0,t) + \dot{H} \frac{\partial }{\partial H} \mathbf{g}(H, \mathbf{x}_0,t) \nonumber \\
    & + D_{\mathbf{x}_0} \mathbf{g}(H, \mathbf{x}_0,t) \cdot \dot{\mathbf{x}_0}\Bigg)  = \dot{\mathbf{x}}, \label{EQT56}
    \end{align}
    where 
    \begin{equation}
        D_{\mathbf{x}_0} \mathbf{g}(H, \mathbf{x}_0,t),
    \end{equation}    
is the  Jacobian matrix of $\mathbf{g}(H, \mathbf{x}_0,t)$ for the vector  $\mathbf{x}_0$.  The function $\mathbf{g}(H, \mathbf{x}_0,t)$ is conveniently chosen. 
\newline By substituting \eqref{equx} and \eqref{Appquasilinear211} in \eqref{EQT56} we obtain 
\begin{align}
       & \Bigg(\mathbf{I}_5 + H D_{\mathbf{x}_0} \mathbf{g}(H, \mathbf{x}_0,t)\Bigg) \cdot \dot{\mathbf{x}_0}= H \mathbf{f}(\mathbf{x}_0 + H \mathbf{g}(H, \mathbf{x}_0,t),t) \nonumber \\
       & -H \frac{\partial }{\partial t} \mathbf{g}(H, \mathbf{x}_0,t) -\dot{H} \mathbf{g}(H, \mathbf{x}_0,t) \nonumber \\
       & -H \dot{H} \frac{\partial }{\partial H} \mathbf{g}(H, \mathbf{x}_0,t), 
\end{align}
where 
$\mathbf{I}_5$ is the $5\times 5$ identity matrix.
                
Then we obtain 
  \begin{small}  
  \begin{align}
 & \dot{\mathbf{x}_0} = \Bigg(\mathbf{I}_5 + H D_{\mathbf{x}_0} \mathbf{g}(H, \mathbf{x}_0,t)\Bigg)^{-1} \nonumber \\
 & \cdot \Bigg(H \mathbf{f}(\mathbf{x}_0 + H \mathbf{g}(H, \mathbf{x}_0,t),t)-H \frac{\partial }{\partial t} \mathbf{g}(H, \mathbf{x}_0,t) \nonumber\\
 & -\dot{H} \mathbf{g}(H, \mathbf{x}_0,t) -H \dot{H} \frac{\partial }{\partial H} \mathbf{g}(H, \mathbf{x}_0,t)\Bigg).  
\end{align}
\end{small}
Using eq. \eqref{EQ:81b}, we have $ \dot{H}= \mathcal{O}(H^2)$. Hence,
\begin{align}
    & \dot{\mathbf{x}_0} = \underbrace{\Bigg(\mathbf{I}_5 - H D_{\mathbf{x}_0} \mathbf{g}(0, \mathbf{x}_0,t) +  \mathcal{O}(H^2)\Bigg)}_{5\times 5 \: \text{matrix}} \nonumber \\
    & \cdot \underbrace{\Bigg(H \mathbf{f}(\mathbf{x}_0, t)-H \frac{\partial }{\partial t} \mathbf{g}(0, \mathbf{x}_0,t) +   \mathcal{O}(H^2)\Bigg)}_{5\times 1 \; \text{vector}}  \nonumber \\
		& \implies \dot{\mathbf{x}_0}= \underbrace{H \mathbf{f}(\mathbf{x}_0, t)-H \frac{\partial }{\partial t} \mathbf{g}(0, \mathbf{x}_0,t) +   \mathcal{O}(H^2)}_{5\times 1 \; \text{vector}}.\label{eqT59}
    \end{align} 
\noindent The strategy is to use eq. \eqref{eqT59} for choosing conveniently $\frac{\partial }{\partial t} \mathbf{g}(0, \mathbf{x}_0,t)$ to prove that 
\begin{align}
 & \dot{\Delta\mathbf{x}_0}= -H G(\mathbf{x}_0, \bar{\mathbf{x}}) +   \mathcal{O}(H^2), \label{EqY60}
  \end{align}
where $\bar{\mathbf{x}}=(\bar{\Omega}_{1}, \bar{\Omega}_{2}, \bar{\Omega}, \bar{\Phi}_{1},  \bar{\Phi}_{2})^T$ and  $\Delta\mathbf{x}_0=\mathbf{x}_0 - \bar{\mathbf{x}}$. The function $G(\mathbf{x}_0, \bar{\mathbf{x}})$ is unknown at this stage and it is independent of periods $T_1$ and $T_2$ due to it is independent of $t$. 

By construction we neglect dependence of $\partial g_i/ \partial t$ and $g_i$ on $H$, i.e., assume $\mathbf{g}=\mathbf{g}(\mathbf{x}_0,t)$ because dependence of $H$ is dropped out along with higher order terms eq. \eqref{eqT59}. Next, we solve a partial differential equation  for $\mathbf{g}(\mathbf{x}_0,t)$ given by:  
\begin{align}
     & \frac{\partial }{\partial t} \mathbf{g}(\mathbf{x}_0,t) = \mathbf{f}(\mathbf{x}_0, t) - \bar{\mathbf{f}}(\bar{\mathbf{x}}) + G(\mathbf{x}_0, \bar{\mathbf{x}}) \nonumber \\
		& = \mathbf{f}_1(\mathbf{x}_0, t) - \bar{\mathbf{f}}_1(\bar{\mathbf{x}}) + \mathbf{f}_2(\mathbf{x}_0, t) - \bar{\mathbf{f}}_2(\bar{\mathbf{x}}) +G(\mathbf{x}_0, \bar{\mathbf{x}}), \label{eqT60}
\end{align}
where we have considered $\mathbf{x}_0$ and  $t$ as independent variables and  we have assumed that the function in the  left hand side denoted by $ \mathbf{f}(\mathbf{x}_0, t) - \bar{\mathbf{f}}(\bar{\mathbf{x}})$ can be separated in the  sum of two vector functions with independent periods $T_1$ and $T_2$. Hence, the right hand side of \eqref{eqT60} is the sum of two almost periodic functions $\mathbf{f}_1(\mathbf{x}_0, t) - \bar{\mathbf{f}}_1(\bar{\mathbf{x}})$  and $\mathbf{f}_2(\mathbf{x}_0, t) - \bar{\mathbf{f}}_2(\bar{\mathbf{x}})$ of independent periods $L_1=\frac{2\pi}{\omega_1}$  and $L_2=\frac{2\pi}{\omega_2}$ for large times, respectively. Then, implementing the average process \eqref{timeavrg} on right hand side of \eqref{eqT60}, where slow-varying dependence of quantities $\mathbf{x}_0$ and $\bar{\mathbf{x}}$  on $t$ are ignored through  averaging process, we obtain \begin{align}
    & \frac{1}{L_1}\int_0^{L_1} \Bigg[\mathbf{f}_1(\mathbf{x}_0, s) - \bar{\mathbf{f}}_1(\bar{\mathbf{x}}) +G(\mathbf{x}_0, \bar{\mathbf{x}}) \Bigg] ds  \nonumber  \\
		& + \frac{1}{L_2}\int_0^{L_2} \Bigg[\mathbf{f}_2(\mathbf{x}_0, s) - \bar{\mathbf{f}}_2(\bar{\mathbf{x}})  \Bigg] ds\\
    & = \bar{\mathbf{f}}_1( {\mathbf{x}}_0)-\bar{\mathbf{f}}_1(\bar{\mathbf{x}} ) + \bar{\mathbf{f}}_2( {\mathbf{x}}_0)-\bar{\mathbf{f}}_2(\bar{\mathbf{x}} )+ G(\mathbf{x}_0, \bar{\mathbf{x}}). \label{newaverage}
\end{align}
Defining 
\begin{align}
 &  G(\mathbf{x}_0, \bar{\mathbf{x}}):=-\Big(\bar{\mathbf{f}}_1( {\mathbf{x}}_0)-\bar{\mathbf{f}}_1(\bar{\mathbf{x}} ) + \bar{\mathbf{f}}_2( {\mathbf{x}}_0)-\bar{\mathbf{f}}_2(\bar{\mathbf{x}} )\Big)\nonumber \\
  & =  -\left(\bar{\mathbf{f}}( {\mathbf{x}}_0)-\bar{\mathbf{f}}(\bar{\mathbf{x}})\right),
\end{align} the average \eqref{newaverage} is zero so that $\mathbf{g}(\mathbf{x}_0,t)$ is bounded.
\newline 
Finally, eq. \eqref{EqY60} transforms to 
\begin{align}
 & \dot{\Delta\mathbf{x}_0}= H \left(\bar{\mathbf{f}}( {\mathbf{x}}_0)-\bar{\mathbf{f}}(\bar{\mathbf{x}})\right) +   \mathcal{O}(H^2),  \label{EqY602}
  \end{align}
and eq. \eqref{eqT60} 
is simplified to 
\begin{align}
     & \frac{\partial }{\partial t} \mathbf{g}(\mathbf{x}_0,t) = \mathbf{f}(\mathbf{x}_0, t) - \bar{\mathbf{f}}( \mathbf{x}_0). \label{eqT602}
\end{align}

Theorem \ref{LFZ1} establishes the existence of the vector \eqref{eqT55}. 

\begin{thm}
\label{LFZ1} Let $\bar{\Omega}_1, \bar{\Omega}_2, \bar{\Omega}, \bar{\Phi}_1, \bar{\Phi}_2$,
 and $H$ be  defined  functions that  satisfy  averaged  equations \eqref{avergsystem}. Then, there exist continuously differentiable functions  $g_1, g_2, g_3$, $g_4$ and $g_5$,  such that   $\Omega_{1}, \Omega_{2}, \Omega, {\Phi}_{1}, {\Phi}_{2}$ are locally given by \eqref{Appquasilinear211}, where ${\Omega}_{10}, {\Omega}_{20}, {\Omega}_0, {\Phi}_{10},  {\Phi}_{20}$ are order zero approximations of them as $H\rightarrow 0$. Then,  functions   ${\Omega}_{10}, {\Omega}_{20}, {\Omega}_0, {\Phi}_{10}$, and ${\Phi}_{20}$ and averaged solution \newline $\bar{\Omega}_1, \bar{\Omega}_2, \bar{\Omega}, \bar{\Phi}_1$, and  $\bar{\Phi}_2$ have the same limit as $t\rightarrow \infty$. 
\end{thm}

Theorem \ref{LFZ1} implies that  ${\Omega}_i, \Omega, \Phi_i, i=1,2$  evolves according to the averaged equations \eqref{avergsystem} as $H\rightarrow 0$.

\subsection{Phase space analysis of the time-averaged systems}
\label{SSection5}

Introducing the time variable $\tau= \tau$, we obtain the guiding system: 
\begin{subequations}\label{XXsispaXX}
\begin{align}
& \frac{d\bar{\Omega}_1}{d \tau}= -3 \bar{\Omega}_1 \left(1- \bar{\Omega}_1- \bar{\Omega}_2 -\bar{\Omega}\right), \\
& \frac{d\bar{\Omega}_2}{d \tau}= -3 \bar{\Omega}_2 \left(1- \bar{\Omega}_1- \bar{\Omega}_2 -\bar{\Omega}\right), \\
&  \frac{d\bar{\Omega}}{d \tau}= -3 \bar{\Omega} \left(1- \bar{\Omega}_1- \bar{\Omega}_2 -\bar{\Omega}\right).
\end{align}
\end{subequations}
The equilibrium points of the guiding system \eqref{XXsispaXX} are the origin, with eigenvalues ${-3,-3,-3}$ and the points in the plane $\bar{\Omega}_1+ \bar{\Omega}_2+ \bar{\Omega}=1$ with eigenvalues 
$\{3,0,0\}$. Therefore, that are nonhyperbolic, with a 1D unstable manifold and a 2D center manifold. 

Taking an arbitrary point \newline $\left(\bar{\Omega}_1, \bar{\Omega}_2, \bar{\Omega}\right)= \left(\bar{\Omega}_{1c}, \bar{\Omega}_{2c}, 1-\bar{\Omega}_{1c}-\bar{\Omega}_{2c}\right)$, $\Omega>0$ lying on this plane,  and defining the new variables 
\begin{align}
  &U= (1- \bar{\Omega}_{1c}- \bar{\Omega}_{2c}) ( \bar{\Omega}_1+ \bar{\Omega}_2+ \bar{\Omega}-1),\\
  & V_1 = ( \bar{\Omega}_{1}+ \bar{\Omega}_{2}) ( \bar{\Omega}_{1c}+ \bar{\Omega}_{2c}-1)+ \bar{\Omega} ( \bar{\Omega}_{1c}+\bar{\Omega}_{2c}), \\
  & V_2= - \bar{\Omega}_{2c} ( \bar{\Omega}_{1}+\bar{\Omega})- \bar{\Omega}_{2}   \bar{\Omega}_{2c}+ \bar{\Omega}_{2},  
\end{align} that translates the equilibrium point to the origin, 
we obtain the system 
\begin{align*}
 & \frac{dU}{d\tau}= 3 U, \;  \frac{dV_{1}}{d\tau}= \frac{3 V_{1} U}{1- \bar{\Omega}_{1c}- \bar{\Omega}_{2c}}, \; \frac{dV_2}{d\tau}= \frac{3 V_2 U}{1- \bar{\Omega}_{1c}- \bar{\Omega}_{2c}}.
\end{align*}
Hence, the unstable manifold is given locally by the graph 
\begin{align}
    & \Big\{(U,V_1,V_2): V_1= h_1(U), V_2= h_2(U), \nonumber\\
    & h_1(0)=0, h_2(0)=0, \nonumber\\
    & h_1^{\prime}(0)=0, h_2^{\prime}(0)=0, |U|<\delta\Big\}. 
\end{align}
Using the invariance of the unstable manifold we obtain for $U\neq 0$ that
\begin{align}
&\frac{ h_1(U)}{1- \bar{\Omega}_{1c}- \bar{\Omega}_{2c}}- h_1^{\prime}(U)=0,\\
& \frac{ h_2(U)}{1- \bar{\Omega}_{1c}- \bar{\Omega}_{2c}}-  h_2^{\prime}(U)=0.
\end{align}
The solutions are the trivial $h_1(U)=h_2(U)=0$ or
\begin{equation}
h_1(U)= c_1 e^{\frac{U}{1- \bar{\Omega}_{1c}- \bar{\Omega}_{2c}}}, \;  h_2(U)= c_2 e^{\frac{U}{1- \bar{\Omega}_{1c}- \bar{\Omega}_{2c}}}. 
\end{equation}
The last solutions do not satisfy the condition $h_1(0)=0, h_2(0)=0, h_1^{\prime}(0)=0, h_2^{\prime}(0)=0$. Hence, the unstable manifold is the $U$-axis and  the dynamics on the unstable manifold is given by $ U^{\prime}= 3 U$.  
\begin{figure}
    \centering
    \includegraphics[width=0.4\textwidth]{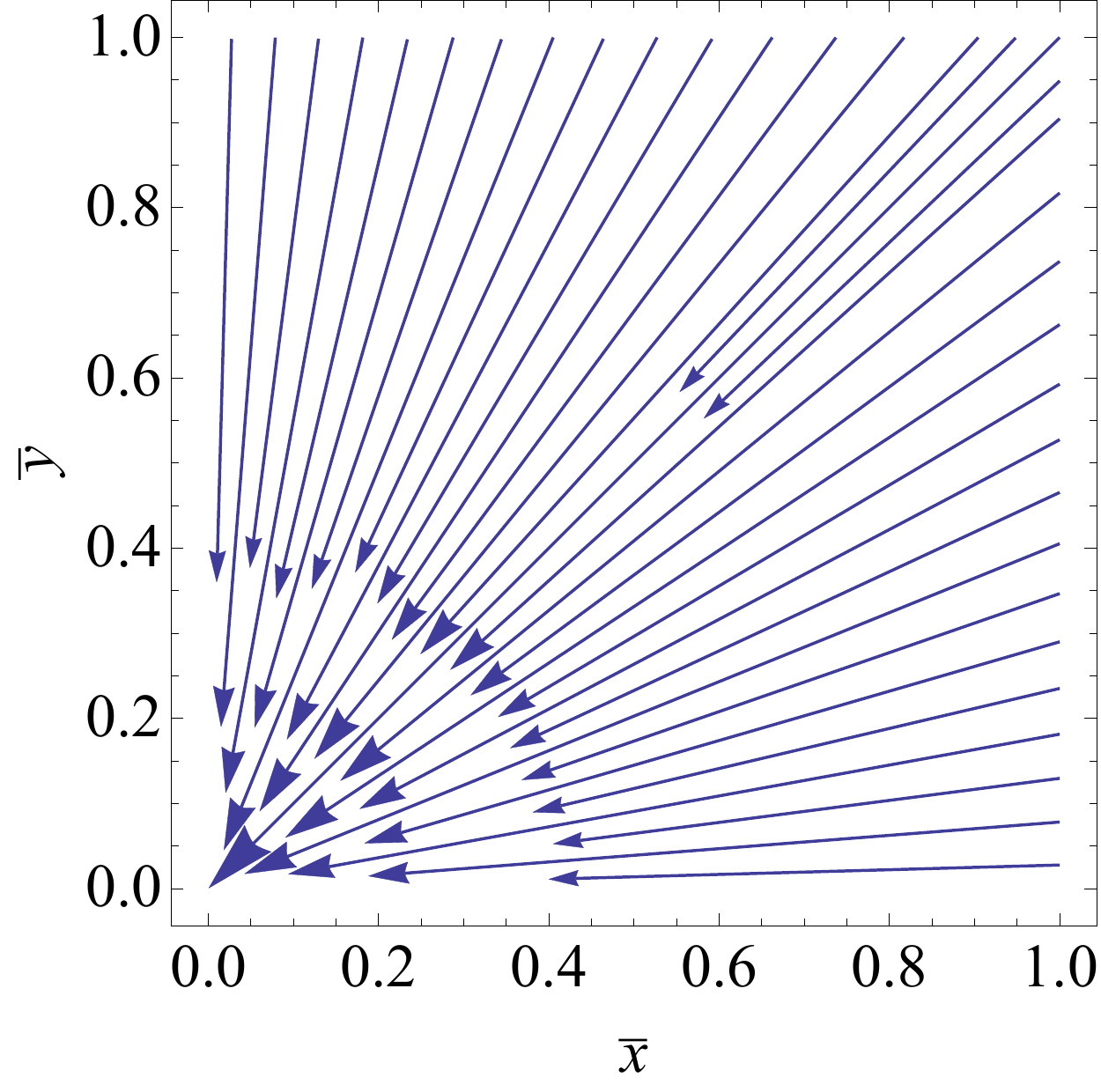}
    \caption{Orbits of the phase space of system \eqref{averaged2D}}
    \label{fig:my_label2}
\end{figure}
The center  manifold is given locally by the graph 
\begin{align}
   &  \Big\{(U,V_1,V_2): U=h_3(V_1,  V_2), \nonumber\\ 
   & h_3(0,0)=0,  \;  Dh_3(0,0)=0, |(V_1, V_2)|<\delta \Big\}. 
\end{align}
Using the invariance of the center manifold we obtain  that $h_3(V_1,  V_2)$ satisfies the quasilinear partial differential equation 
\begin{align}
& \frac{V_2  h_3^{(0,1)}(V_1,V_2)  h_3(V_1,V_2)}{1- \bar{\Omega}_{1c}- \bar{\Omega}_{2c}}\nonumber \\
& +\frac{ V_1  h_3^{(1,0)}(V_1,V_2)  h_3(V_1,V_2)}{1- \bar{\Omega}_{1c}- \bar{\Omega}_{2c}}-
    h_3(V_1,V_2)=0.
\end{align}
This equation have the trivial solution $ h_3(V_1,V_2)= 0$ and the general solution 
\begin{equation}
 h_3(V_1,V_2)= c_1\left(\frac{V_2}{V_1}\right)+\ln (V_1) (1- \bar{\Omega}_{1c}- \bar{\Omega}_{2c}).
\end{equation}
which do not satisfy the conditions $h_3(0,0)=0$, and $Dh_3(0,0)=0$. 
Then, the center manifold is the plane $U=0$. 

The guiding equation  can be studied by defining $\bar{x}= \bar{\Omega}$ and  $\bar{y}=\bar{\Omega}_1+\bar{\Omega}_2+\bar{\Omega}$,  we obtain the reduced 2D system 
\begin{align}\label{averaged2D}
&  \frac{d\bar{x}}{d \tau} = -3 \bar{x} \left(1- \bar{y}\right),\;  \frac{d \bar{y}}{d \tau}= -3 \bar{y} \left(1-\bar{y}\right).
\end{align}
This system \eqref{averaged2D} have the equilibrium point $(\bar{x}, \bar{y})$ with eigenvalues $-3, -3$. Then it is a sink.
Additionally, we have the line of equilibrium points $\bar{y}=1$  with eigenvalues $0,3$ which is normally hyperbolic and a source.

In  figure \ref{fig:my_label2} some orbits of the phase space of system \eqref{averaged2D} are presented. 

\section{Discussions}
\label{discussionsA}

In section \ref{model} we presented a coupled effective axion-like model in flat FLRW cosmology.  In particular, we discussed the mechanism to make the effective axion decay constant $f_{2}$ arbitrary large, and to make the real scalar field $\psi$  a heavy field, whose evolution is dominated only by the first term in the potential \eqref{potentialrealfull} if $\mu_1\gg \mu_2$ and $f_2\gg 1$, while the real scalar field $\xi$ to be a light field, whose evolution is dominated only by the second term in the potential \eqref{potentialrealfull} with $\psi\approx 0$. 

For $\mu_1\gg \mu_2$
we have 
\begin{align*}
   & \ddot{\psi} +  3 H \psi+   m^{2}_{\psi} \psi =0, \;  \ddot{\xi} +3 H \xi+m^{2}_{\xi}  \xi=0,  \; m^{2}_{\psi} \gg 
 m^{2}_{\xi}.
\end{align*}
Neglecting $\psi$ in the second term of \eqref{potentialrealfull} and renaming $\psi=\phi_1$ and $\xi=\phi_2$, we obtain the effective potential \eqref{pot_phi}.
This potential was generalized to \eqref{pot}, 
where the heavy field is $\phi_1$  and the light field is  $\phi_2$, which interact through the third term presented in the above potential. The interaction is turned on when $n\neq 0$. When $n=0$ the first and third terms are merged by replacing $\mu_1^4+\mu_3^4 \rightarrow \mu_1^4$ and potential \eqref{pot_phi} is recovered. Then, in section \ref{SECT_III_a} we have discussed the theorem \ref{local-estimate} based on energy density estimates. Using results \eqref{theo1_a} and \eqref{theo1_b} of theorem \ref{local-estimate}, we generically have 
\begin{small}
\begin{align*}
  \lim_{t\rightarrow \infty} \left( \rho(t), \dot{\phi_1}(t),  \dot{\phi_2}(t),\frac{V(\phi_1(t), \phi_2(t))}{3H(t)^2}\right)=(0, 0, 0, 1).  
 \end{align*}
\end{small}
That is, the kinetic terms in $\Omega_{\phi}$ and the matter-energy density tend to zero. Then, the cosmological solution is dominated by the potential term.  
Numerical simulations of the system \eqref{Raych}-\eqref{KG} were discussed in section \ref{SECT_III_b}. Initial conditions $\phi_{1}|_{z=100}=0.155$,  $\phi_{2}|_{z=100}=0.7835$, $\frac{d\phi_{1}}{d z}_{z=100}=0$ and $\frac{d\phi_{2}}{d z}_{z=100}=0$ were considered. The initial value $H|_{z=100}$  is estimated from expression \eqref{HLambdaCDM}. We see that the $\Lambda$CDM is recovered and in this case a local (not zero) minimum of the potential is approached. 

As shown in figure \ref{Fig2c}, a local minimum with minimum value $\lim_{z\rightarrow -1} E(z)^2=\frac{\Lambda_{\text{eff}}}{3 H_0^2}=0.682603$ is obtained.  By evaluating the solution which converges to $\phi^*$, the parametric curve  $\frac{V(\phi_1(z), \phi_2(z))}{3H_0^2}, z\in[-1,100]$, was obtained  and attached to the surface. 

The parameter values and initial conditions were chosen for obtaining a ratio of the DM and DE densities at $z=0$ equal to 
  $r|_{z=0}=\frac{0.315}{0.685}= \frac{63}{137} \approx 0.459854$. 
  
  A late-times $r\rightarrow 0$, $\Omega_\phi$ dominates, in particular $\sqrt{\frac{V(\phi_1^*,\phi_2^*)}{3 \lim_{z\rightarrow -1}H(z)^2}}= \sqrt{\frac{\Lambda_{\text{eff}}}{3 H_0^2 \lim_{z\rightarrow -1} E(z)}}=1$. That is, for this choice of parameters and initial conditions, our model it suffers of the CCP, since it is not indistinguishable from $\Lambda$CDM. A subtle difference, is in that $\lim_{z \rightarrow -1} W_{\text{int}}=  0.101566$, which means that $\approx 1\%$ of the total dimensionless energy density corresponds to interaction between the two fields $\phi_1$ and $\phi_2$. 
  
  It is plausible to think of the light scalar field $\phi_2$ as DE, and the heavy field $\phi_1$, as an axion-like DM which interacts through the interaction term $V_{\text{int}}\left(\phi_1,\phi_2\right)$ defined by \eqref{V_int}. 
  
  Increasing the contribution of $V_{\text{int}}$ at late-times, or introducing an explicit interaction term between the axion-like part and CDM, the CCP can be alleviated or solved (e.g., as in quintessence and phantom field scenarios in  \cite{Chimento:2000kq,Zimdahl:2001ar,Chimento:2003iea,Chimento:2003sb,Cai:2004dk,Guo:2004vg,Curbelo:2005dh}. If $V_{\text{int}}\left(\phi_1,\phi_2\right)\equiv 0$, and introducing an effective interaction $Q= 2 \ln \chi(\phi_1(\tau), \phi_2(\tau))$, we obtain, from the balance of energy,  the modified KG equations, and modified continuity equation given by:  
  \begin{subequations}
  \label{interacting-scheme}
  \begin{align}
&\dot{\rho_{\phi_1}}+3H(\rho_{\phi_1} + p_{\phi_1})= -\frac{\partial Q(\phi_1, \phi_2)}{\partial \phi_1} \rho \dot{\phi_1}, \label{2IntKG1}\\
&\dot{\rho_{\phi_2}}+3H(\rho_{\phi_2} + p_{\phi_2})= -\frac{\partial Q(\phi_1, \phi_2)}{\partial \phi_2} \rho  \dot{\phi_2}, \label{2IntKG2}
\\
& \dot{\rho}+3H \rho= \rho\frac{d Q(\phi_1(t), \phi_2(t))}{d t}. 
\end{align}
\end{subequations}
Then, defining $\tau=\ln(a/a_0)$, we have 
\begin{align}\label{exactR}
    \frac{d \ln r}{d \tau}= 3+  (r+1)  \frac{d Q}{d \tau}- 6 (r+1)\frac{V}{3 H^2}.
\end{align}
As $\frac{V}{3 H^2}\rightarrow 1$ (according to theorem \ref{local-estimate} the potential energy dominates at late-times), we have 
\begin{align}
\label{approxR}
    \frac{d \ln r}{d \tau}= -3- 6 r + 2(r+1)  \frac{d \ln \chi}{d \tau}.
\end{align}
\begin{figure}[h!]
    \centering
    \includegraphics[scale=0.6]{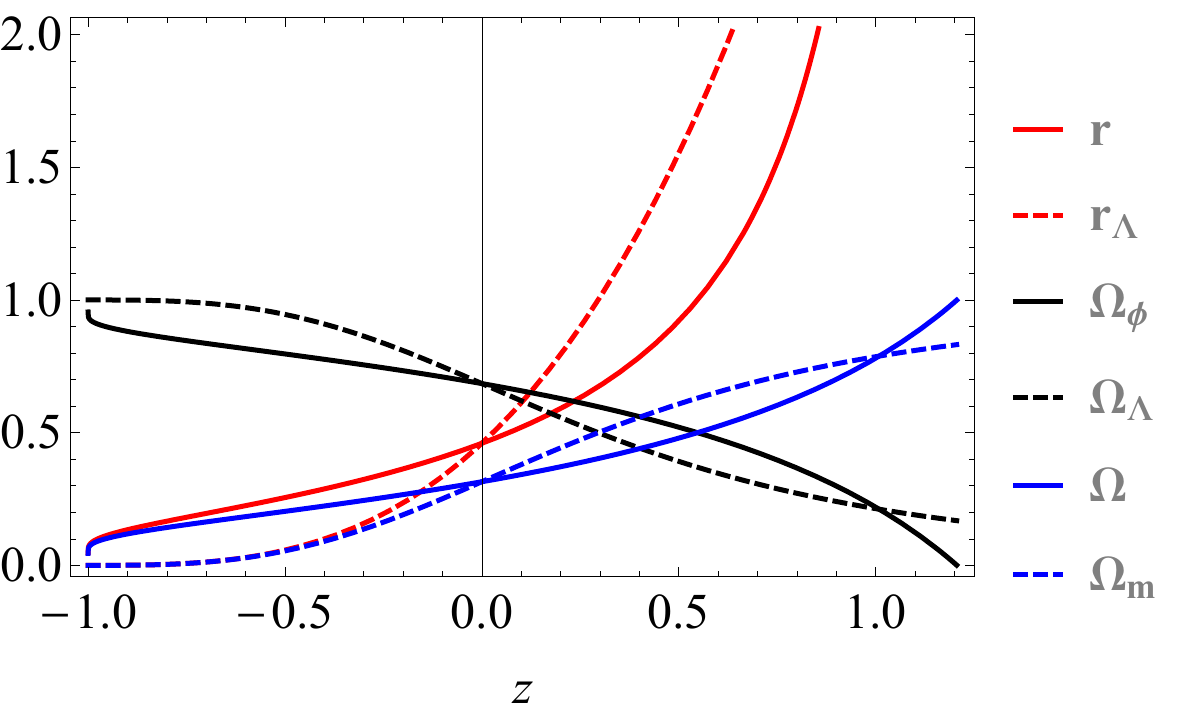}
    \caption{Evolution of \eqref{128} in terms of redshift, which give the asymptotic values of the ratio $r$ and $\Omega_\phi$, $\Omega_m$ under the interacting scheme \eqref{interacting-scheme} with $Q= - 3 (\delta^2+1)\ln(1+z)$. The exact solutions for the $\Lambda$CDM model  are superimposed (dashed lines) for a comparison}
    \label{fig:128}
\end{figure}
Then, 
\begin{align}
    r(\tau )=\frac{e^{-3 \tau } \chi ^2(\tau )}{c_1-2\int_0^{\tau } e^{-3 s} \chi (s) \left( \chi '(s)-3 \chi (s)\right) \, ds},
\end{align}
where $c_1=\frac{\Omega_{\Lambda 0}\chi^2(0)}{ \Omega _{m0}}$ is chosen to have $r(0)=\frac{\Omega_{m0}}{ \Omega_{\Lambda0}}=63/137$. 
In the non-minimal coupling case $\chi=1$, we obtain
$r(\tau )=\frac{1}{\left(c_1+2\right) e^{3 \tau }-2} \rightarrow 0$, 
as $\tau\rightarrow \infty$. 
However, if we chose, for example, $\chi(\tau)=  e^{\frac{3}{2} \left(\delta ^2+1\right) \tau }$, 
then for $z< z_{\text{crit}}$, 
\begin{align*}
r(z) = \frac{\delta ^2 \Omega_{m 0}}{\left(1-\delta ^2\right)
   \Omega_{m 0}+\left(\delta ^2-\Omega_{m 0}\right) (z+1)^{3
   \delta ^2}}, 
\end{align*}
where  $z_{\text{crit}}+1= \left(\frac{\Omega_{m 0}-\delta ^2}{\left(1-\delta ^2\right)
   \Omega_{m 0}}\right)^{-\frac{1}{3 \delta ^2}}$.
Therefore, at late-times $r\rightarrow r_0=\frac{\delta ^2}{1-\delta ^2}$. Then, the CCP is solved due to \begin{align}
    & \Omega_{\phi}= \frac{1}{1+r} \rightarrow 1 - \delta^2, \nonumber \\
    & \Omega= \frac{r}{1+r} \rightarrow \delta^2 \implies \frac{\Omega}{\Omega_{\phi}}= \mathcal{O}(1). 
\end{align}  
Using Planck 2018 results \cite{Planck2018} we have the current values $\Omega_{m0}\approx 0.315, \Omega_{\Lambda0}\approx 0.685$, and, at late times, 
\begin{subequations}
\label{128}
\begin{align}
   & r(z)= \frac{63 \delta ^2}{\left(200 \delta ^2-63\right) (z+1)^{3 \delta
   ^2}-63 \left(\delta ^2-1\right)}, \\
   & \Omega_\phi(z)= 1-\frac{63 \delta ^2}{\left(200
   \delta ^2-63\right) (z+1)^{3 \delta ^2}+63},\\
   &\Omega(z)= \frac{63 \delta
   ^2}{\left(200 \delta ^2-63\right) (z+1)^{3 \delta ^2}+63}. 
\end{align}
\end{subequations}
Setting, $\delta = 0.2$ we obtain the equality DM-DE epoch approximately at $z_{eq}\approx 0.547797$.  The approximations are valid for $z<z_{\text{crit}}=1.2067$. If we integrate the full equations \eqref{interacting-scheme} and \eqref{exactR}, the exact expressions for $r, \Omega_\phi$ and $\Omega_m$ ares less sharp than displayed in figure \ref{fig:128} for $z \approx 1$, but the approximated solutions and the numerical exact ones matches very well as $z\rightarrow -1$ where the potential energy start dominating. Then,  \eqref{approxR} becomes an accurate approximation of \eqref{exactR}.

Figure \ref{fig:128} shows the evolution of \eqref{128} in terms of redshift, which give the asymptotic values of the ratio $r$ and $\Omega_\phi$, $\Omega_m$ under the interacting scheme \eqref{interacting-scheme} with $Q= 3 (\delta^2+1)\ln(a/a_0) =- 3 (\delta^2+1)\ln(1+z)$.  This kind of interaction indicates a phenomenological solution
to the CCP. Actually, once the universe reaches the stable $\Omega_\phi$-dominated state with constant ratio $r = r_0$, it will live in
this state for a very long time. Then, it is not a coincidence to live in this long-living state where  $r_0 \sim  1$ implies $\rho \sim \rho_\phi$.
The full analysis is not covered in the present research. In a forthcoming paper we will develop arguments like in  \cite{Chimento:2000kq,Zimdahl:2001ar,Chimento:2003iea,Chimento:2003sb,Cai:2004dk,Guo:2004vg,Curbelo:2005dh} for the two-field scenario. 

In other regard, taking Taylor expansions near  $\phi_1=0$, $\phi_2=0$ we have 
$ V(\phi_1, \phi_2)\approx  \phi_1^2 \left(\frac{\mu_1^4}{2
   f_1^2}+\frac{\mu_3^4}{2
   f_1^2}\right)-\frac{\mu_3^4 n \phi_1
   \phi_2}{f_1 f_2}    +\phi_2^2
   \left(\frac{\mu_2^4}{2 f_2^2}+\frac{\mu_3^4
   n^2}{2 f_2^2}\right)$. 
Using the transformation \eqref{lin-phi-Psi}, with $c$ defined by \eqref{c}, we recover the potential $V(\Psi_1, \Psi_2)\approx \frac{\omega_1^2}{2}\Psi_1^2 + \frac{\omega_2^2}{2}\Psi_2^2$,  where $\omega_{1,2}^2$ are given by \eqref{omega}. 
In section \ref{SECTT:4} we shown that the solutions typically  tends to the global minimum, where $\eta=0, \Psi_1=\phi_1=0, \Psi_2=\phi_2=0$, corresponding to Minkowski solution.
For the non-zero local minima, we have de Sitter solutions which are all saddle points. 

Moreover, we introduced in section \ref{DSA} novel dynamical variables and dimensionless time variables, which have not been used in analyzing these cosmological dynamics where expansion normalized dynamical variables are usually adopted. In the uncoupled case the equilibrium points can be fully identified and characterized. However, in the coupled case we need to solve transcendental equations and a little progress is made. The main difficulties that arise using standard dynamical systems approaches are due to the oscillations entering the nonlinear system through the Klein-Gordon (KG) equations. This motivated the analysis of the oscillations using averaging techniques in section \ref{SECTT:4}. In particular, integrating \eqref{avergsystem} we use the definition $\bar{\Omega}_{t}:= \bar{\Omega}_1+\bar{\Omega}_2+\bar{\Omega}$. 

Then, we acquire the equations 
\begin{align}
& \frac{ d \bar{\Omega}_{t}}{d\tau}=    -3  (1- \bar{\Omega}_{t})  \bar{\Omega}_{t}, \;  \frac{ d H}{d\tau}= -\frac{3}{2} H \bar{\Omega}_{t},
\\
& \frac{d\bar{\Omega}_1}{d \tau}= -3 \bar{\Omega}_1 \left(1- \bar{\Omega}_{t}\right),\;  \frac{d\bar{\Omega}_2}{d \tau}= -3 \bar{\Omega}_2 \left(1- \bar{\Omega}_{t}\right).
\end{align}
Then, assuming $\bar{\Omega}_{t}(0)<1$, we obtain
\begin{small}
\begin{subequations}
\label{eq:111}
\begin{align}
& \bar{\Omega}_{t}(\tau )= \frac{\bar{\Omega}_{t}(0)}{\bar{\Omega}_{t}(0)+e^{3 \tau } (1-\bar{\Omega}_{t}(0))},
\\
   & H(\tau )=
 H_0 e^{-\frac{3 \tau}{2}} \sqrt{\bar{\Omega}_{t}(0)+e^{3 \tau }
   (1-\bar{\Omega}_{t}(0))},
\\
& \bar{\Omega} (\tau )= \frac{\bar{\Omega}_{t}(0) -\bar{\Omega}_1(0)-\bar{\Omega}_2(0)}{\bar{\Omega}_{t}(0)+e^{3 \tau } (1-\bar{\Omega}_{t}(0))},
\end{align}
\begin{align}
&\bar{\Omega}_1(\tau )= 
   \frac{\bar{\Omega}_1(0)}{\bar{\Omega}_{t}(0)+e^{3 \tau } (1-\bar{\Omega}_{t}(0))},
\\
& \bar{\Omega}_2(\tau )= 
   \frac{\bar{\Omega}_2(0)}{\bar{\Omega}_{t}(0)+e^{3 \tau } (1-\bar{\Omega}_{t}(0))}.
\end{align}
\end{subequations}
\end{small}
Hence, on average, $H$ is decreasing from $H=\infty $ as $\tau\rightarrow -\infty$ to $H=H_0\sqrt{1-\bar{\Omega}_{t}(0)}$ as $\tau\rightarrow +\infty$. 
Furthermore, on average, we have 
\begin{small}
\begin{align*}
& \lim_{\tau\rightarrow -\infty} \left(H, \bar{\Omega}_{t},  \bar{\Omega},  \bar{\Omega}_{1},  \bar{\Omega}_{2}\right) \nonumber \\
 & = \Big(\text{sgn}(H_0)\infty , 1,\frac{\bar{\Omega}_{t}(0)-\bar{\Omega}_1(0)- \bar{\Omega}_2(0)}{\bar{\Omega}_{t}(0)},  \frac{\bar{\Omega}_1(0)}{\bar{\Omega}_{t}(0)},\frac{\bar{\Omega}_2(0)}{\bar{\Omega}_{t}(0)}\Big),  
\end{align*}
\end{small}
and
\begin{small}
\begin{align*}
&  \lim_{\tau\rightarrow +\infty} \left(H, \bar{\Omega}_{t},  \bar{\Omega},  \bar{\Omega}_{1},  \bar{\Omega}_{2}\right) =
\left( H_0 \sqrt{1-\bar{\Omega}_{t}(0)}, 0,0,0,0\right).
\end{align*}
\end{small}
That is, as far as $\bar{\Omega}_{t}(0)<1$ the universe ends on a de Sitter inflationary phase.

Using the definitions \eqref{eq:59},   we obtain 
\begin{equation*}
r_i= \sqrt{6} \omega_i^{-2} H  \sqrt{\Omega_i}, \; \rho= 3 H^2\Omega. 
\end{equation*}
Hence, 
\begin{equation*}
     \Psi_i= \sqrt{6} \omega_i^{-2} H  \sqrt{\Omega_i} \sin \left(t \omega_i-\Phi_{i}\right).
\end{equation*}
But 
\begin{equation}
\label{eq:120}
    \frac{d t}{d \tau}= \frac{1}{H(\tau)}. 
\end{equation}
Then,
\begin{align*}
     & \Psi_i(\tau)= \sqrt{6} \omega_i^{-2} H(\tau)  \sqrt{\Omega_i(\tau)} \nonumber \\
     & \times  \sin \left(\omega_i\int_0^\tau H^{-1}(s) ds -\Phi_{i}(\tau)\right).
\end{align*}
The key step is assuming that $H(\tau)$, $\Omega_i(\tau)$ and $\Omega(\tau)$ can be approximated (for $\bar{\Omega}_t(0)<1$) by the averaged expressions  \eqref{eq:111}, and from $ \partial_t{\bar{\Phi}_1}=
     \partial_t{\bar{\Phi}_2}=0$, we replace $\Phi_{i}(\tau)$ by constants $\bar{\Phi}_{i}(0)$. These are accurate approximations as $\tau\rightarrow \infty$ according to Theorem \ref{LFZ1}. 
Then, 
     \begin{widetext}
     \begingroup\makeatletter\def\f@size{9}\check@mathfonts
     \begin{align*}
&\omega_1^2 \Psi_1(\tau)=- {\sqrt{6} H_0 e^{-\frac{3 \tau}{2}} \sqrt{\bar{\Omega}_1(0)} \sin \scriptscriptstyle
   \left(\frac{2 \omega_1 \left(\ln \left(-\bar{\Omega}_t(0)+\sqrt{1-\bar{\Omega}_t(0)}+1\right)-\ln
   \left(\sqrt{(\bar{\Omega}_t(0)-1) \left(e^{3 \tau } (\bar{\Omega}_t(0)-1)-\bar{\Omega}_t(0)\right)}-e^{3 \tau /2} (\bar{\Omega}_t(0)-1)\right)\right)}{3 H_0 \sqrt{1-\bar{\Omega}_t(0)}}+\bar{\Phi}_{10}\right)},
\\
& \omega_2^2\Psi_2(\tau)=  - {\sqrt{6} H_0 e^{-\frac{3 \tau}{2}} \sqrt{\bar{\Omega}_2(0)} \sin \scriptscriptstyle
   \left(\frac{2 \omega_2 \left(\ln \left(-\bar{\Omega}_t(0)+\sqrt{1-\bar{\Omega}_t(0)}+1\right)-\ln
   \left(\sqrt{(\bar{\Omega}_t(0)-1) \left(e^{3 \tau } (\bar{\Omega}_t(0)-1)-\bar{\Omega}_t(0)\right)}
   -e^{3 \tau /2} (\bar{\Omega}_t(0)-1)\right)\right)}{3 H_0 \sqrt{1-\bar{\Omega}_t(0)}}+\bar{\Phi}_{20}\right)}{}. 
     \end{align*}
     \endgroup
 Finally, using the inverse of \eqref{lin-phi-Psi}, 
we have 
     \begingroup\makeatletter\def\f@size{8}\check@mathfonts
     \begin{align*}
& \phi_1(\tau)=\frac{\sqrt{6} H_0 e^{-\frac{3 \tau}{2}} \sqrt{\bar{\Omega}_1(0)}
   \sin \scriptscriptstyle \left(\frac{2 \omega_1 \left(\ln \left(-\bar{\Omega}_t(0)+\sqrt{1-\bar{\Omega}_t(0)}+1\right)-\ln
   \left(\sqrt{(\bar{\Omega}_t(0)-1) \left(e^{3 \tau } (\bar{\Omega}_t(0)-1)-\bar{\Omega}_t(0)\right)}-e^{3 \tau /2} (\bar{\Omega}_t(0)-1)\right)\right)}{3 H_0 \sqrt{1-\bar{\Omega}_t(0)}}+\bar{\Phi}_{10}\right)}{\sqrt{c^2+1} \omega_1^2} \nonumber\\
   & -\frac{\sqrt{6} c H_0 e^{-\frac{3 \tau}{2}} \sqrt{\bar{\Omega}_2(0)} \sin \scriptscriptstyle \left(\frac{2 \omega_2 \left(\ln
   \left(-\bar{\Omega}_t(0)+\sqrt{1-\bar{\Omega}_t(0)}+1\right)-\ln
   \left(\sqrt{(\bar{\Omega}_t(0)-1) \left(e^{3 \tau } (\bar{\Omega}_t(0)-1)-\bar{\Omega}_t(0)\right)}-e^{3 \tau /2} (\bar{\Omega}_t(0)-1)\right)\right)}{3 H_0 \sqrt{1-\bar{\Omega}_t(0)}}+\bar{\Phi}_{20}\right)}{\sqrt{c^2+1} \omega_2^2},
\\
& \phi_2(\tau)= -\frac{\sqrt{6} c H_0 e^{-\frac{3 \tau}{2}} \sqrt{\bar{\Omega}_1(0)} \sin \scriptscriptstyle\left(\frac{2 \omega_1 \left(\ln
   \left(-\bar{\Omega}_t(0)+\sqrt{1-\bar{\Omega}_t(0)}+1\right)-\ln
   \left(\sqrt{(\bar{\Omega}_t(0)-1) \left(e^{3 \tau } (\bar{\Omega}_t(0)-1)-\bar{\Omega}_t(0)\right)}-e^{3 \tau /2} (\bar{\Omega}_t(0)-1)\right)\right)}{3 H_0 \sqrt{1-\bar{\Omega}_t(0)}}+\bar{\Phi}_{10}\right)}{\sqrt{c^2+1} \omega_1^2}\nonumber \\
   & -\frac{\sqrt{6} H_0 e^{-\frac{3 \tau}{2}} \sqrt{\bar{\Omega}_2(0)} \sin \scriptscriptstyle\left(\frac{2 \omega_2 \left(\ln
   \left(-\bar{\Omega}_t(0)+\sqrt{1-\bar{\Omega}_t(0)}+1\right)-\ln
   \left(\sqrt{(\bar{\Omega}_t(0)-1) \left(e^{3 \tau } (\bar{\Omega}_t(0)-1)-\bar{\Omega}_t(0)\right)}-e^{3 \tau /2} (\bar{\Omega}_t(0)-1)\right)\right)}{3 H_0 \sqrt{1-\bar{\Omega}_t(0)}}+\bar{\Phi}_{20}\right)}{\sqrt{c^2+1} \omega_2^2}. 
     \end{align*}
     \endgroup
     
   If $\bar{\Omega}_{t}(0)\ll 1$, we have the approximation
      \begingroup\makeatletter\def\f@size{8}\check@mathfonts
\begin{align*}
&   \phi_1(\tau)=  \frac{\sqrt{6} H_0 e^{-3 \tau /2}   \left(\sqrt{\bar{\Omega}_{1}(0)} \omega_2^2 \sin \left(\bar{\Phi}_{10}-\frac{\tau 
   \omega_1}{H_0}\right)-c \omega_1^2
   \sqrt{\bar{\Omega}_{2}(0)} \sin \left(\bar{\Phi}_{20}-\frac{\tau 
   \omega_2}{H_0}\right)\right)}{\sqrt{c^2+1}
   \omega_1^2 \omega_2^2} \nonumber \\
   & +\frac{\left(3 e^{-3 \tau /2}
   \tau +e^{-9 \tau /2}-e^{-3 \tau /2}\right) \bar{\Omega}_{t}(0) \left(c
   \omega_1 \sqrt{\bar{\Omega}_{2}(0)} \cos \left(\bar{\Phi}_{20}-\frac{\tau  \omega_2}{H_0}\right)-\sqrt{\bar{\Omega}_{1}(0)} \omega_2
   \cos \left(\bar{\Phi}_{10}-\frac{\tau  \omega_1}{H_0}\right)\right)}{\sqrt{6} \sqrt{c^2+1} \omega_1
   \omega_2},
\\
   &    \phi_2(\tau)=-\frac{\sqrt{6} H_0 e^{-3
   \tau /2} \left(c \sqrt{\bar{\Omega}_{1}(0)} \omega_2^2 \sin
   \left(\bar{\Phi}_{10}-\frac{\tau  \omega_1}{H_0}\right)+\omega_1^2 \sqrt{\bar{\Omega}_{2}(0)}
   \sin \left(\bar{\Phi}_{20}-\frac{\tau  \omega_2}{H_0}\right)\right)}{\sqrt{c^2+1} \omega_1^2
   \omega_2^2} \nonumber \\
  &  + \frac{\left(3 e^{-3 \tau /2} \tau +e^{-9 \tau
   /2}-e^{-3 \tau /2}\right) \bar{\Omega}_{t}(0) \left(c
   \sqrt{\bar{\Omega}_{1}(0)} \omega_2 \cos \left(\bar{\Phi}_{10}-\frac{\tau  \omega_1}{H_0}\right)+\omega_1 \sqrt{\bar{\Omega}_{2}(0)} \cos \left(\bar{\Phi}_{20}-\frac{\tau 
   \omega_2}{H_0}\right)\right)}{\sqrt{6} \sqrt{c^2+1}
   \omega_1 \omega_2}.
\end{align*}
\endgroup
     \end{widetext}
On the other hand, given that $\bar{\Omega}_{t}(0)=1$  is an invariant set, then $\bar{\Omega}=\bar{\Omega}(0), \bar{\Omega}=\bar{\Omega}(0), \bar{\Omega}=\bar{\Omega}(0)$ remains constant with $\bar{\Omega}_{t}(0)=1$. 

Asymptotically we have 
$ \frac{dH}{d\tau}=-\frac{3}{2}H$. 
Defining $\tau_0$ as the current value associated to $t=t_a$ (``age of the universe''), which as before can be set to $\tau_0=0$, we have  $H(\tau )= H_0 e^{-\frac{3}{2}\tau }$.
Substituting this expression for $H$ in \eqref{eq:120} and integrating, we obtain  $t(\tau )= \frac{3 H_0 t_a+2 e^{3 \tau /2}-2}{3 H_0}.$
Finally, combining all  we have $H= \frac{2 H_0}{2 +3 H_0(t- t_a)}$
which tends to zero as $t\rightarrow \infty$.
That is, the universe passes through a matter-dominated phase before reaching a Minkowski stage. 

We have shown the application of methods from the theory of averaging nonlinear dynamical systems allows us to prove that time-dependent systems and their corresponding time-averaged versions have the same late-time dynamics. Therefore,  simple time-averaged systems determine the future asymptotic behavior. Then, we can study the time-averaged system using standard techniques of dynamical systems.

\section{Conclusions}
\label{conclusions}

In this paper, we have analyzed the coupled axion-like model following Ref. \cite{DAmico:2016jbm} consisting of   two canonical scalar fields $\phi_1,\,\phi_2$ interacting via the potential \eqref{V_int}. We have introduced dimensionless dynamical variables and dimensionless time variables. In the uncoupled case the equilibrium points were fully identified and characterized. However, in the coupled case we need to solve transcendental equations and a little progress is made. The main difficulties are due to the oscillations entering the nonlinear system through KG equations.  

Using local energy estimates, we have proved  theorem \ref{local-estimate}. This result shows that if $O^+(x_0)$ is the positive orbit that passes  through the regular point  $x_0$ defined as in \eqref{interior}, then, since $H$ is positive and decreases along $O^+(x_0)$, the limit $\lim_{t\rightarrow \infty} H(t)$ exists, and it is a non-negative number $\eta$. Then, we have $\lim_{t\rightarrow \infty} \left( \rho, \chi_1, \chi_2\right)=(0,0, 0)$ and 
$3\eta^2  
= \lim_{t\rightarrow \infty}  V\left(\phi_1(t),\phi_2(t)\right)$.
Depending on initial conditions, the solution tends to a constant value of $H^*= \sqrt{\frac{V(\phi_1^*, \phi_2^*)}{3}}$, related with a local minimum $\phi^{*}=(\phi_1^*, \phi_2^*)$ with non zero minimum value of $V(\phi_1, \phi_2)$ satisfying eqs. \eqref{maxima-minima}. They correspond to de Sitter solutions $a\propto \exp\left[-\sqrt{\frac{V(\phi_1^*, \phi_2^*)}{3}} t\right]$. However, choosing initial conditions with a small enough value of $3H^2(t_0)$ the orbit is trapped by the basin of attraction of the global minimum $\phi^{*}=(0,0)$, $H^*=0$. Then, the Minkowski solution is the late-time attractor.

The main result in this paper is our Theorem \ref{LFZ1}. It states that in the first-order approximations of normalized scalar field densities, and the values of two scalar fields $\Phi_1$ and $\Phi_2$, defined as $\Phi_i= t \omega_i -\tan^{-1}\left(\frac{\omega_i \Psi_i}{\dot \Psi_i}\right)$ (called phase variables, where $\Psi_1$ and $\Psi_2$ are functions of $\phi_1$ and $\phi_2$ through \eqref{lin-phi-Psi}) near $H=0$, the full systems and their averaged values (with an averaged function to be properly defined)  have the same limit as $H\rightarrow 0$ (as $\tau\rightarrow \infty$). The averaged values of the phases, denoted by $\bar{\Phi}_i$, are zero, such that, on average, $\bar{\Psi}_i(t)= r_i \sin \left(t \omega_i\right)$ for some constants $r_i$. Therefore, with this approach, oscillations entering the nonlinear system through the KG equation can be controlled and smoothed out as the Hubble factor $H$, acting as a time-dependent perturbation parameter, tends monotonically to zero. We have studied the time-averaged system using standard techniques of dynamical systems and numerical simulations are presented as evidence of such behavior. 

This approach has potential applications in physical situations where it is necessary to consider the time variation of the fields in the vicinity of the potential minimum. One such situation is the reheating after inflation. During reheating the inflaton scalar field oscillates around the potential minimum. For nonzero $H$ this gives rise to time-dependent oscillatory dynamics, which is responsible for particle production via quantum field theory. The relevant calculations depend on the form of the potential, and in particular, are quite complicated for harmonic potentials. The result presented here shows that one can ``average out" the oscillations arising due to the harmonic functions, thus simplifying the problem. Indeed, by using some inverse transformations, one can find from the $\bar{\Omega}_i$ to $\bar{\phi}_i$, i.e., the averaged version of the original field variables. We hope that this approach may help calculations of reheating in the context of $N$-inflation model \cite{Dimopoulos:2005ac}. 

This approach can also be useful if one wants to consider linear cosmological perturbations for coupled axion cosmological models near the potential minimum, which is an attractor. In the cosmological perturbation theory, cosmological perturbations at the linear level are governed by equations whose coefficients are composed of background quantities. Therefore, proper knowledge of the background dynamics is necessary for further perturbative analyses. The result presented in this paper simplifies the dynamical analysis of the background near this attractor, which in turn facilitates a subsequent analysis of cosmological perturbations using procedures similar to those used in \cite{Wainwright:2005} (chapt. 14)  and in \cite{amebook, Amendola:1999dr, Basilakos:2019dof, Alho:2019jho, Alho:2020cdg, Paliathanasis:2021egx}.
\section*{Acknowledgements}
Genly Leon and Esteban   Gonz\'alez have the financial support of Agencia Nacional de Investigaci\'on y Desarrollo - ANID 
through the program FONDECYT Iniciaci\'on grant no.
11180126. Additionally,  this research is funded by Vicerrector\'ia de Investigaci\'on y Desarrollo Tecnol\'ogico at Universidad Católica del Norte. The work of Bin Wang was partially supported by the key project of NNSFC under contract 11835009.  Samuel Lepe is acknowledged for discussions. We thank an anonymous referee for valuable comments which helped improve our work.

\appendix

\section{Proof of Theorem \ref{LFZ1}}
\label{apppa}

\begin{lem}[\textbf{Gronwall's Lemma (Integral form)}]
\label{Gronwall}
 Let be $\xi(t)$ a nonnegative function, summable over  $[0,T]$ which satisfies almost everywhere the integral inequality $$\xi(t)\leq C_1 \int_0^t \xi(s)ds +C_2, \;  C_1, C_2\geq 0.$$
       Then, 
      $\xi(t)\leq C_2  e^{C_1 t},$
        almost everywhere for $t$ in $0\leq t\leq T$. In particular, if    
     $$\xi(t)\leq C_1 \int_0^t \xi(s)ds, \;  C_1\geq 0,$$
        almost everywhere for $t$ in $0\leq t\leq T$. Then,  $
           \xi \equiv 0$  
        almost everywhere for $t$ in $0\leq t\leq T$.
 \end{lem}
\begin{lem}[Mean value theorem]
\label{lemma6}
 Let $U \subset \mathbb{R}^n$ be open, $\mathbf{f}: U \rightarrow \mathbb{R}^m$ continuously differentiable and  $\mathbf{z}\in U$, $\mathbf{h}\in \mathbb{R}^m$ vectors such that the line segment $\mathbf{z}+\eta \; \mathbf{h}$,  $0 \leq \eta \leq 1$ remains in $U$. Then we have:
\begin{equation}
    \mathbf{f}(\mathbf{z}+\mathbf{h})-\mathbf{f}(\mathbf{z}) = \left (\int_0^1 \mathbb{D}\mathbf{f}(\mathbf{z}+\eta \; \mathbf{h})\,d\eta\right)\cdot \mathbf{h},
\end{equation} where  $\mathbb{D} \mathbf{f}$ denotes the Jacobian matrix of $\mathbf{f}$ and the integral of a matrix is understood as a componentwise.
\end{lem}

\subsection{Proof of Theorem \ref{LFZ1}}

Defining $Z= 1-\Omega -{\Omega_1}- {\Omega_2}$ it follows 
from \begin{align*}
   & \dot{Z}=3 Z H \Big(\Omega_1 \cos (2 (\Phi_1-t  \omega_{1}))+ \Omega_2 \cos (2 ( \Phi_{2}-t  \omega_{2})) \nonumber \\
   & +\Omega +\Omega_1+ \Omega_{2}\Big)+\mathcal{O}\left(H^2\right),\end{align*}
that the sign of $1-\Omega -{\Omega_1}- {\Omega_2}$
 is invariant as $H\rightarrow 0$. 

From the equation  \eqref{expdotepsilon}
or its averaged version, it follows $H$ is a  monotonic decreasing function of  $t$, if $\Omega +{\Omega_1}+ {\Omega_2}<1$ due to $0\leq \Omega, {\Omega}_1, \Omega_2$. 
This allow to define recursively the sequences 
\begin{align}
   & \left\{\begin{array}{c}
       t_0=t_0   \\ \\
        H_0=H(t_0) 
    \end{array}\right., \;  \left\{\begin{array}{c}
       {t_{n+1}}= {t_{n}} +\frac{1}{H_n}   \\ \\
       H_{n+1}= H(t_{n+1})  
    \end{array}\right.,
\end{align}
such that  $\lim_{n\rightarrow \infty}H_n=0$ y $\lim_{n\rightarrow \infty}\eta_n=\infty$.

Given the expansions \eqref{Appquasilinear211}, equations 
\eqref{eqT602} become
\begin{widetext}
\begin{small}
\begin{align}
\label{g1-g5}
\left(\begin{array}{c}
	\frac{\partial g_1}{\partial t}\\ \frac{\partial g_2}{\partial t}\\ \frac{\partial g_3}{\partial t} \\ \frac{\partial g_4}{\partial t} \\  \frac{\partial g_5}{\partial t}	\\
	\end{array}\right)= 	
\left(
\begin{array}{c}
  6 (\Omega_{10}-1) \Omega_{10} \cos ^2(\Phi_{10}-t
    \omega_{1})+6  \Omega_{10} \Omega_{20} \cos
   ^2(\Phi_{20}-t \omega_{2})-3  \Omega_{10}
   (\Omega_{10}+\Omega_{20}-1) \\
 3  \Omega_{20} (\Omega_{10} \cos (2 (\Phi_{10}-t
   \omega_{1}))-\Omega_{20}+1)+6  (\Omega_{20}-1)
   \Omega_{20} \cos ^2(\Phi_{20}-t \omega_{2}) \\
	3 \Omega_{0} \Omega_{10} \cos (2 (\Phi_{10}-t
    \omega_{1}))+3  \Omega_{0} \Omega_{20} \cos (2
   (\Phi_{20}-t  \omega_{2}))\\
 \frac{3}{2}   \sin (2 (\Phi_{10}-t  \omega_{1})) \\
 \frac{3}{2}   \sin (2 (\Phi_{20}-t  \omega_{2}))  \\
\end{array}
\right),
\end{align}
\end{small}
with solution 
\begin{small}
\begin{align*}
& g_1(H, \Omega_{i0}, \Omega_{0},\Phi_{i0}, t)=
-\frac{3 (\Omega_{10}-1) \Omega_{10} \sin (2 (\Phi_{10}-t  \omega_{1}))}{2  \omega_{1}}-\frac{3 \Omega_{10} \Omega_{20} \sin (2 ( \Phi_{20}-t  \omega_{2}))}{2 \omega_{2}},\\
& g_2(H, \Omega_{i0}, \Omega_{0},\Phi_{i0}, t)= -\frac{3 \Omega_{10} \Omega_{20}
   \sin (2 (\Phi_{10}-t \omega_1))}{2  \omega_{1}}-\frac{3 (\Omega_{20}-1) \Omega_{20} \sin (2
   (\Phi_{20}-t \omega_2))}{2  \omega_{2}},
\\
& g_3(H, \Omega_{i0}, \Omega_{0},\Phi_{i0}, t)= -\frac{3 \Omega_{0} \Omega_{10}
   \sin (2 (\Phi_{10}-t \omega_1))}{2  \omega_{1}}-\frac{3 \Omega_{0} \Omega_{20} \sin (2 ( \Phi_{20}-t \omega_2))}{2 \omega_2},\\
     & g_4(H, \Omega_{i0}, \Omega_{0},\Phi_{i0}, t)=
\frac{3 \cos (2 ({\Phi}_{10}-t \omega_{1}))}{4 \omega_{1}}, \\
     & g_5(H, \Omega_{i0}, \Omega_{0},\Phi_{i0}, t)= 
\frac{3 \cos (2 ({\Phi}_{20}-t \omega_{2}))}{4 \omega_{2}}, 
\end{align*}
\end{small}
where we set five integration functions to zero. 

On the other hand,  substituting  \eqref{g1-g5} in 
equations  \eqref{EqY602} the resulting equations can be simplified to 
\begin{small}
\begin{subequations}
\label{NEWEqY602}
\begin{align}
&\frac{d\Omega_{10}}{d t}= 3 \Omega_{10} (\Omega_{0}+\Omega_{10}+\Omega_{20}-1) H \nonumber \\
& -\frac{9 \Omega_{10}  H^2}{8 (\omega_1 \omega_2)} \Bigg(2 \omega_2 ((\Omega_{10}-1) (\Omega_{0}+\Omega_{10}+\Omega_{20})+(5
   \Omega_{10}-3) \Omega_{20} \cos (2 (\Phi_{20}-t \omega_2))) \sin (2 (\Phi_{10}-t \omega_1)) \nonumber \\
	& +5 (\Omega_{10}-1) \Omega_{10} \omega_2 \sin (4 (\Phi_{10}-t \omega_1)) \nonumber \\
	& +2 \omega_1 \Omega_{20} (\Omega_{0}+\Omega_{10}+\Omega_{20}+(5 \Omega_{10}-2) \cos (2 (\Phi_{10}-t \omega_1))+5 \Omega_{20} \cos (2
   (\Phi_{20}-t \omega_2))) \sin (2 (\Phi_{20}-t \omega_2))\Bigg)+\mathcal{O}\left(H^3\right),
\\
&\frac{d\Omega_{20}}{d t}= 3 \Omega_{20} (\Omega_{0}+ \Omega_{10}+\Omega_{20}-1) H \nonumber \\
	& +\frac{9 \Omega_{20}  H^2}{8 \omega_1 \omega_2} \Bigg(-5 \omega_2 \sin (4 (\Phi_{10}-t \omega_1)) \Omega_{10}^2-2 \omega_2 (\Omega_{0}+\Omega_{10}+ \Omega_{20}) \sin (2 (\Phi_{10}-t \omega_1)) \Omega_{10} \nonumber \\
	& +(\omega_2 (2-5 \Omega_{20})+\omega_1 (5 \Omega_{20}-3)) \sin (2 (\Phi_{10}-\Phi_{20}+t ( \omega_{2}-\omega_1))) \Omega_{10} \nonumber \\
	& -(-3 \omega_1-2 \omega_2+5 (\omega_1+\omega_2) \Omega_{20}) \sin (2 (\Phi_{10}+\Phi_{20}-t (\omega_1+ \omega_{2}))) \Omega_{10} \nonumber \\
	& -2 \omega_1 (\Omega_{20}-1) (\Omega_{0}+\Omega_{10}+\Omega_{20}+5 \Omega_{20} \cos (2 (\Phi_{20}-t \omega_2))) \sin (2 (\Phi_{20}-t
   \omega_2))\Bigg)+\mathcal{O}\left(H^3\right),
\\
&\frac{d\Omega_{0}}{d t}=  3 \Omega_{0} (\Omega_{0}+\Omega_{10}+\Omega_{20}-1) H \nonumber \\
& -\frac{9 \Omega_{0} H^2}{4 (\omega_1 \omega_2)}\Bigg( ( \Omega_{0}+\Omega_{10}+\Omega_{20}+5 \Omega_{10} \cos (2 (\Phi_{10}-t \omega_1))+5 \Omega_{20} \cos (2 (\Phi_{20}-t \omega_2))) \nonumber \\
	& \times (\Omega_{10} \omega_2 \sin (2
   (\Phi_{10}-t \omega_1))+\omega_1 \Omega_{20} \sin (2 (\Phi_{20}-t \omega_2)))\Bigg)+\mathcal{O}\left(H^3\right),
\\
&\frac{d \Phi_{10}}{d t}=  \frac{9 H^2 \cos (2 (\Phi_{10}-t \omega_1)) (\Omega_{0}+\Omega_{10}+\Omega_{20}+(\Omega_{10}+2) \cos (2 (\Phi_{10}-t \omega_1))+\Omega_{20} \cos (2 (\Phi_{20}-t \omega_2)))}{8 \omega_1}+\mathcal{O}\left(H^3\right), \label{eqC13}\\
&\frac{d \Phi_{20}}{d t}=   \frac{9 H^2 \cos (2 (\Phi_{20}-t \omega_2)) (\Omega_{0}+\Omega_{10}+\Omega_{20}+\Omega_{10} \cos (2 (\Phi_{10}-t
   \omega_1))+(\Omega_{20}+2) \cos (2 (\Phi_{20}-t \omega_2)))}{8 \omega_2}+\mathcal{O}\left(H^3\right).  \label{eqC14}
\end{align}
\end{subequations}
\end{small}
Furthermore, eqs.
 \eqref{EqY602} become 
 \begin{small}
\begin{align}
& \dot{\Delta \Omega_{10}}= 3 H \left(\Omega_{10} (\Omega_{0}+\Omega_{10}+\Omega_{20}-1)-\bar{\Omega}_1 \left(\bar{\Omega }+\bar{\Omega}_1+\bar{\Omega}_2-1\right)\right) +\mathcal{O}\left(H^2\right),\\
& \dot{\Delta \Omega_{20}}=3 H
   \left(\Omega_{20} (\Omega_{0}+\Omega_{10}+\Omega_{20}-1)-\bar{\Omega}_2 \left(\bar{\Omega }+\bar{\Omega}_1+\bar{\Omega}_2-1\right)\right) +\mathcal{O}\left(H^2\right),\\
& \dot{\Delta \Omega_{0}}= 3 H
   \left(\Omega_{0} (\Omega_{0}+\Omega_{10}+\Omega_{20}-1)-\bar{\Omega } \left(\bar{\Omega }+\bar{\Omega}_1+\bar{\Omega}_2-1\right)\right) +\mathcal{O}\left(H^2\right).
\end{align}
\end{small}
\end{widetext}
Denoting $\mathbf{z}_0=(\Omega_{10}, \Omega_{20}, \Omega_0)^T$, $\bar{\mathbf{z}}=(\bar{\Omega}_{1}, \bar{\Omega}_{2}, \bar{\Omega})^T$ and $\Delta\mathbf{z}_0= \mathbf{z}_0 - \bar{\mathbf{z}}$ with \newline $0\leq |\Delta\mathbf{z}_0|:=\max \left\{|\Omega_{10}-\bar{\Omega}_{1}|, |\Omega_{20}-\bar{\Omega}_{2}|, |\Omega_{0}-\bar{\Omega}| \right\}< \infty$ in the closed interval $[t_n,t]$ equations \eqref{NEWEqY602} are reduced to a 3-dimensional system which can be written symbolically as:
\begin{align}
 & \dot{\Delta\mathbf{z}_0}= H \left(\bar{\mathbf{f}}( {\mathbf{z}}_0)-\bar{\mathbf{f}}(\bar{\mathbf{z}})\right) +  H^2 \mathbf{g} ^{[2]}({\mathbf{z}}_0, 
\bar{\mathbf{z}}, \Phi_{10}, \Phi_{20}), 
  \end{align}
plus equations \eqref{eqC13} and  \eqref{eqC14},  
where the vector function $\bar{\mathbf{f}}$ in eq. \eqref{EqY602}
is explicitly given by: 
\begin{align}\label{eqbarf} 
    &\bar{\mathbf{f}}(z_1, z_2, z_3) = \left(\begin{array}{c}  -3z_1 \left(1- z_1- z_2 -z_3\right) \\
 -3z_2 \left(1- z_1- z_2 -z_3 \right) \\
 -3z_3 \left(1- z_1- z_2 -z_3 \right)\\
\end{array}\right). 
\end{align}
The last two rows corresponding to $\frac{d \Phi_{10}}{d t}$ and $\frac{d \Phi_{20}}{d t}$ were omitted. 
The vector function \eqref{eqbarf}  with polynomial components in variables $(z_1, z_2, z_3)$ is continuously differentiable in all its components. 
The higher order terms 
$\mathbf{g} ^{[2]}({\mathbf{z}}_0, 
\bar{\mathbf{z}}, \Phi_{10}, \Phi_{20})$
are bounded in its components on the interval $[t_{n},t_{n+1}]$.

Using same initial conditions for $\mathbf{z}_0$ and $\bar{\mathbf{z}}$ we obtain by integration: 
\begin{align}
\label{int}
 & \Delta\mathbf{z}_0(t) = \int_{t_n}^t \dot{\Delta\mathbf{z}_0}(s) d s \nonumber \\
 & =  \int_{t_n}^t \left(H(s) \Big(\bar{\mathbf{f}}( {\mathbf{z}}_0(s))-\bar{\mathbf{f}}(\bar{\mathbf{z}}(s))\right) \nonumber \\
 & +  H^2(s) \mathbf{g} ^{[2]}({\mathbf{z}}_0(s), 
\bar{\mathbf{z}}(s), \Phi_{10}(s), \Phi_{20}(s))\Big) ds. 
\end{align}
Using Lemma \ref{lemma6} we have 
\begin{align}
   & \bar{\mathbf{f}}( {\mathbf{z}}_0(s))-\bar{\mathbf{f}}(\bar{\mathbf{z}}(s))=  {\mathbf{A}(s)}  \cdot \left({\mathbf{z}}_0(s) - \bar{\mathbf{z}}(s)\right),
\end{align} 
with 
\begin{align}
{\mathbf{A}(s)}  ={\left (\int_0^1 D   \bar{\mathbf{f}}\left(\bar{\mathbf{z}}(s)+ \eta \; \left({\mathbf{z}}_0(s) - \bar{\mathbf{z}}(s)\right)\right)\,d \eta\right)},
\end{align} where  $D\bar{\mathbf{f}}$ denotes the Jacobian matrix  of $\bar{\mathbf{f}}$, and the integral of a matrix is to be understood componentwise.
Omitting the dependence on $s$ we calculate   the components of  
\begin{equation}
 \mathbf{A}=   \left(
\begin{array}{ccc}
 a_{11} & a_{12} & a_{13} \\
 a_{21} & a_{22} & a_{23} \\
 a_{31} & a_{32} & a_{33} \\
\end{array}
\right),
\end{equation}
which are
\begin{small}
\begin{subequations}
\label{A-compts}
\begin{align}
  a_{11}=& \frac{3}{2} \left(\bar{\Omega }+\Omega _0+2 \Omega _{10}+\Omega _{20}+2 \bar{\Omega }_1+\bar{\Omega }_2-2\right), \\  
    a_{12}=& \frac{3}{2} \left(\Omega _{10}+\bar{\Omega }_1\right),\\
   a_{13}=& \frac{3}{2} \left(\Omega _{10}+\bar{\Omega }_1\right),\\
   a_{21}=& \frac{3}{2} \left(\Omega _{20}+\bar{\Omega }_2\right),\\
   a_{22}=& \frac{3}{2} \left(\Omega _{20}+\bar{\Omega }_2\right),\\
   a_{23}=& \frac{3}{2} \left(\Omega _{20}+\bar{\Omega }_2\right),
\\
  a_{31}=&  \frac{3}{2} \left(\bar{\Omega }+\Omega _0\right), \\
  a_{32}=&  \frac{3}{2} \left(\bar{\Omega }+\Omega _0\right), 
\\
  a_{33}= &  \frac{3}{2} \left(2 \bar{\Omega }+2 \Omega _0+\Omega _{10}+\Omega _{20}+\bar{\Omega }_1+\bar{\Omega }_2-2\right).
\end{align}
\end{subequations}
\end{small}
Taking   sup norm
$ |\mathbf{v}|=\max \left\{v_1, v_2, v_3 \right\}$ of a vector function and  the sup norm of a matrix
${|} \mathbf{A} {|}$ defined by $\max\{|a_{ij}|:  i=1,2,3, j=1,2,3\}$, where $a_{ij}$ are given in \eqref{A-compts} 
we have 
\begin{equation*}
    \Big{|} \mathbf{A}(s) \cdot \Delta\mathbf{z}_0(s) \Big{|}\leq 3 \Big{|} \mathbf{A}(s) \Big{|} \Big{|}\Delta\mathbf{z}_0(s)\Big{|}, \;  \forall s\in [t_n, t_{n+1}].
\end{equation*}
By continuity of polynomials $a_{i j} \left({\Omega}_{10}, {\Omega}_{20}, {\Omega}_0, \bar{\Omega}_1, \bar{\Omega}_2, \bar{\Omega}\right)$ given in \eqref{A-compts} and by continuity of  functions functions  ${\Omega}_{10}, {\Omega}_{20}, {\Omega}_0, {\Phi}_{10},  {\Phi}_{20}$, $\bar{\Omega}_1, \bar{\Omega}_2, \bar{\Omega}, \bar{\Phi}_1, \bar{\Phi}_2$ in $[t_n, t_{n+1}]$ the following finite constants are found:
\begin{align*}
&   L= 3 \max_{t\in[t_n,t_{n+1}]} \Big{|} \mathbf{A} (t)\Big{|},
\\
& M= \max_{t\in[t_{n},t_{n+1}]}   \Big{|}  \mathbf{g}^{[2]}({\mathbf{z}}_0(t), 
\bar{\mathbf{z}}(t), \Phi_{10}(t), \Phi_{20}(t))\Big{|},
\end{align*}
such that for all $t\in[t_n, t_{n+1}]$: 
\begin{align*}
 & \Big{|} \dot{\Delta\mathbf{z}_0} \Big{|} = \Bigg{|} \int_{t_n}^t \dot{\Delta\mathbf{z}_0} d s \Bigg{|} \\
 & = \Bigg{|} \int_{t_n}^t \Bigg( H \left(\bar{\mathbf{f}}( {\mathbf{z}}_0)-\bar{\mathbf{f}}(\bar{\mathbf{z}})\right) +  H^2 \mathbf{g} ^{[2]}({\mathbf{z}}_0, 
\bar{\mathbf{z}}, \Phi_{10}, \Phi_{20})\Bigg) ds  \Bigg{|}\nonumber \\
 & \leq H_n \int_{t_n}^t \Big{|}  \bar{\mathbf{f}}( {\mathbf{z}}_0)-\bar{\mathbf{f}}(\bar{\mathbf{z}}) \Big{|} ds +   M H_n^2 (t-t_n) \nonumber \\
 & \leq H_n \int_{t_n}^t  \Big{|} \mathbf{A}(s) \cdot \Delta\mathbf{z}_0(s) \Big{|} ds  +   M H_n^2 (t-t_n)\nonumber \\
& \leq L H_n \int_{t_n}^t  \Big{|}
 \Delta\mathbf{z}_0(s) \Big{|} ds +   M H_n^2 (t-t_n) \nonumber \\
 & \leq  L H_n \int_{t_n}^t  \Big{|}  \Delta\mathbf{z}_0(s) \Big{|} ds +   M H_n, 
\end{align*}
due to $t-t_n\leq {t_{n+1}}- {t_{n}} =\frac{1}{H_n}$.

Using   Gronwall's Lemma \ref{Gronwall}, we have for $t \in[t_n, t_{n+1}]$: 
\begin{align*}
 & \Big{|} \Delta \mathbf{z}_0(t)  \Big{|} \leq   M  H_n   e^{L  H_n(t-t_n)} \leq    M  {H_n}e^{L},
 \end{align*} 
 due to $t-t_n\leq {t_{n+1}}- {t_{n}} =\frac{1}{H_n}$.
 Then, 
 \begin{small}
 \begin{align*}
& \Big{|} \Delta \Omega_{10}(t) \Big{|} \leq    M_1 e^{L_1} {H_n}, \\ & \Big{|} \Delta \Omega_{20}(t) \Big{|} \leq     M_1 e^{L_1} {H_n}, \\ & \Big{|} \Delta \Omega_{0}(t) \Big{|} \leq   M_1 e^{L_1} {H_n}.
\end{align*}
\end{small}
Combining the previous results, we find constants $N_1$ and $N_2$ such that 
 \begin{align}
    & \Big{|} \Delta \Phi_i(t) \Big{|} \leq  \bigints_{t_n}^t  \Bigg|   \frac{d \Delta\Phi_{i0}}{d s}(s) \Bigg|  ds   +\mathcal{O}\left({H_n}^3\right)
    \nonumber\\
    & \leq N_i H_n^2 (t-t_n) +\mathcal{O}\left({H_n}^3\right), i=1,2, 
\\
  &   \Big{|} \Delta \Phi_i(t) \Big{|} \leq  N_i H_n^2 (t_{n+1}-t_n)  +\mathcal{O}\left({H_n}^3\right) \nonumber \\
     & =  N_i H_n +\mathcal{O}\left({H_n}^3\right).
\end{align}

Finally, taking the limit as $n\rightarrow \infty$, we obtain $H_n\rightarrow  0$. Then,
as $H_n \rightarrow 0$, it follows
\\ {{\textbf{Result 1:} The functions $\Omega_{10}, \Omega_{20}, \Omega_{0}$ and  $\bar{\Omega}_1, \bar{\Omega}_2, \bar{\Omega}$  have the same limit as $\tau\rightarrow \infty$.}} 
\\ {{\textbf{Result 2:} The functions $\Phi_{i}$ and  $\bar{\Phi}_i$,  $i=1,2$ have the same asymptotic behavior as $\tau\rightarrow \infty$.}} $\qed$

\subsection{Alternative proof}

An alternative approach is to define $x=\Omega_0, y=\Omega_{10}+\Omega_{20}+\Omega_0$,  $\bar{x}=\bar{\Omega}, \bar{y}=\bar{\Omega}_{1}+\bar{\Omega}_{2}+\bar{\Omega}$, $\Delta x= x-\bar{x}$, $\Delta y=y-\bar{y}$ and $\Delta \Phi_{i0}(t)=\Phi_{i0}(t)-\bar{\Phi}_i(t), i=1,2$,  and taking the same initial conditions at $t=t_n$,  $x(t_n)=\bar{x}(t_n)$,  $y(t_n)=\bar{y}(t_n)$,  and $\Phi_{i}(t_n)=\bar{\Phi}_i(t_n)=\bar{\Phi}_i$. Hence
\begin{align}
&  \frac{d \Delta x}{d t}=-3 H\left[x(1-y) -\bar{x} \left(1- \bar{y}\right)\right], \\
& \frac{d \Delta y}{d t} =-3 H\left[(y -\bar{y})(1-y - \bar{y})\right].
\end{align}
 Let $t \in[t_n, t_{n+1}]$ such that $t- t_n \leq {t_{n+1}}- {t_{n}}=\frac{1}{H_n}$. 
Then,  we obtain: 
\begin{small}
 \begin{align}
    &\displaystyle{\Big{|}\Delta x(t)\Big{|}=\Big{|}\int_{t_n}^{t}[x^{\prime}(s)-\bar{x}^{\prime}(s)]ds \Big{|}} \nonumber \\
    & = \Big{|}\int_{t_n}^{t} \Big[{3H(s) \left[ x(s) (1-y(s))- \bar{x}(s) (1-\bar{y}(s)) \right]}+\mathcal{O}\left(H_n ^2\right)\Big] ds \Big{|} \nonumber\\
    & \leq 3\int_{t_n}^{t} \Big[\underbrace{H_n \max_{s\in [t_n,t_{n+1}]}\left\{\Big{|}1- y(s)\Big{|} ,\Big{|}1- \bar{y}(s)\Big{|}\right\}}_{|\cdot|\leq H_n L_1}  \Big{|} \Delta x(s) \Big{|}+\Big{|}\mathcal{O}\left(H_n ^2\right)\Big{|}\Big] ds \nonumber\\
    & \leq  3 H_n L_1 \int_{t_n}^{t} \Big{|} \Delta x(s) \Big{|} ds  + K_1 H_n ^2 (t-t_n) \nonumber \\
    & =3 H_n L_1 \int_{t_n}^{t} \Big{|} \Delta x(s) \Big{|} ds  + K_1 H_n.
\end{align}
\end{small}
Applying Gronwall's Lemma \ref{Gronwall}, it follows 
\begin{align}
 & \Big{|} \Delta x(t) \Big{|} \leq   K_1 H_n   e^{3H_n L_1 (t-t_n)} \leq K_1 H_n   e^{3  L_1 }.
\end{align} for all $t \in[t_n, t_{n+1}]$.
In a similar way we obtain: 
\begin{small}
 \begin{align}
    &\displaystyle{\Big{|}\Delta y(t)\Big{|}=\Big{|}\int_{t_n}^{t}[y^{\prime}(s)-\bar{y}^{\prime}(s)]ds \Big{|}}\nonumber\\
    & = \Big{|}\int_{t_n}^{t} \Big[{3H(s) \left(1-y(s)-  \bar{y}(s)\right)  \Delta y(s)}+\mathcal{O}\left(H_n ^2\right)\Big] ds \Big{|} ,\nonumber\\
     & \leq 3\int_{t_n}^{t} \Big[\underbrace{H_n \max_{s\in [t_n,t_{n+1}]}\left\{\Big{|}1-y(s)-  \bar{y}(s)\Big{|}\right\} }_{|\cdot|\leq H_n L_2}  \Big{|} \Delta y(s) \Big{|}+\Big{|}\mathcal{O}\left(H_n ^2\right)\Big{|}\Big] ds \nonumber\\
    & \leq  3 H_n L_2 \int_{t_n}^{t} \Big{|} \Delta y(s) \Big{|} ds  + K_2 H_n ^2 (t-t_n) \nonumber \\
    & \leq 3 H_n L_2 \int_{t_n}^{t} \Big{|} \Delta y(s) \Big{|} ds  + K_2 H_n,
\end{align}
\end{small}
for all  $t \in[t_n, t_{n+1}]$. 

Applying Gronwall's Lemma \ref{Gronwall}, it follows
\begin{align*}
 & \Big{|} \Delta y(t) \Big{|} \leq   K_2 H_n   e^{3H_n L_2 (t-t_n)} \leq K_2 H_n   e^{3  L_2 }.
\end{align*} for all $t \in[t_n, t_{n+1}]$ such that $t- t_n \leq {t_{n+1}}- {t_{n}}=\frac{1}{H_n}$. Then 
\begin{small}
 \begin{align*}
  & \Big{|} \Delta x(t) \Big{|}   \leq  K_1 {H_n}, \\ &  \Big{|} \Delta \Omega_2(t) \Big{|}  \leq  \Big{|} \Delta y(t) \Big{|}   \leq  K_2 {H_n}, \\
  & \Big{|} \Delta \Omega_1(t) \Big{|}     \leq  K_3 {H_n},\\
    &  \Big{|} \Delta \Phi_i(t) \Big{|} \leq  M_i H_n^2 (t_{n+1}-t_n)  +\mathcal{O}\left({H_n}^3\right)  =  M_i H_n +\mathcal{O}\left({H_n}^3\right).
\end{align*}
\end{small}
Finally, taking the limit as $n\rightarrow \infty$, we obtain $H_n\rightarrow  0$ and {\textbf{Result 1}} and {\textbf{Result 2}}  follow. $\qed$

\begin{figure*}
    \centering
    \subfigure[\label{fig:Averagedflow3D} Projections in the space $(\Omega,H,\Omega_{t})$, where $\Omega_{t}=\Omega_{1}+\Omega_{2}+\Omega$.]{\includegraphics[scale = 0.45]{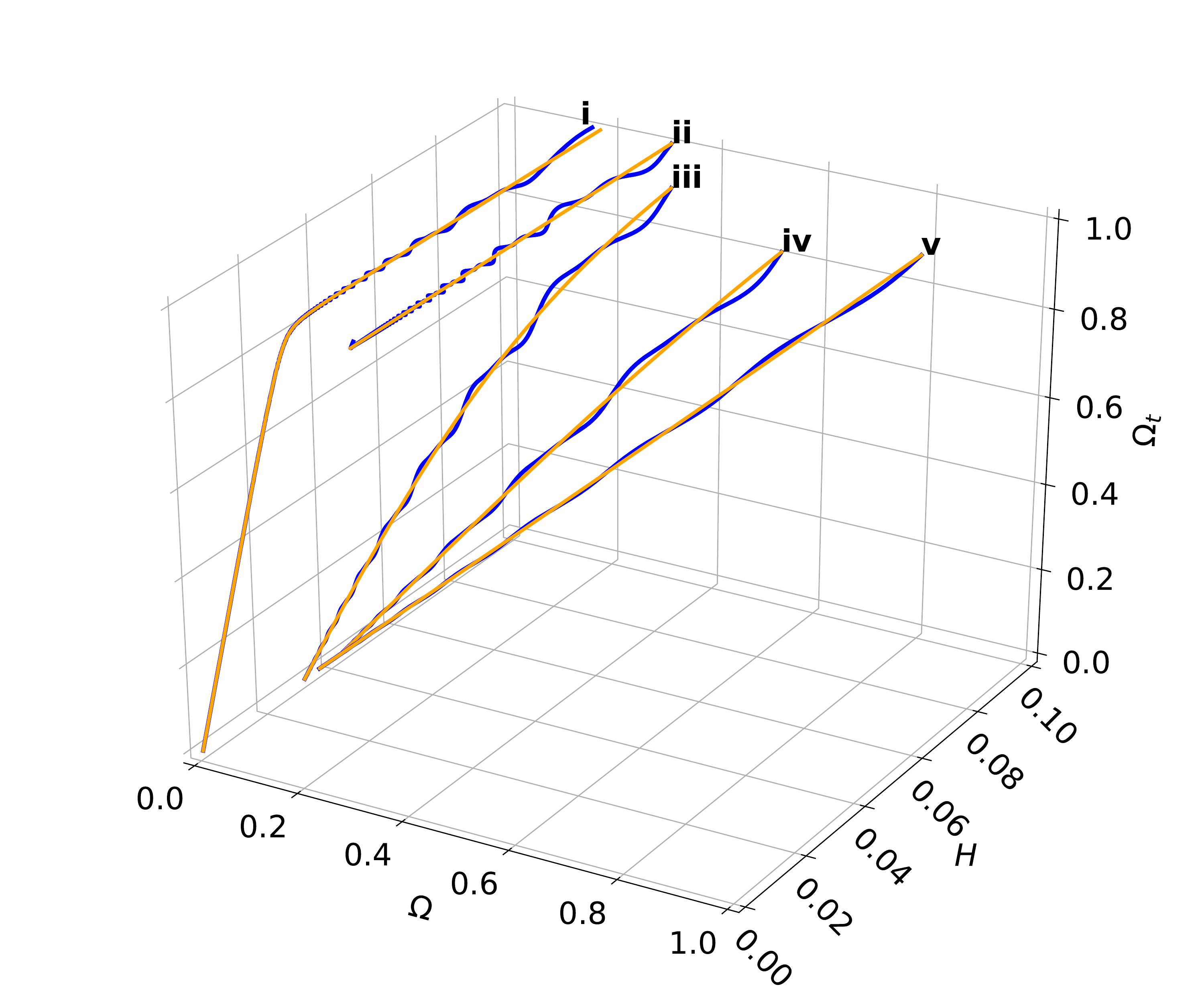}}
    \subfigure[\label{fig:Averagedflow2D} Projections in the space $(\Omega,\Omega_{t})$, where $\Omega_{t}=\Omega_{1}+\Omega_{2}+\Omega$. The plot on the right represent a zoomed region of the plot on the left ]{\includegraphics[scale = 0.65]{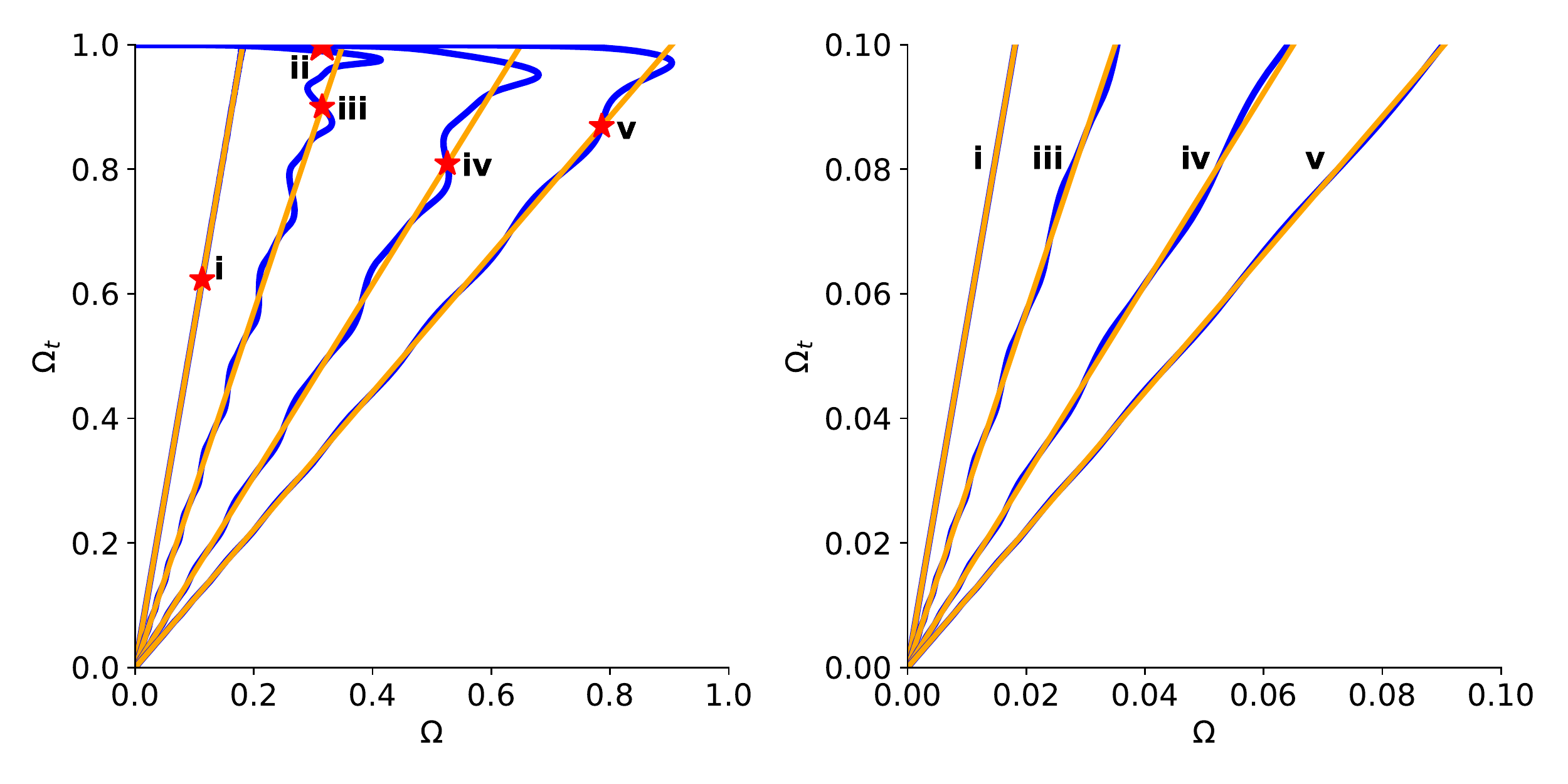}}
    \caption{Some solutions of the truncated system \eqref{truncatedsystem} (blue) and time-averaging system \eqref{eq93} (orange) for the fixed values of $\omega_{1}=\sqrt{2}$ and $\omega_{2}=\sqrt{2}/2$. We use as initial condition for both system the five data sets presented in Table \ref{TableCI}. These plots are numerical evidence that the main Theorem \ref{LFZ1} is fulfilled. That is, the solution of the truncated system follow the track of the solutions of the time-averaged system and the oscillations experimented by the truncated system are smoothed out as $H\rightarrow 0$.}
    \label{fig:Averagedflow}
\end{figure*}

\section{Numerical procedure}
\label{NUmerical}
In this section, we present the numerical evidence that supports the main Theorem \ref{LFZ1} presented in section \ref{SECT:II}. For this purpose, an algorithm in the programming language \textit{Python} was implemented for solving numerically the systems of differential equations obtained for the model under study. The systems of differential equations were solved using the \textit{solve\_ivp} code provided by the \textit{SciPy} open-source \textit{Python}-based ecosystem. The integration method used was \textit{Radau} that is an implicit Runge-Kutta method of the Radau IIa family of order $5$  with a relative and absolute tolerances of $10^{-4}$ and $10^{-7}$, respectively. 
In particular, we integrate the  perturbed system truncated at second order in $H$:
   \begingroup\makeatletter\def\f@size{7}\check@mathfonts
 \begin{equation}
\label{truncatedsystem}
\left\{\begin{array}{cc}
 \frac{dH}{d \tau} = -H \left[3 {\Omega_1} \cos ^2(t {\omega_1}-{\Phi_1})+3 {\Omega_2} \cos ^2(t {\omega_2}-{\Phi_2})+\frac{3 \Omega }{2}\right], \; 
 \frac{d t}{d \tau}  = 1/H,  \\\\
\frac{d\Omega_1}{d \tau} = 3 {\Omega_1} \left(2 ({\Omega_1}-1) \cos ^2(t \omega_{1}-\Phi_1)+2 {\Omega_2} \cos ^2(t \omega_{2}-\Phi_2)+\Omega \right), \\\\
\frac{d\Omega_2}{d \tau} =3  {\Omega_2} \left(2 {\Omega_1} \cos ^2(t \omega_{1}-\Phi_1)+2 ({\Omega_2}-1) \cos ^2(t \omega_{2}-\Phi_2)+\Omega \right), \\\\
\frac{d\Omega}{d \tau} =3 \Omega  (\Omega +{\Omega_1}+{\Omega_2}+{\Omega_1} \cos (2 (t \omega_{1}-\Phi_1))+{\Omega_2} \cos (2 (t \omega_{2}-\Phi_2))-1),\\\\
\frac{d\Phi_1}{d \tau} =-\frac{3}{2}\sin (2 (t \omega_{1}-\Phi_1)),
\;
\frac{d\Phi_2}{d \tau} =-\frac{3}{2}\sin (2 (t \omega_{2}-\Phi_2)).
\end{array}
\right. 
\end{equation} 
\endgroup
and the averaged system:
   \begingroup\makeatletter\def\f@size{9}\check@mathfonts
\begin{equation}
\left\{\begin{array}{cc}
     \frac{d H}{d \tau}=  -\frac{3}{2}H \left(\bar{\Omega}_1+ \bar{\Omega}_2 +\bar{\Omega} \right), \;  
     \frac{dt}{d \tau} = 1/H, \\\\
      \frac{d\bar{\Omega}_1}{d \tau}= -3 \bar{\Omega}_1 \left(1- \bar{\Omega}_1- \bar{\Omega}_2 -\bar{\Omega}\right), \\\\
\frac{d\bar{\Omega}_2}{d \tau}= -3 \bar{\Omega}_2 \left(1- \bar{\Omega}_1- \bar{\Omega}_2 -\bar{\Omega}\right), \\\\
\frac{d\bar{\Omega}}{d \tau}= -3 \bar{\Omega} \left(1- \bar{\Omega}_1- \bar{\Omega}_2 -\bar{\Omega}\right), \\\\
     \frac{d{{\bar{\Phi}_1}}}{d \tau}=0, \;   \frac{d{{\bar{\Phi}_1}}}{d \tau}=0.  
\end{array}\right. \label{eq93} 
\end{equation}
\endgroup
The system \eqref{truncatedsystem} was integrated in the interval  $-40\leq\tau\leq 10$, and the system \eqref{eq93}  was integrated in the interval $-40\leq\tau\leq 40$, partitioned in $10000$ data points. The values of $\omega_{1}=\sqrt{2}$ and $\omega_{2}=\sqrt{2}/2$, and the initial conditions presented in Table \ref{TableCI}, were considered.
   \begin{table}[h!]
\caption{\label{TableCI} Five initial data sets for the simulation of the truncated system \eqref{truncatedsystem} and time-averaged system \eqref{eq93}. All the conditions are chosen in order to fulfill the inequality $\bar{\Omega}_{1}+\bar{\Omega}_{2}+\bar{\Omega}\leq 1$.}
\footnotesize\setlength{\tabcolsep}{4.5pt}
    \begin{tabular}{lcccccccc}\hline
Sol.  & \multicolumn{1}{c}{$H(0)$} & \multicolumn{1}{c}{$\bar{\Omega}_{1}(0)$} & \multicolumn{1}{c}{$\bar{\Omega}_{2}(0)$} & \multicolumn{1}{c}{$\bar{\Omega}(0)$} & \multicolumn{1}{c}{$\bar{\Phi}_{1}(0)$} & \multicolumn{1}{c}{$\bar{\Phi}_{2}(0)$}  & \multicolumn{1}{c}{$t(0)$}  \\\hline
        i & $0.001$ & $0.255$ & $0.255$ & $0.113$ & $0$ &  $0$ & $0$ \\
        ii & $0.1$ & $0.424$ & $0.261$ & $0.315$ & $0$ &  $0$ & $0$ \\
        iii & $0.1$ & $0.243$ & $0.342$ & $0.315$ & $0$ &  $0$ & $0$ \\
        iv & $0.1$ & $0.105$ & $0.178$ & $0.526$ & $0$ & $0$ & $0$ \\
        v & $0.1$ & $0.005$ & $0.078$ & $0.786$ & $0$ & $0$ & $0$ \\ \hline
    \end{tabular}
\end{table}

In figure \ref{fig:Averagedflow} are presented the numerical results of the integration of the truncated system \eqref{truncatedsystem} (blue lines) and time-averaged system \eqref{eq93} (orange lines),  for the initial conditions presented in the Table \ref{TableCI}. In figure \ref{fig:Averagedflow3D} are presented the solutions in the projection $(\Omega,H,\Omega_{t})$ while in figure \ref{fig:Averagedflow2D} are presented the solution in the projection $(\Omega,\Omega_{t})$, where $\Omega_{t}=\Omega_{1}+\Omega_{2}+\Omega$. These figures represent the numerical evidence that the main Theorem \ref{LFZ1}, presented in section \ref{SECT:II}, is fulfilled. That is, the solution of the truncated system follows the track of the solutions of the time-averaged system, and the oscillations experimented by the truncated system are smoothed out as $H\rightarrow 0$, having both systems of differential equations the same late-time dynamics. 
On average, $H$ is decreasing from $H=\infty $ as $\tau\rightarrow -\infty$ to $H=H_0\sqrt{1-\bar{\Omega}_{t}(0)}$ as $\tau\rightarrow +\infty$.  That is, as far as $\bar{\Omega}_{t}(0)<1$ the universe ends on a de Sitter inflationary phase. Then, depending on initial conditions, both the non-oscillating curve and the oscillating one tends to the same constant value of $H$, and it is related to a de Sitter solution. Recall, each local minimum $\phi^{*}=(\phi_1^*, \phi_2^*)$ with non zero minimum value of $V(\phi_1, \phi_2)$, satisfying eqs. \eqref{maxima-minima}, corresponds to de Sitter solutions $a\propto \exp\left[-\sqrt{\frac{V(\phi_1^*, \phi_2^*)}{3}} t\right]$. 
On the other hand, given that $\bar{\Omega}_{t}(0)=1$  is an invariant set, then $\bar{\Omega}=\bar{\Omega}(0), \bar{\Omega}=\bar{\Omega}(0),  \bar{\Omega}=\bar{\Omega}(0)$ remains constant with $\bar{\Omega}_{t}(0)=1$. Finally, if we chose initial conditions for a small enough value of $3H^2(t_0)$ such that the orbit is trapped by the basin of attraction of the global zero minimum, we obtain a Minkowski solution.

\end{document}